\documentclass[3p,times]{elsarticle}

\usepackage{ecrc}
\usepackage{dcolumn}
\usepackage{float}
\usepackage{amssymb}
\usepackage{booktabs}
\usepackage{graphicx}      
\usepackage{natbib}        
\usepackage{array}

\volume{00}

\firstpage{1}

\journalname{Nuclear Physics A}

\runauth{Tabassum Naz et al.}

\jnltitlelogo{Nuclear Physics A}

\CopyrightLine{2015}{Published by Elsevier Ltd.}

\usepackage{amssymb}

\begin{document}

\begin{frontmatter}




\title {Microscopic description of structural evolution in Pd, Xe, Ba, Nd, Sm, Gd and Dy isotopes\\}

\author{Tabassum Naz$^{1,*}$} \ead{tabasumnaz321@gmail.com}
\author{G. H. Bhat$^{2,\dag}$} \ead{gwhr.bhat@gmail.com}
\author{S. Jehangir$^{2,\dag}$} \ead{jehangir@nitsri.net}
\author{Shakeb Ahmad$^{3,+}$} \ead{physics.sh@gmail.com(Corresponding author)}
\author{J. A. Sheikh$^{4,++}$} \ead{sjaphysics@gmail.com}

\address{$^1$Department of Physics, Aligarh
Muslim University, Aligarh, UP - 202 002, India}
\address{$^2$Department of Physics, S P college, Cluster University Srinagar, 190001,  India.}
\address{$^3$Physics Section, Women's College, 
Aligarh Muslim University, Aligarh- 202002, India.}
\address{$^4$Cluster University Srinagar, Jammu and Kashmir, 190 001, India}

\begin{abstract}
Aiming to understand the role of triaxiality and the evolution of the ground state nuclear shapes, 
we have carried out a microscopic study for a series of 
chains of Pd, Xe, Ba, Nd, Sm, Gd, and Dy isotopes.  This is done within the 
self-consistent Relativistic-Hartree-Bogoliubov (RHB) formalism, and supported by 
the Triaxial Projected Shell Model (TPSM) approach. Pairing interaction separable 
in the momentum space with DD-ME2 force parameter is used to generate the potential energy 
surfaces(PESs) under the axial and triaxial symmetry. Shape evolution manifest 
themself in very clear manner in almost all the isotopic chains.
Properties of the global mimima have been found to be in good agreement with 
the available experimental data.
Relatively flat PESs, and $\gamma$-soft nature, have been suggested $^{108,110}$Pd, 
$^{132,134}$Xe and $^{134}$Ba
as good candidates for E(5) symmetry, while $^{102}$Pd is not found suitable for E(5) symmetry.
The PESs with a bump, and rigidity against triaxial variable($\gamma$) suggested 
$^{150}$Nd, $^{152}$Sm and $^{154}$Gd to be good candidates while 
$^{150}$Sm and $^{156}$Dy are poor candidates of X(5) critical-point symmetry.  
The findings of the present RHB calculations supported by TPSM are qualitatively 
in good agreement with the experimental and other theoretical calculations.
\end{abstract}

\begin{keyword}
\PACS 25.70.-z \sep 25.70.Gh \sep 25.70.Jj \sep 25.70.Mn

\end{keyword}

\end{frontmatter}

\section{INTRODUCTION}
Atomic nucleus is one of the most remarkable quantum many-body systems depicting a rich variety of 
shapes (geometric configurations) within an isotopic/isotonic chain. The single-nucleon shell 
structure can be dramatically altered with the addition or subtraction of a few nucleons. In some cases, 
it can also lead to shape transitions within the isotopic/isotonic chain. The transition may occur
either from spherical to $\gamma$-unstable deformed or from spherical to axially deformed. 
Understanding the shape, and its modifications near the critical point of the shape phase transition
is one of the topical issues in nuclear structure studies. The nuclei around the critical point 
of the phase transition are characterized by a certain dynamical symmetry. To understand the 
manifestation of nuclear phase transition and its corresponding critical point dynamical symmetry, 
many theoretical as well as experimental studies have been done~\cite{iachello}-\cite{lm}.
These studies were aimed to study the structural evolution along with the possible existence of the two 
well known dynamical symmetries [E(5) and X(5)] at the critical points of shape phase transitions. 
The E(5) and X(5) symmetries have been introduced in Refs.~\cite{iachello,iachello1} in the 
framework of the Bohr Hamiltonian. 
E(5) does correspond to the second order shape phase transition seen in the Interacting Boson Model (IBM)
between the U(5) and O(6) symmetries, and X(5) does correspond to the first order shape phase transition 
seen in the IBM between the U(5) and SU(3) symmetries, but the terminology has to differ. 
Schematically, there are two symmetry triangles: one for the Bohr Hamiltonian, on which E(5) 
and X(5) appear, and one for the IBM, on which the first and second order transitions appear.
The analysis of the structural evolution based on the potential energy surface(PES) leads to the 
prediction of these symmetries. At the critical point of shape phase transition, the PES of the particular
nucleus is expected to be flat-bottom. If the PES of a particular nucleus at the critical point of shape
transition may be described by an infinite square well in $\beta$-variable, independent of the collective
$\gamma$-variable then, it is supposed to be a possible candidate for E(5).  
For X(5), it is assumed that the PES with a bump, and rigidity against triaxial variable($\gamma$)
may be described by the sum of an infinite square well and harmonic oscillator. 

It is expected and observed that the neutron-rich isotopes of the Pd, Xe, Ba, Nd, Sm, Gd and Dy 
are located within an interesting region of the nuclear chart. 
Furthermore, many interesting structural variation are expected which are sensitive to the number of nucleons.
Studies aiming to the structural evolution of Pd(Z=46), Xe(Z=54) and 
Ba(Z=56) isotopes have predicted the shape transition from spherical to deformed nuclei. 
These studies have also predicted the possible candidates for the critical point 
symmetry~\cite{Casten:2000zz,Arias:2001re,Zamfir:2002dk,Clark:2004xb,mw,uk}\cite{Zhang:2002zu}-\cite{RodriguezGuzman:2007tq}\cite{Li:2010qu}.
The study, done by Casten and Zamfir~\cite{Casten:2000zz}, has shown the shape transition from spherical to deformed
system and proposed $^{134}$Ba as the candidate to exhibit the E(5) characteristics.  
Although, till date, the absolute transition probabilities are not available for a full comparison 
with the calculations. None of the other nuclei have been found to show a better depiction of such 
symmetry and hence $^{134}$Ba is still considered to be the ideal candidate as supported by 
other studies~\cite{Arias:2001re}. Shape transition has also been observed in Pd isotopes, and  
$^{102}$Pd~\cite{Zamfir:2002dk,Clark:2004xb}, $^{108}$Pd~\cite{Zhang:2002zu,Fossion:2006xg,RodriguezGuzman:2007tq}
are proposed to be possible candidates for E(5) symmetry.
On the basis of a systematic analysis done on the energy level data, B(E2) transition rates~\cite{Clark:2004xb,mw}, 
and measurements of E1 and M1 strengths of $^{124-136}$Xe~\cite{uk}, the shape phase transition around
$A\approx$130 has been demonstrated. These studies along with others have suggested 
$^{128,130,132}$Xe~\cite{Clark:2004xb,mw,uk,dlz,Fossion:2006xg,RodriguezGuzman:2007tq,Li:2010qu} 
as the possible candidates exhibiting E(5) character. 
It is known that the isotopes of rare-earth region around N = 90 show transitional properties. 
The isotopes of Nd, Sm, Gd, and Dy are observed to lie in an ideal region of the nuclear chart
for the study of shape transition from spherical nuclei at the closed neutron shell at N=82 to deformed 
nuclei.
Further study has shown that N=90 isotones are the best candidates for X(5) critical-point symmetry~\cite{Clark:2004xb}. 
First candidates to display X(5) symmetry were $^{150}$Nd~\cite{r.kr}, $^{152}$Sm~\cite{Casten:2001zz,Gupta:2017fyl}. 
This was further supported by other studies done for $^{152}$Sm~\cite{ca4,za,cl,rb} and for $^{150}$Nd~\cite{ca4,cl,dlz0}.
Other possible candidates for X(5) symmetry, amongst N=90 isotones are $^{154}$Gd~\cite{dt,dewald} and $^{156}$Dy~\cite{mac,dewald}. 
It is predicted that, the $^{156}$Dy has more $\gamma$-soft nature than any other X(5) 
candidate, yet shows many features of a typical X(5) nucleus~(\cite{dewald,Moller:2006yx} and the references therein).

The purpose of this paper is to investigate the occurrence of the nuclear phase transition, 
and to search the possible nuclei corresponding to the critical point symmetry.
It is known that within the mean-field approach the study of nuclear shape evolution with the 
number of nucleons is usually done through the potential 
energy surfaces  (PESs).   
In the present calculations, we have tried to search for the 
examples of nuclei near the critical-point of the nuclear phase transition based on the PESs. This is done within the 
Relativistic-Hartree-Bogoliubov (RHB) formalism, supported by the Triaxial Projected Shell Model (TPSM) 
approach. These studies have been carried out for $^{96-114}$Pd, $^{128-140}$Xe, $^{126-142}$Ba, $^{142-156}$Nd,  
$^{144-158}$Sm, $^{146-158}$Gd, and $^{148-160}$Dy isotopes.
Within the isotopic chain we search for the shape transition, and around the critical point for the one with relatively flat 
PES or PES with a bump as one of the candidates for the critical point symmetry. Further, the 
triaxial calculation is done to investigate the behaviour of the triaxial parameter $\gamma$ for those isotopes 
where the existence of any of the above symmetries is expected. 
In RHB approach, we have used the density-dependent DD-ME2~\cite{46} parameter set. It provides a successful 
description of ground state properties~\cite{49}-\cite{52} over all the nuclear chart. 
In order to investigate the high-spin behaviour of the nuclei near the
critical points, Triaxial projected shell model (TPSM) approach has been employed.
This manuscript is organized as follows. In section~\ref{theory} a
general overview of the theoretical
formalism is presented. The numerical results of the calculations
are discussed and compared in section~\ref{results}. 
Summary and conclusions are in section~\ref{conclusion}.
\section{Theoretical Approaches}
\label{model}
This present work concerns the microscopic description of the axial and triaxial shapes
along with the corresponding ground state properties of neutron-rich Pd, Xe, Ba, Nd, Sm, Gd 
and Dy isotopes. This has been done within the RHB formalism
with density-dependent finite range meson-exchange model. Further study, designed to 
explore the rotational properties of these systems, is obtained by using the TPSM approach.
\subsection{The Meson-exchange Model}
\label{theory}
In the present calculations the density-dependent finite
range meson-exchange model (DD-ME)~\cite{46}\cite{47}-\cite{61} within the 
Relativistic-Hartree-Bogoliubov (RHB) formalism is used. 
The DD-ME model has been used earlier very successfully and have provided 
an excellent predictions of different ground states and excited state properties
throughout the entire periodic table of nuclei~\cite{46}-\cite{52}\cite{sha}-\cite{67}. 
The present investigation uses the very successful, density-dependent
meson-exchange DD-ME2~\cite{46} parameter set.
The pairing correlation is taken care within a pairing interaction separable in 
momentum space. For the details of the calculations 
see Refs.~\cite{sha}-\cite{63}\cite{68}-\cite{85}.
The potential energy surface (PES) calculation is done
by imposing constraints on both axial and triaxial mass quadrupole moments. 
It is performed by the method of quadratic constrained~\cite{81}
by using an unrestricted variation of the function
\begin{equation}
{\langle{\hat{H}}\rangle}+\sum_{\mu=0,2}{C_{2\mu}}{({\langle{{\hat{Q}_{2\mu}}}
\rangle}-q_{2\mu})}^{2}
\label{eq:W}
\end{equation}
where $\langle{\hat{H}}\rangle$ is the total energy,
${(\langle{\hat{Q_{2\mu}}}\rangle}$ denotes the
expectation values of mass quadrupole operators,
\begin{equation}
\hat{Q}_{20}=2{z}^2-{x}^2-{y}^2 \;\;\; {\mbox{and}} \;\;\; \hat{Q}_{22}={x}^2-{y}^2
\label{eq:X}
\end{equation}
$q_{2\mu}$ is the constrained value of the
multipole moment, and $C_{2\mu}$ is the corresponding stiffness constant.
Moreover, the quadratic constraint adds an extra force term
$\sum_{\mu=0,2}{\lambda_\mu}{{\hat{Q}_{2\mu}}}$ to the system, where
\begin{equation}
\lambda_\mu=2{C_{2\mu}}{({\langle{{\hat{Q}_{2\mu}}}\rangle}-q_{2\mu})}
\end{equation}
for a self consistent solution. This term is necessary to force the
system to a point in deformation space different from a stationary point.
The convergence of the numerical calculation is taken care properly in 
terms of the optimum numbers of oscillator quanta for fermions and bosons.
\subsection{Triaxial Projected Shell Model Approach}
\label{TPSM}
It has been demonstrated recently that multi-quasiparticle triaxial
projected shell model (TPSM) approach provides a coherent
description of yrast, $\gamma$ and multi-quasiparticle band structures
in transitional nuclei~\cite{Js16}-\cite{YK00}. In this approach, three dimensional projection
technique is employed to project out the good angular-momentum states from
triaxially deformed Nilsson + BCS basis. Shell model Hamiltonian is subsequently
diagonalized using these angular-momentum projected basis states~\cite{JY01}-\cite{JG09}. 
As in the earlier PSM calculations, we use the pairing plus
quadrupole-quadrupole Hamiltonian \cite{KY95}
\begin{equation}
\hat H = \hat H_0 - {1 \over 2} \chi \sum_\mu \hat Q^\dagger_\mu
\hat Q^{}_\mu - G_M \hat P^\dagger \hat P - G_Q \sum_\mu \hat
P^\dagger_\mu\hat P^{}_\mu,
\label{hamham}
\end{equation}
Where $\chi$ is the interaction strength of the QQ-force. The monopole pairing
strength $G_M$ is of the standard form
\begin{equation}
G_M = {{G_1 - G_2{{N-Z}\over A}}\over A} ~{\rm for~neutrons,}~~~~
G_M = {G_1 \over A} ~{\rm for~protons.} \label{pairing}
\end{equation}

In the present work, we consider $G_1=20.12$ and $G_2=13.13$,
which approximately reproduce the observed odd-even mass difference.
and this choice of $G_M$ is appropriate for the
single-particle space employed in the model, where three major
shells are used for each type of nucleons ($N=4,5,6$ ($N=3,4,5$) and $3,4,5 (2,3,4)$ for
neutrons (protons) for A $\sim 160$ and A $\sim 130$ regions respectively). The quadrupole pairing strength $G_Q$ is
assumed to be proportional to $G_M$, and the proportionality
constant being set equal to 0.16~\cite{Js16}-\cite{YK00}.
\section{Results and Discussion}
\label{results}
\subsection{ Relativistic-Hartree-Bogoliubov(RHB) with density-dependent finite
range meson-exchange model}
In this section, we present the microscopic description of Pd(Z=46), Xe(Z=54), Ba(Z=56),
Nd(Z=60), Sm(Z=62), Gd(Z=64), and Dy(Z=66) isotopic chains. 
We have performed the constrained calculations to obtain the axial as well
as the triaxial potential energy surfaces(PESs). 
The effective interaction used is density-dependent DD-ME2.
\subsubsection{Axial Symmetry}
\begin{figure}
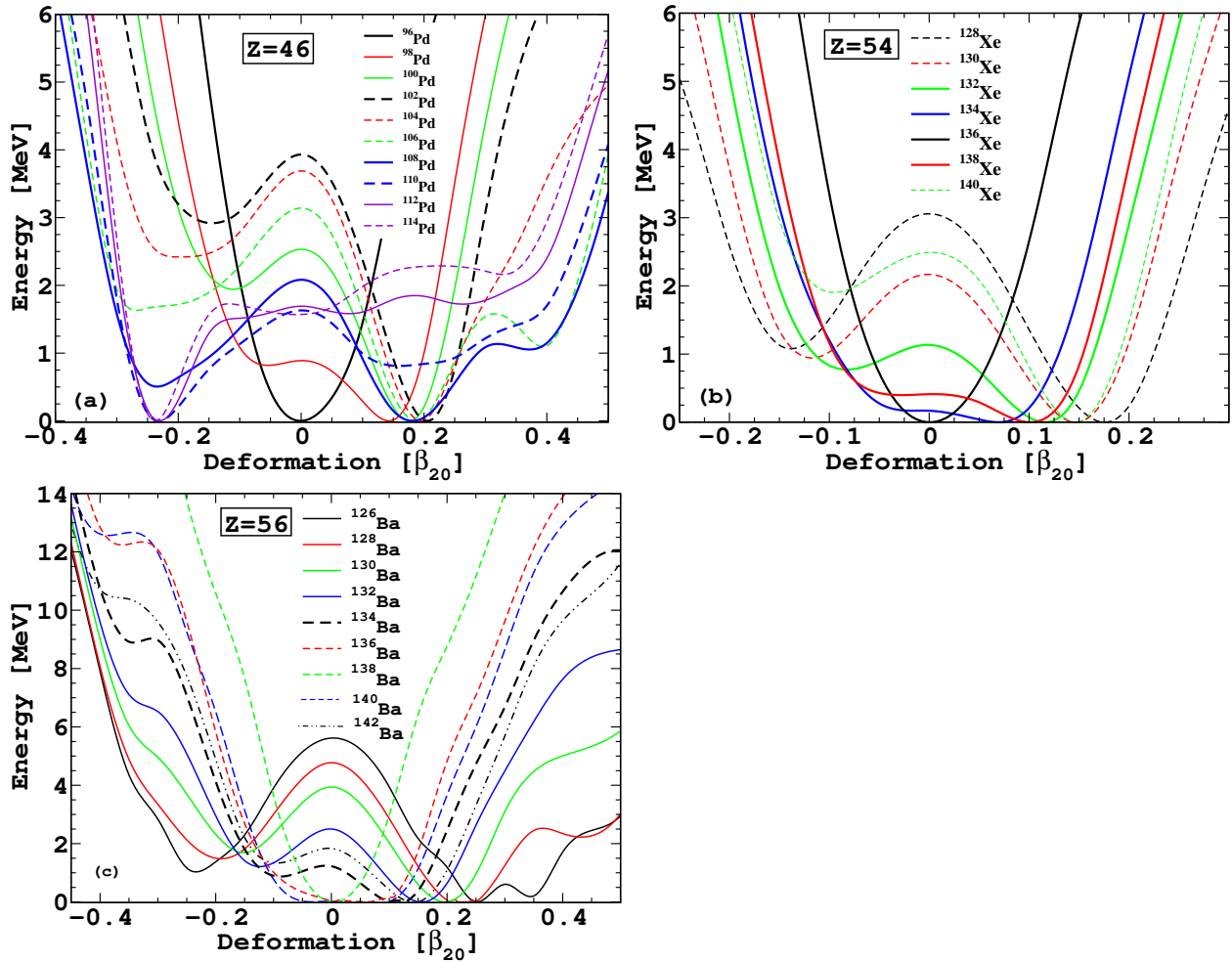

\includegraphics[scale=0.3]{pes_pd.eps}
\includegraphics[scale=0.3]{pes_Xe.eps}
\includegraphics[scale=0.3]{pes_ba.eps}
\caption{\label{fig:figure1}(Color online) Potential energy surfaces(PESs) for (a) $^{96-114}$Pd, (b) $^{128-140}$Xe, 
(c) and $^{126-142}$Ba, calculated using the RHB theory with the DD-ME2 force. Thick lines corresponds to the
possible isotopes that have been suggested to show E(5) critical-point symmetry.}
\end{figure}

In Fig.~\ref{fig:figure1}, we display the PESs of $^{96-114}$Pd, $^{128-140}$Xe, and $^{126-142}$Ba    
as a function of the quadrupole deformation $\beta_2$.
From these figures, one can observe the shape transition from the spherical $^{96}$Pd(N=50)
to the $\gamma$-unstable(prolate) $^{108}$Pd(N=62), then to the $\gamma$-unstable(oblate) $^{110-114}$Pd(N=62-68) isotopes.
Similar shape phase transition can be seen from Xe isotopes. But, in this case the shape transition can be seen for either side of 
the spherical $^{136}$Xe(N=82) to the $\gamma$-unstable(prolate) $^{128}$Xe(N=62) and towards neutron increasing $^{140}$Xe(N=86).
The same behaviour is there in case of Ba isotopes, from spherical $^{138}$Ba(N=82)
to the $\gamma$-unstable(prolate) $^{126}$Ba(N=70).
In Fig.~\ref{fig:figure1}(a), we can see the shape coexistence in almost all the isotopes of Pd(Z=46) with an energy difference
between prolate and oblate around 0.5MeV(minimum) for $^{108}$Pd, and about 3MeV(maximum) for $^{102}$Pd.  
Just at the critical point of shape transition from prolate ($^{108}$Pd) to oblate ($^{110}$Pd), the PESs of 
$^{108}$Pd and $^{110-114}$Pd are quite flat among all the Pd isotopes. 
In Fig.~\ref{fig:figure1}(b), at $^{132,134}$Xe and $^{138}$Xe, the PESs has a flat bottom ($\leq$0.75MeV energy difference), 
and transition occurs from deformed to spherical and from spherical to deformed respectively. 
The PES of $^{134,138}$Xe is more flat than $^{132}$Xe.
In case of Ba isotopes, shown in Fig.~\ref{fig:figure1}(c),
the shape coexistence is found in all isotopes except in $^{138}$Ba. 
A relative flat PES is found in $^{134}$Ba, covering $0.1\leq|\beta_2|\leq 0.15$. 

\begin{figure}
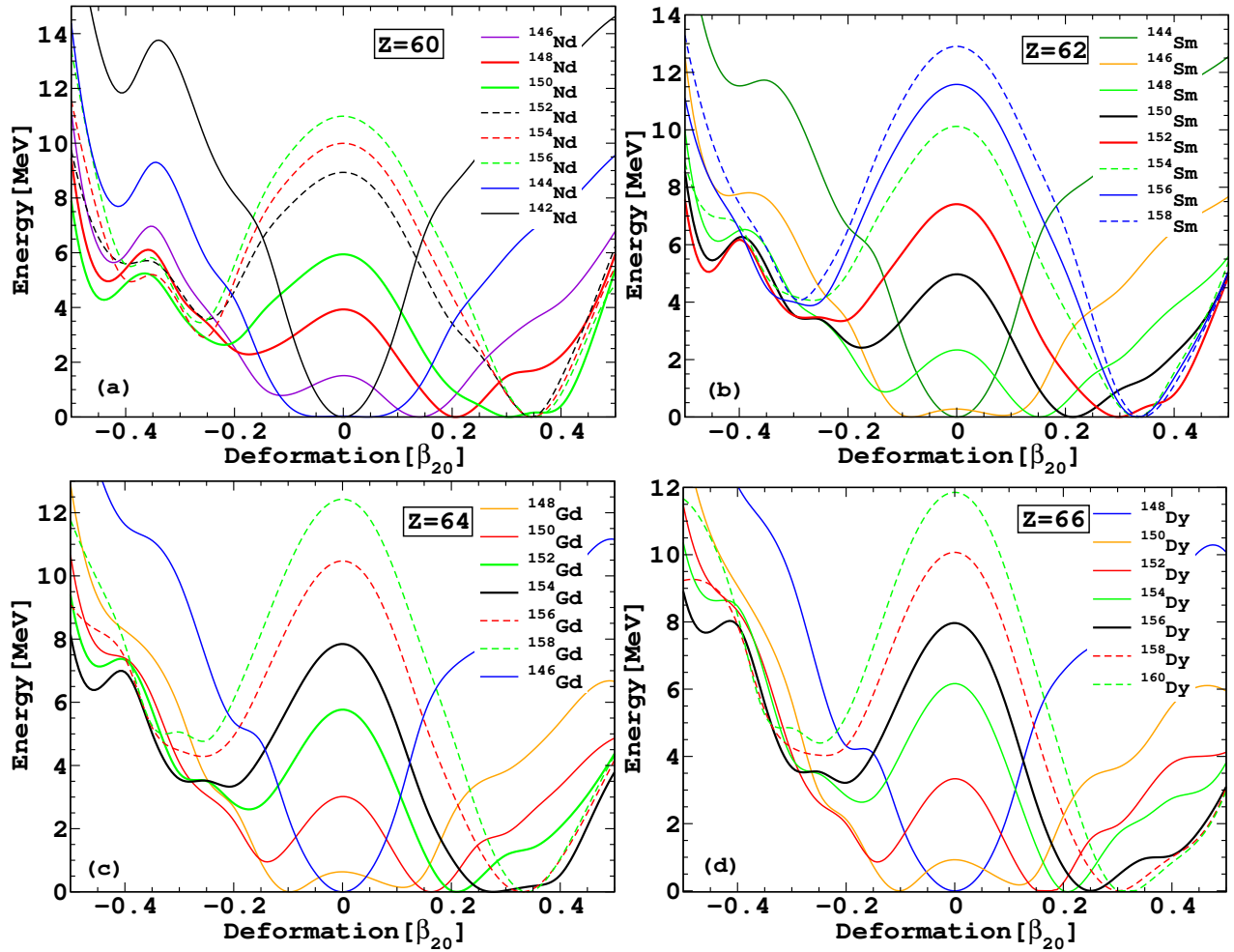

\includegraphics[scale=0.3]{pes_nd.eps}
\includegraphics[scale=0.3]{pes_sm.eps}
\includegraphics[scale=0.3]{pes_gd.eps}
\includegraphics[scale=0.3]{pes_dy.eps}
\caption{\label{fig:figure3}(Color online) Potential energy surfaces(PESs) for (a) $^{142-156}$Nd, (b) $^{144-158}$Sm, 
(c) $^{146-158}$Gd, and (d) $^{148-160}$Dy, calculated using the RHB theory with the DD-ME2 force. Thick lines corresponds to the
possible isotopes that have been suggested to show X(5) critical-point symmetry.}
\end{figure}

Now, we present our results for the isotopic chains of some rare-earth nuclei, Nd(Z=60), Sm(Z=62), Gd(Z=64), and Dy(Z=66).
The PESs for $^{142-156}$Nd, $^{144-158}$Sm, $^{146-158}$Gd, and $^{148-160}$Dy isotopes
are shown in Fig.~\ref{fig:figure3}.
In Fig.~\ref{fig:figure3}(a), the $^{142}$Nd is spherical 
and as the neutron number increases, shape transition towards a well deformed prolate minimum can be seen.
The isotopes $^{152-156}$Nd show a well-deformed prolate minimum. The nucleus $^{144}$Nd is 
having a flat minimum within $-0.1\leq|\beta_2|\leq 0.1$. Shape coexistence with a small potential barrier of $\sim$1.5MeV
and excitation energy of $\sim$0.70MeV can be seen in $^{146}$Nd nucleus.
The isotopes $^{148}$Nd(N=88) and $^{150}$Nd(N=90) both lie in the transition region from spherical shape to
a well-deformed shape. 
But, $^{150}$Nd is exhibiting a rather flat minimum than $^{148}$Nd in the prolate regime, and shallower minimum
in the oblate regime.
$^{148}$Nd has a deep prolate minimum and a shallow oblate minimum with 2.3MeV excitation energy,
and, the energy barrier of 4MeV. 
However, $^{150}$Nd has a rather flat potential energy surface on the prolate side $0.2\leq\beta_2\leq 0.4$,
and shallow oblate minimum at the excitation energies 2.6MeV with 6MeV of the energy barrier.
Similar transitional behavior can be seen in PESs for $^{144-158}$Sm isotopes (Z=62) shown in Fig.~\ref{fig:figure3}(b).
We have transition from spherical $^{144}$Sm to clear prolate shape $^{154-158}$Sm isotopes. The isotopes $^{146,148}$Sm
is showing the shape coexistence, and the transitional behavior appears for $^{150}$Sm(N=88) and $^{152}$Sm(N=90).
For Gd(Z=64) and Dy(Z=66) isotopes, PESs are shown in Figs.~\ref{fig:figure3}(c) and (d), respectively, the same is true.
The isotopes $^{154}$Gd(N=90) and $^{156}$Dy(N=90) show a flat minimum in $\beta_2>$0 regime and shallower minimum
in $\beta_2<$0 regime being in between the transition from spherical to well-deformed prolate shape.
The observation of flat potential energy surfaces within these isotopic chains, motivated us to explore their
triaxial character.
\subsubsection{Triaxial Symmetry}
\begin{figure}
\centering
\includegraphics[scale=0.28]{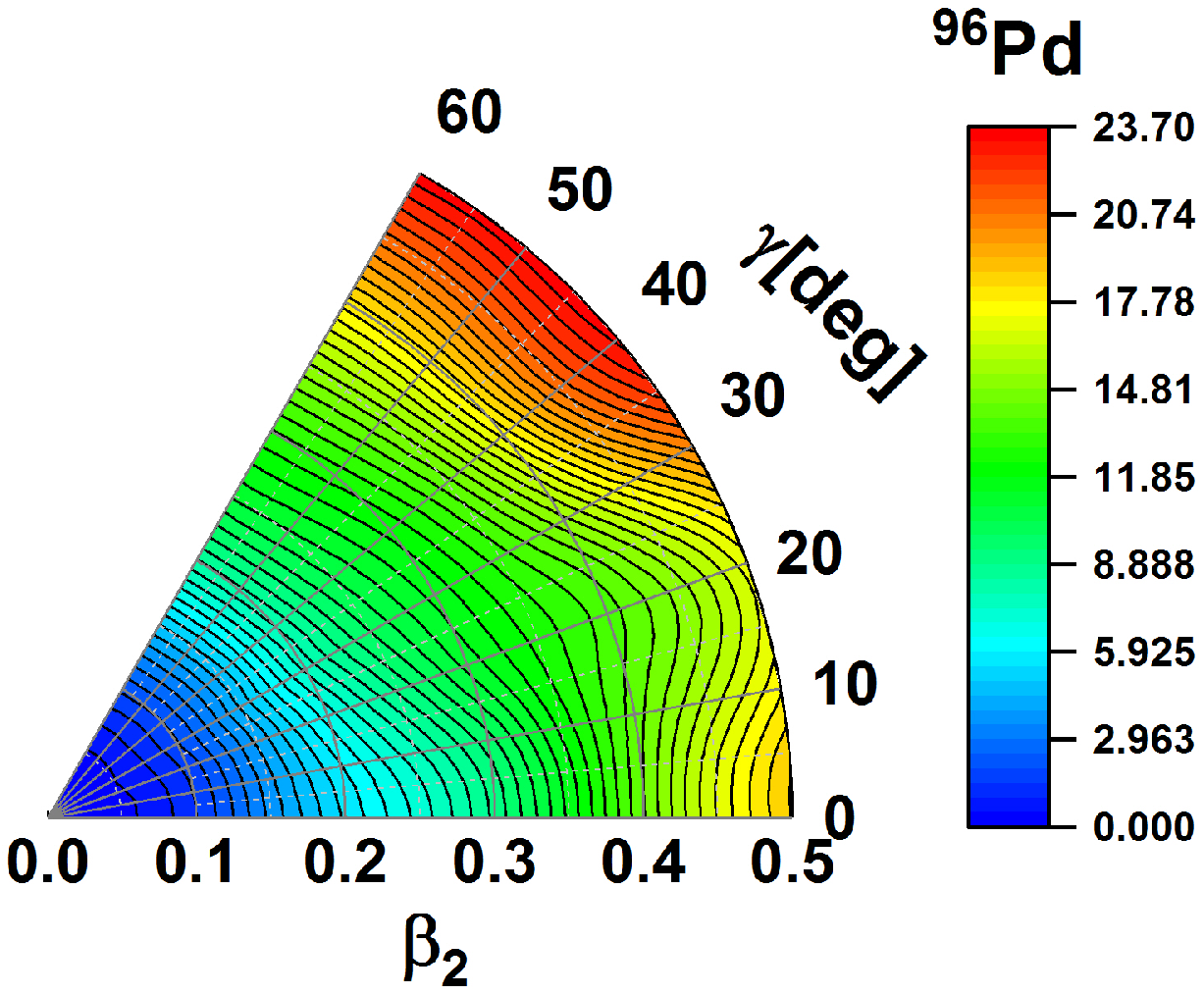}
\includegraphics[scale=0.28]{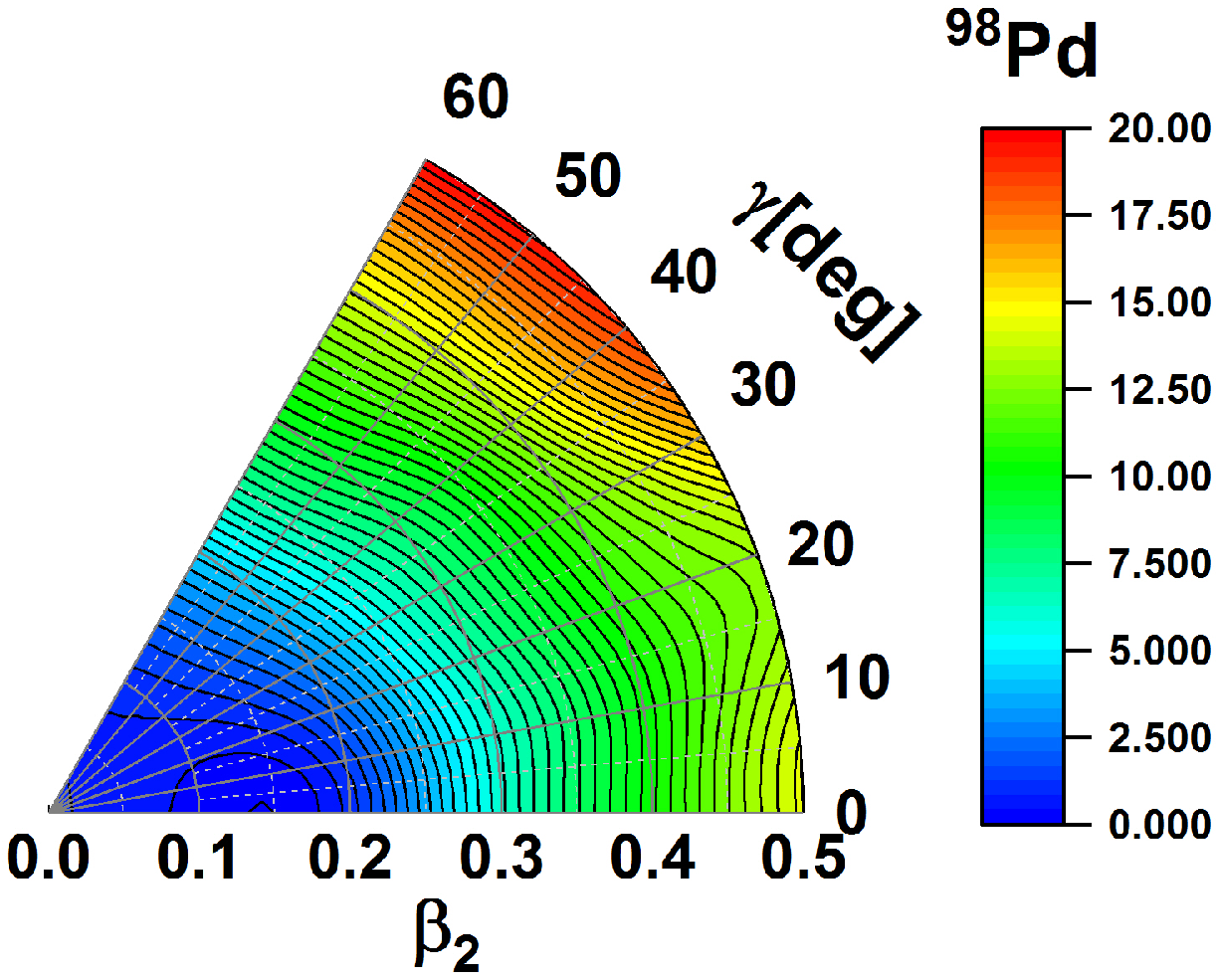}
\includegraphics[scale=0.28]{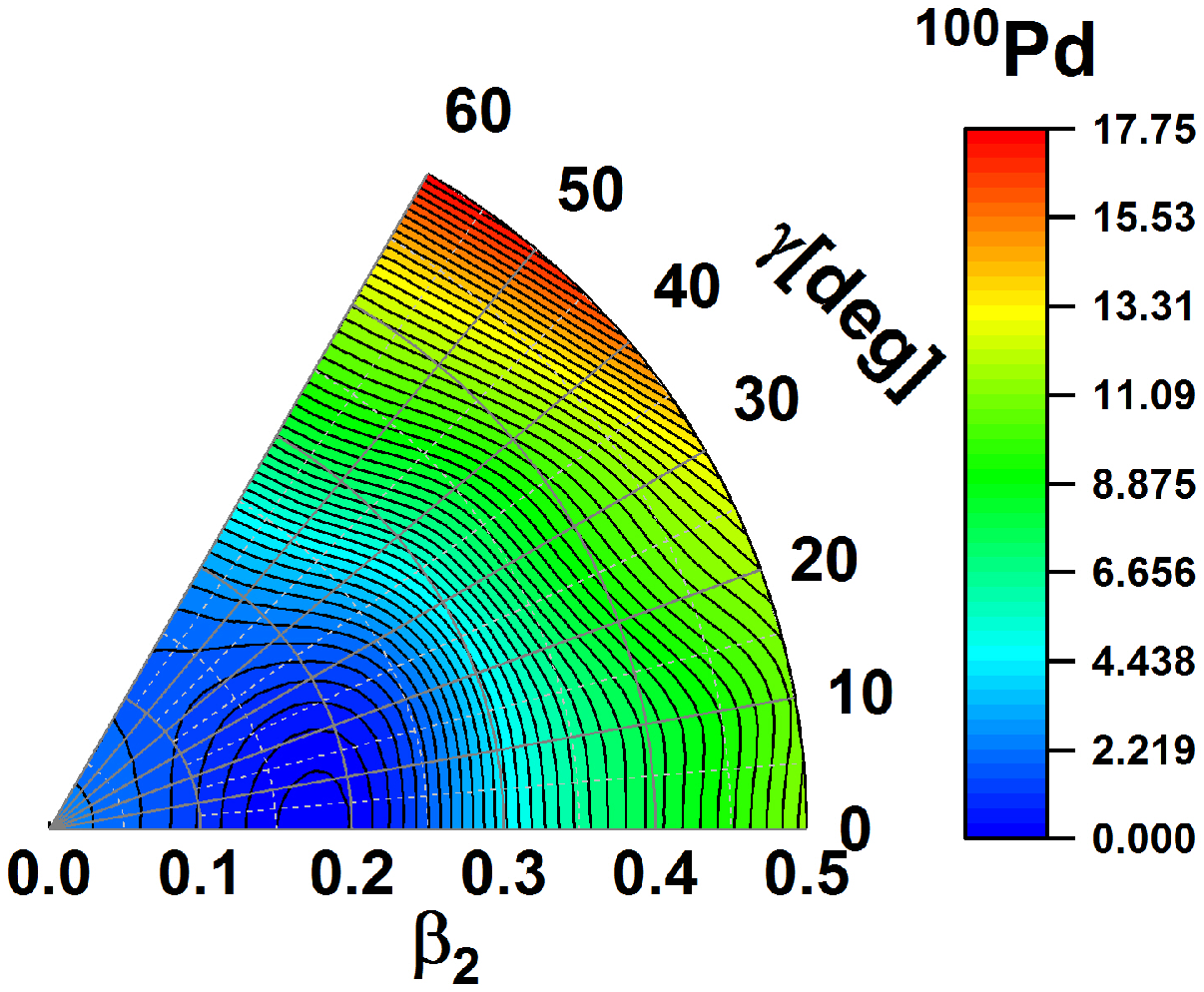}
\includegraphics[scale=0.28]{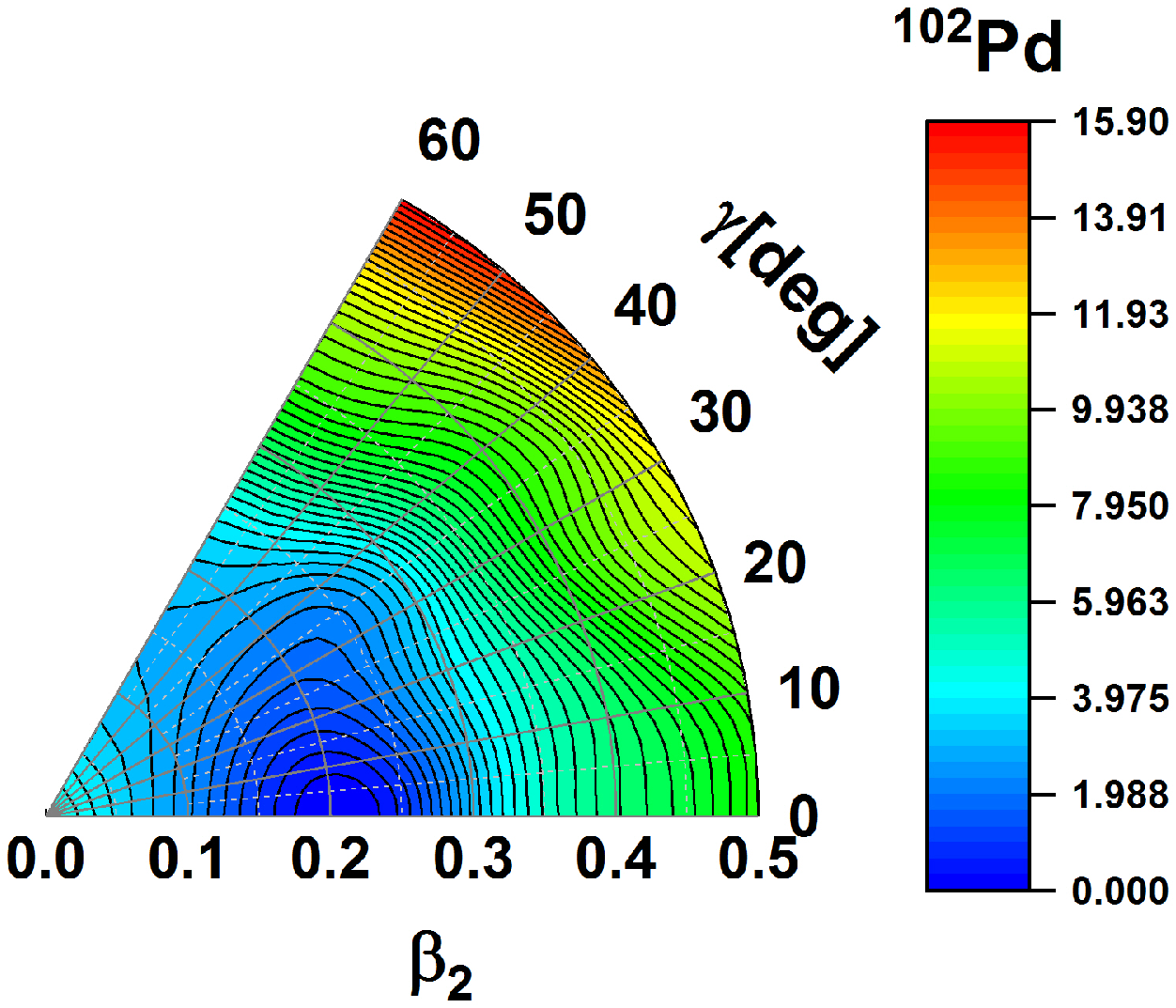}
\includegraphics[scale=0.28]{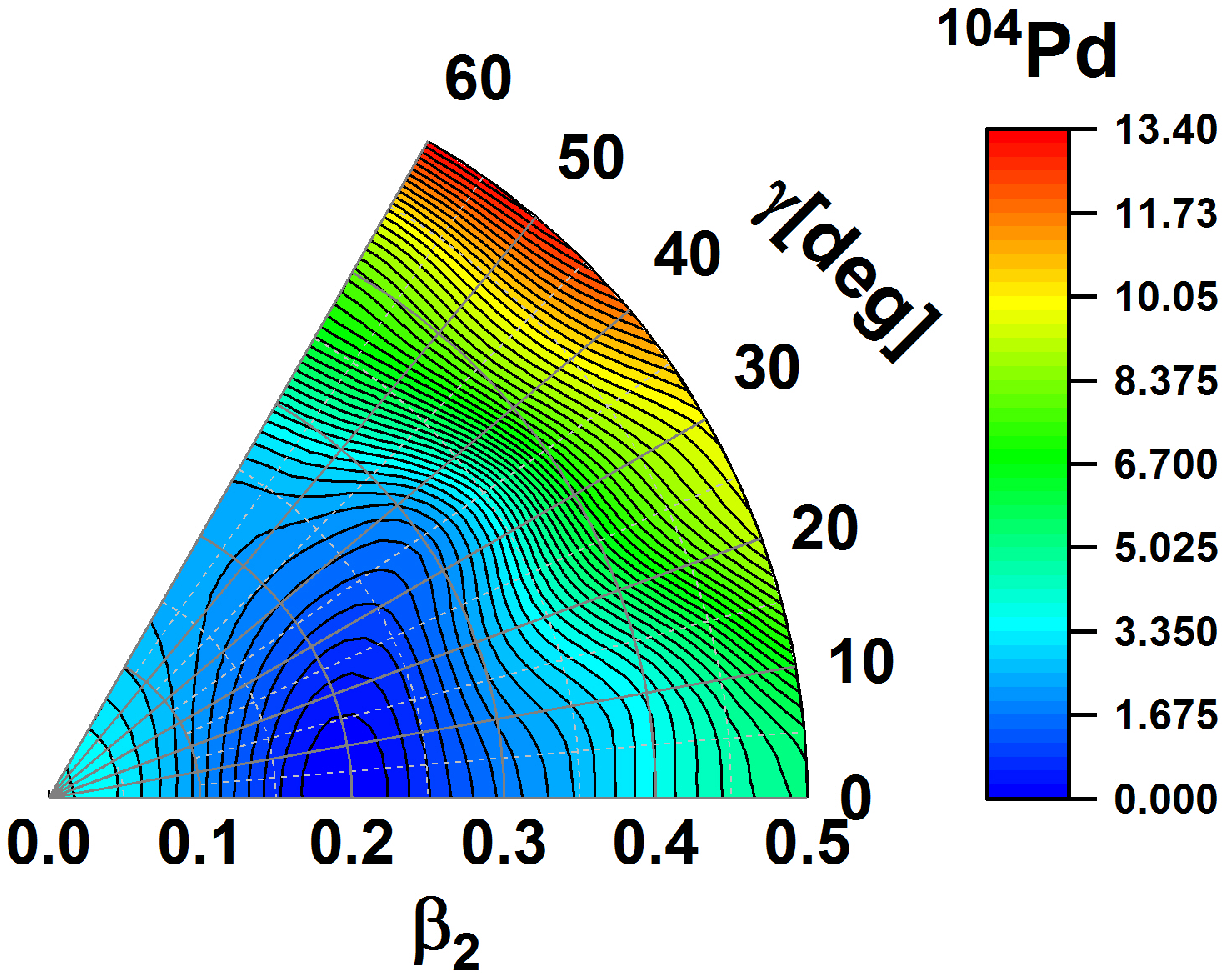}
\includegraphics[scale=0.28]{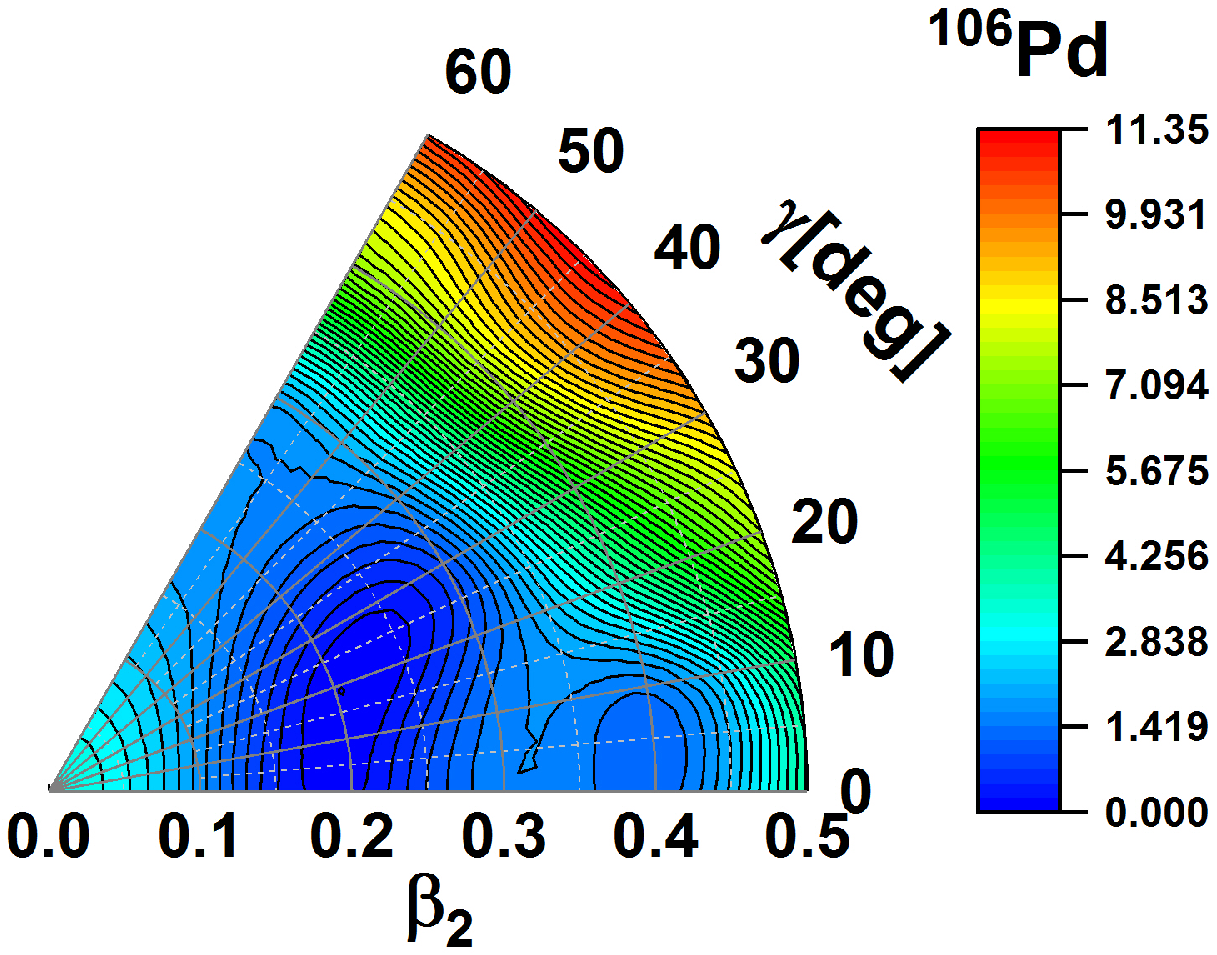}
\includegraphics[scale=0.28]{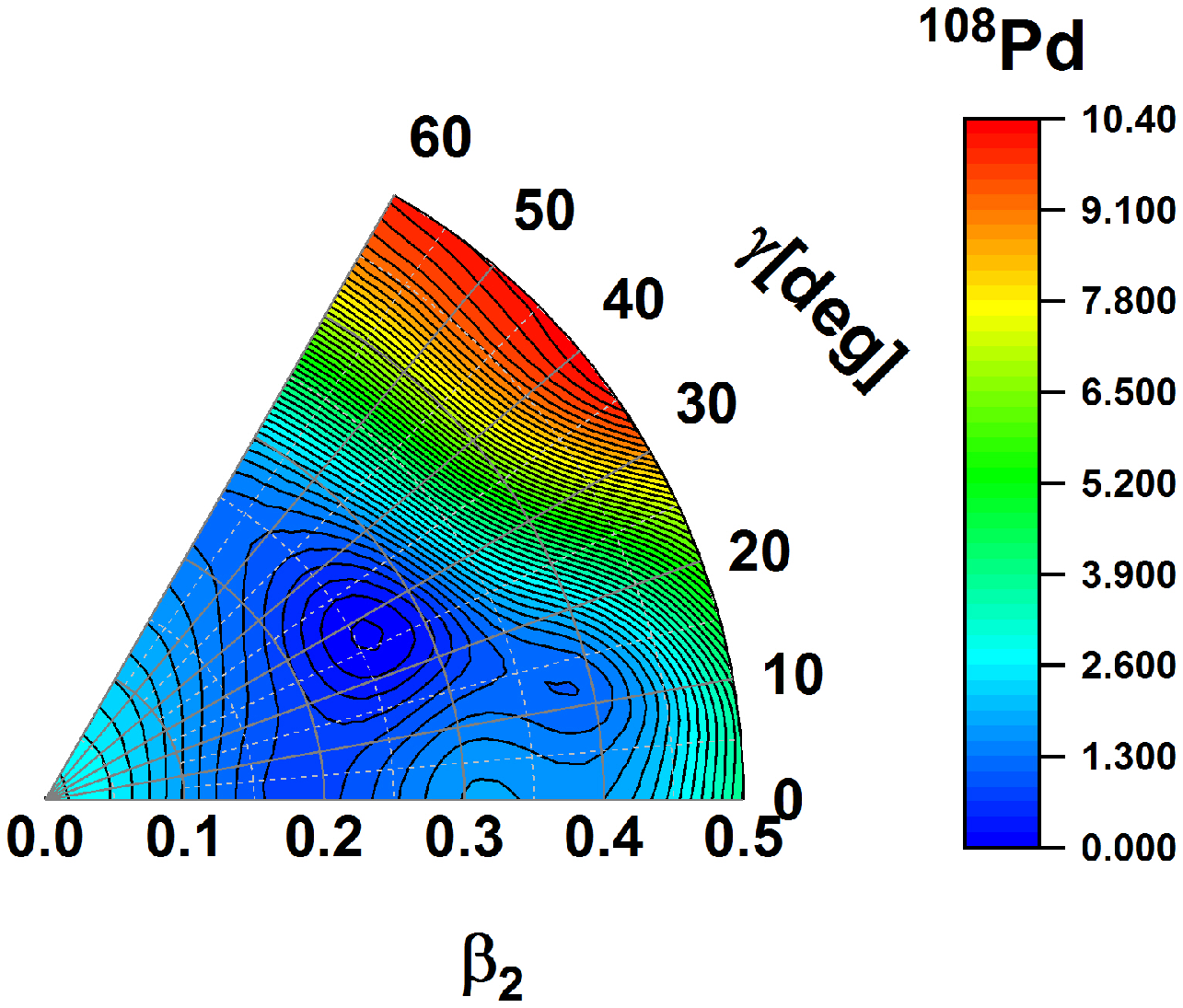}
\includegraphics[scale=0.28]{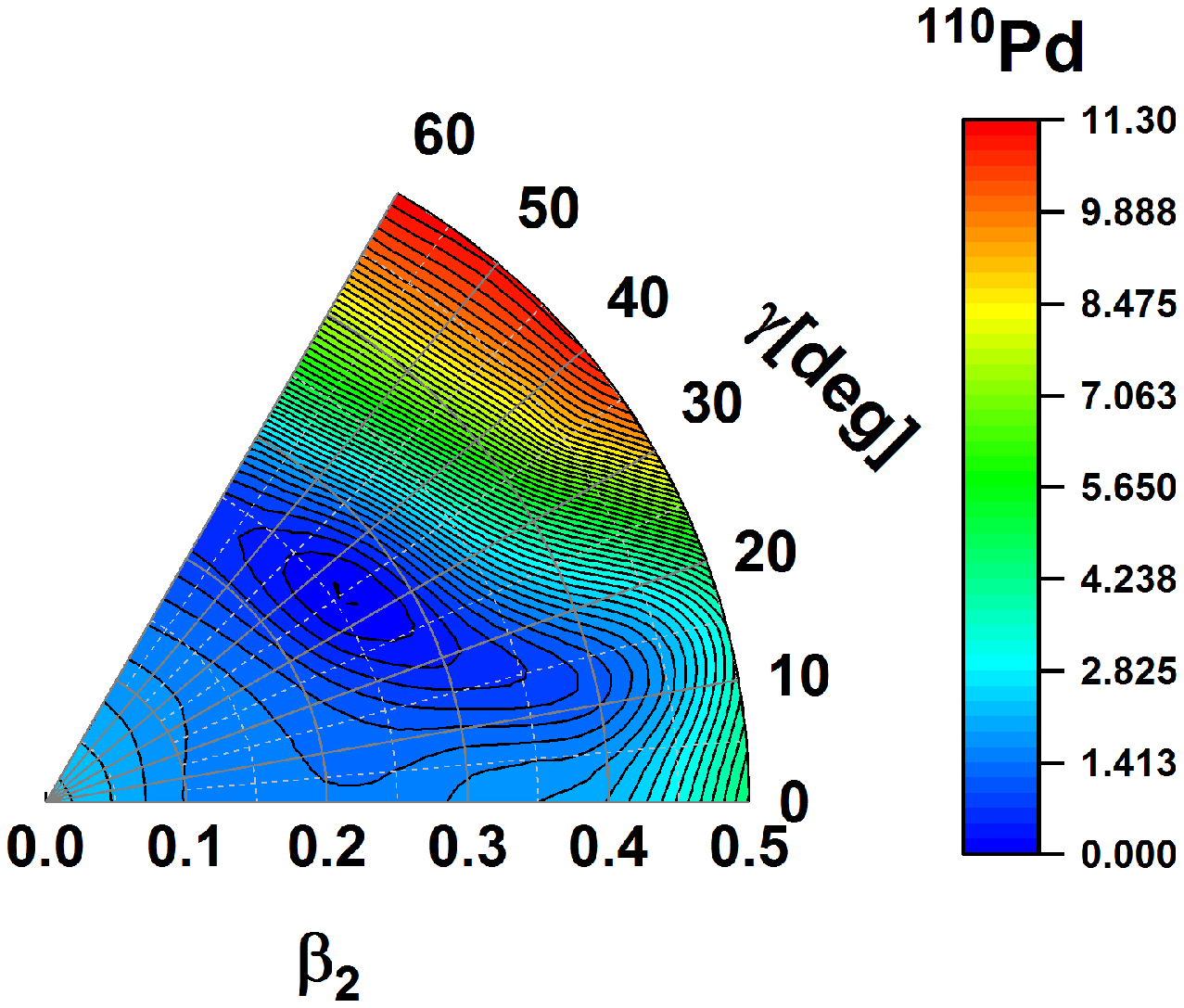}
\includegraphics[scale=0.28]{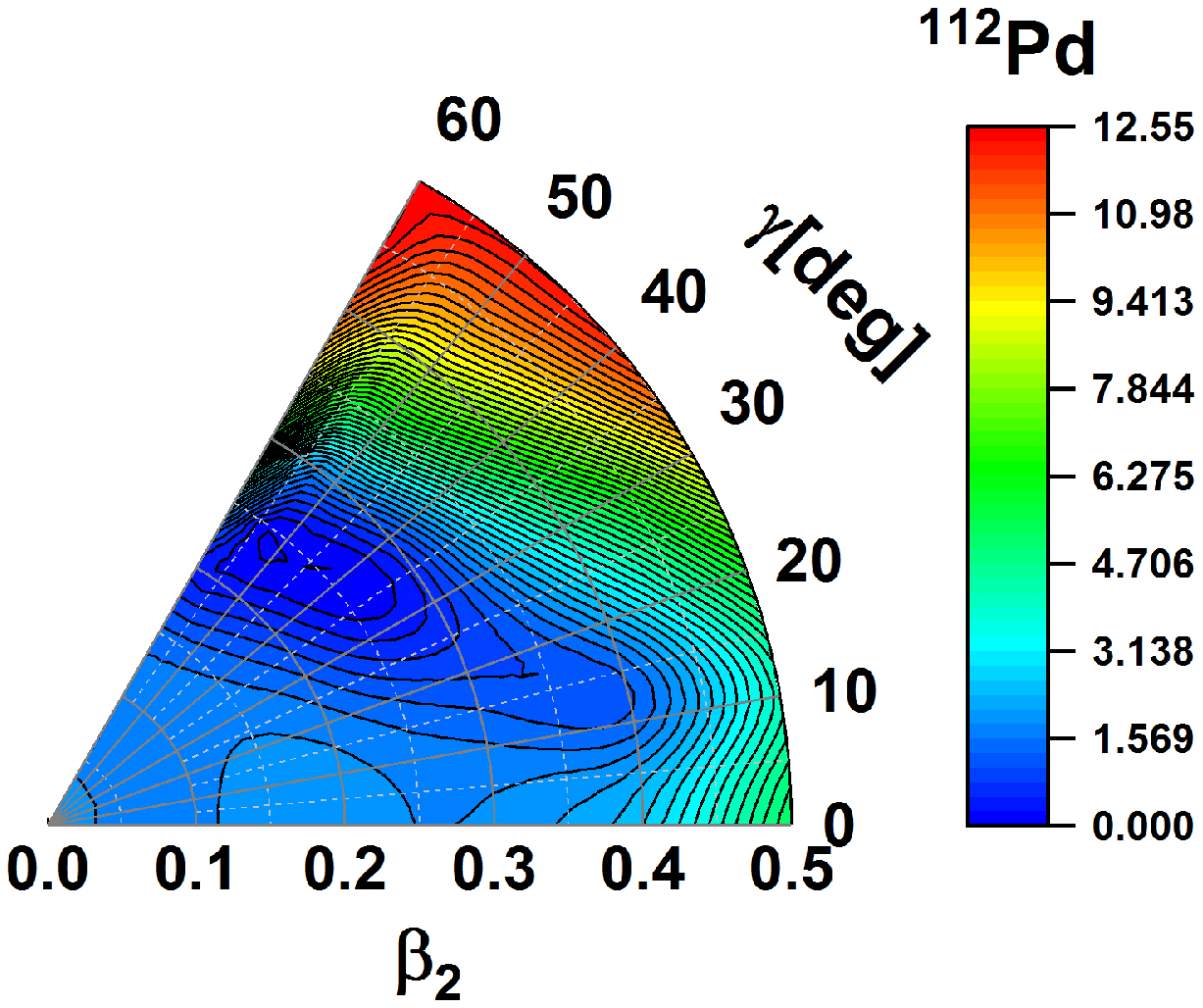}
\includegraphics[scale=0.28]{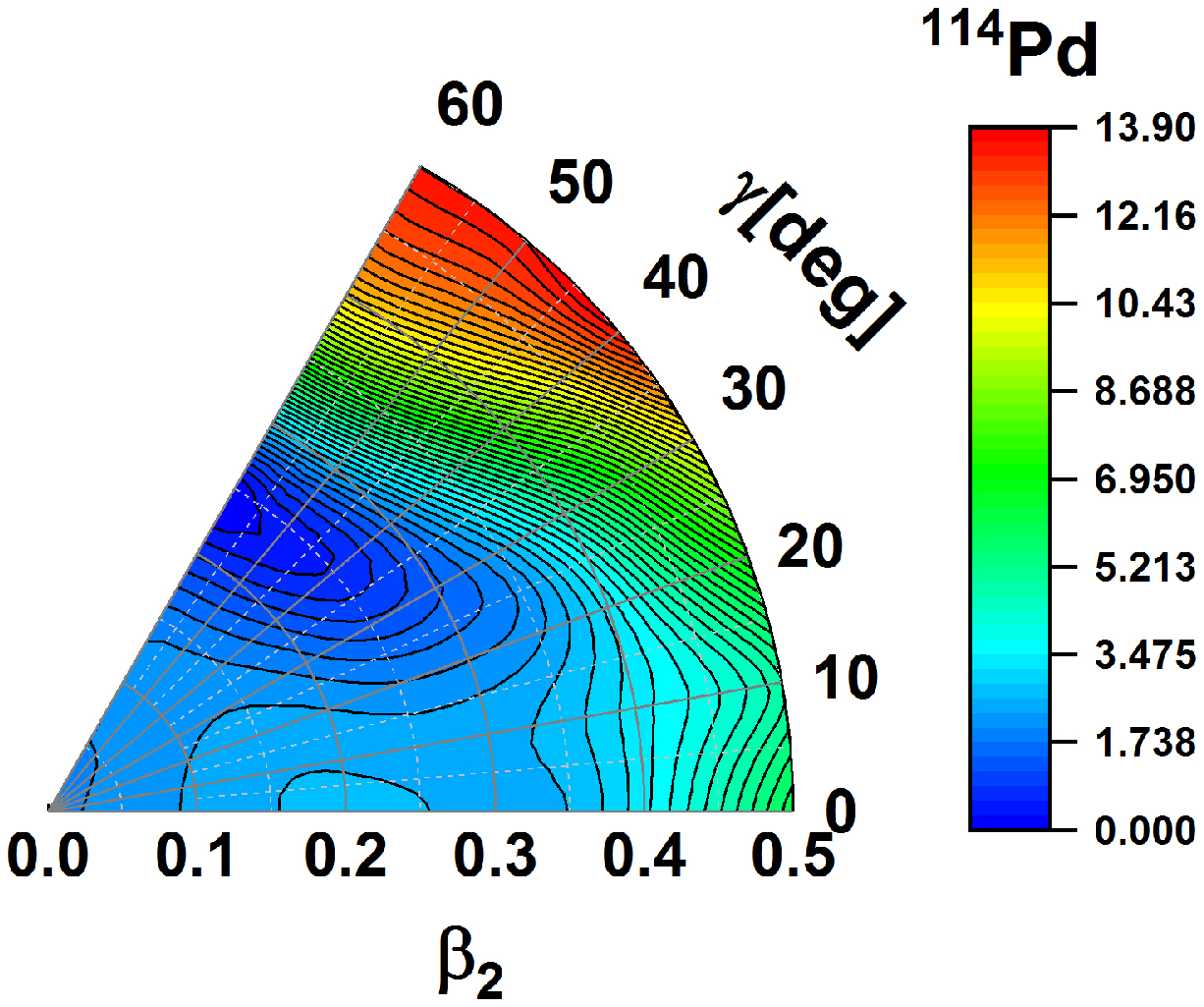}
\caption{\label{fig:figure21}(Color online) Mean filed potential energy surfaces 
for the nuclei $^{96-114}$Pd in the ($\beta, \gamma$) plane, 
obtained from a triaxial RHB calculations with the DD-ME2
parameter set. The color scale shown at the right has the
unit of MeV, and scaled such that the ground state has a zero MeV energy.}
\end{figure}

\begin{figure}
\centering
\includegraphics[scale=0.28]{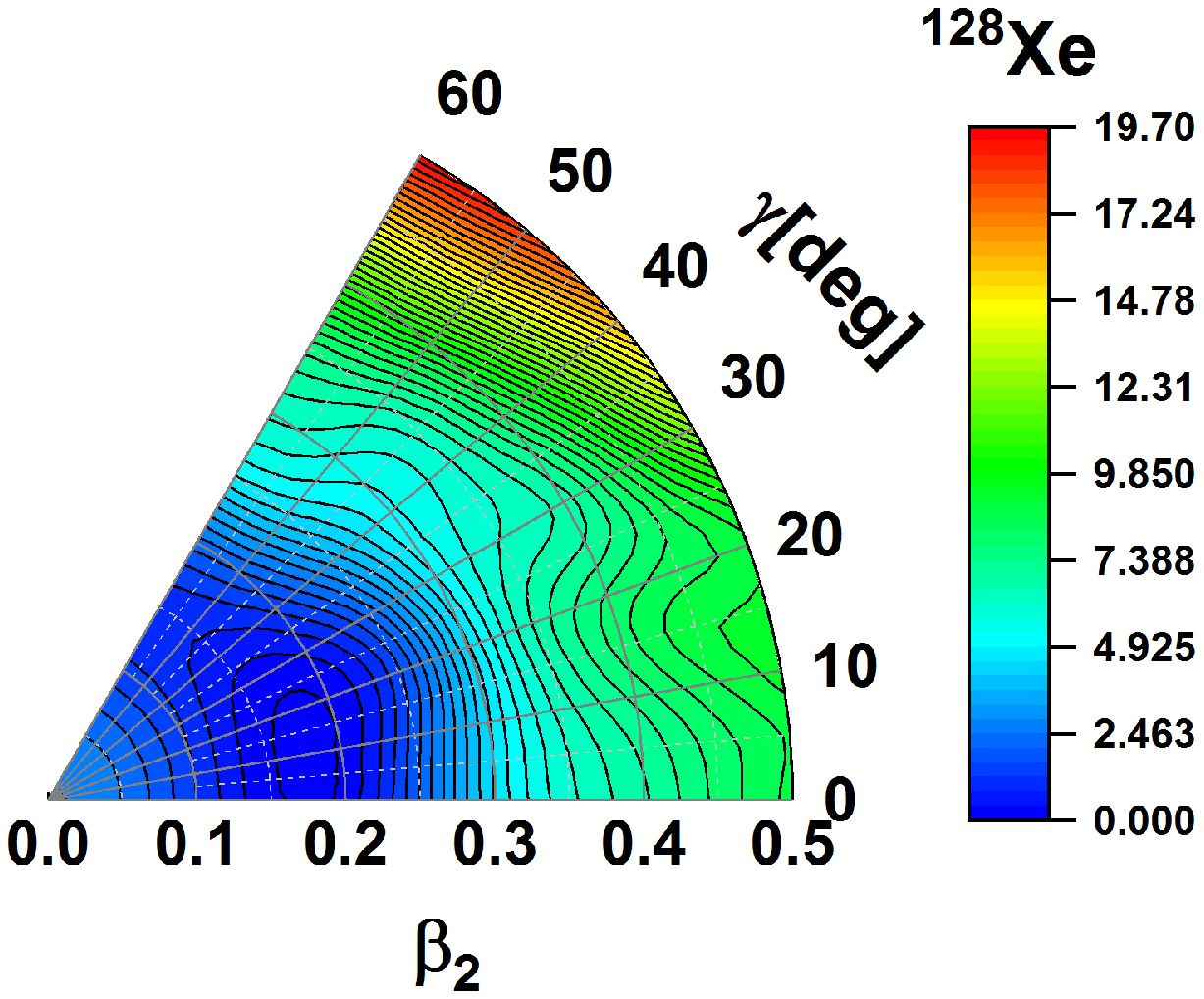}
\includegraphics[scale=0.28]{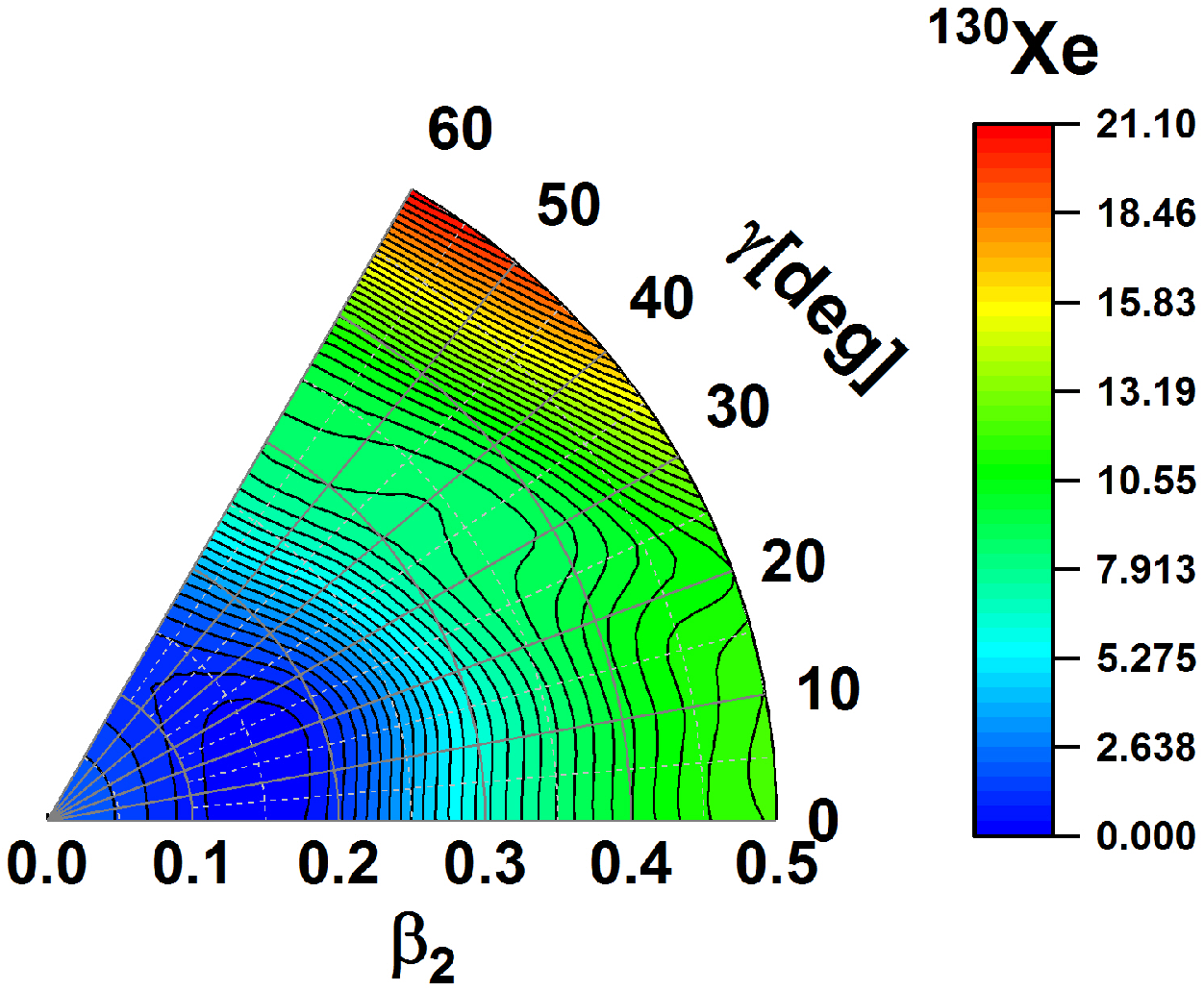}
\includegraphics[scale=0.28]{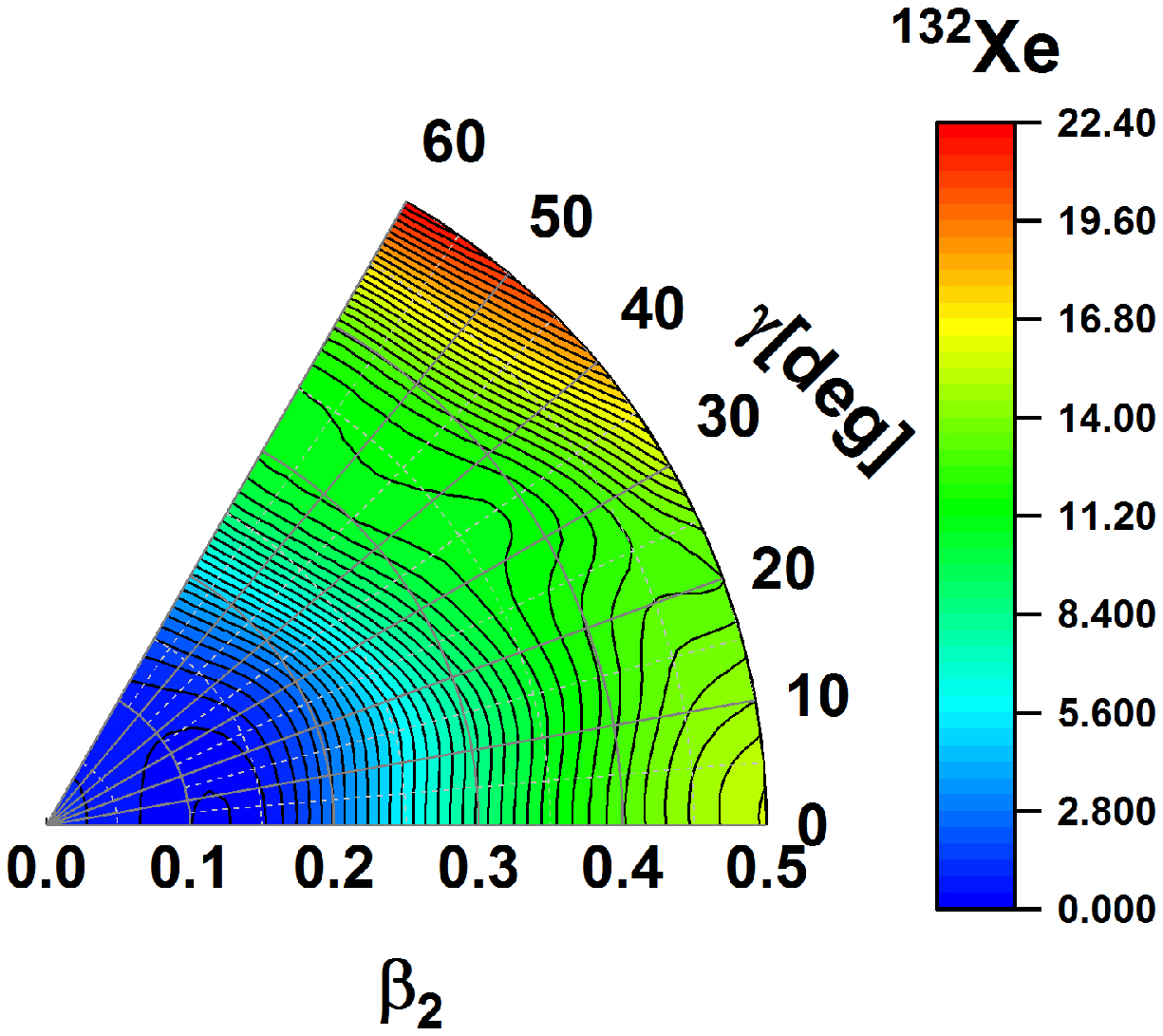}
\includegraphics[scale=0.28]{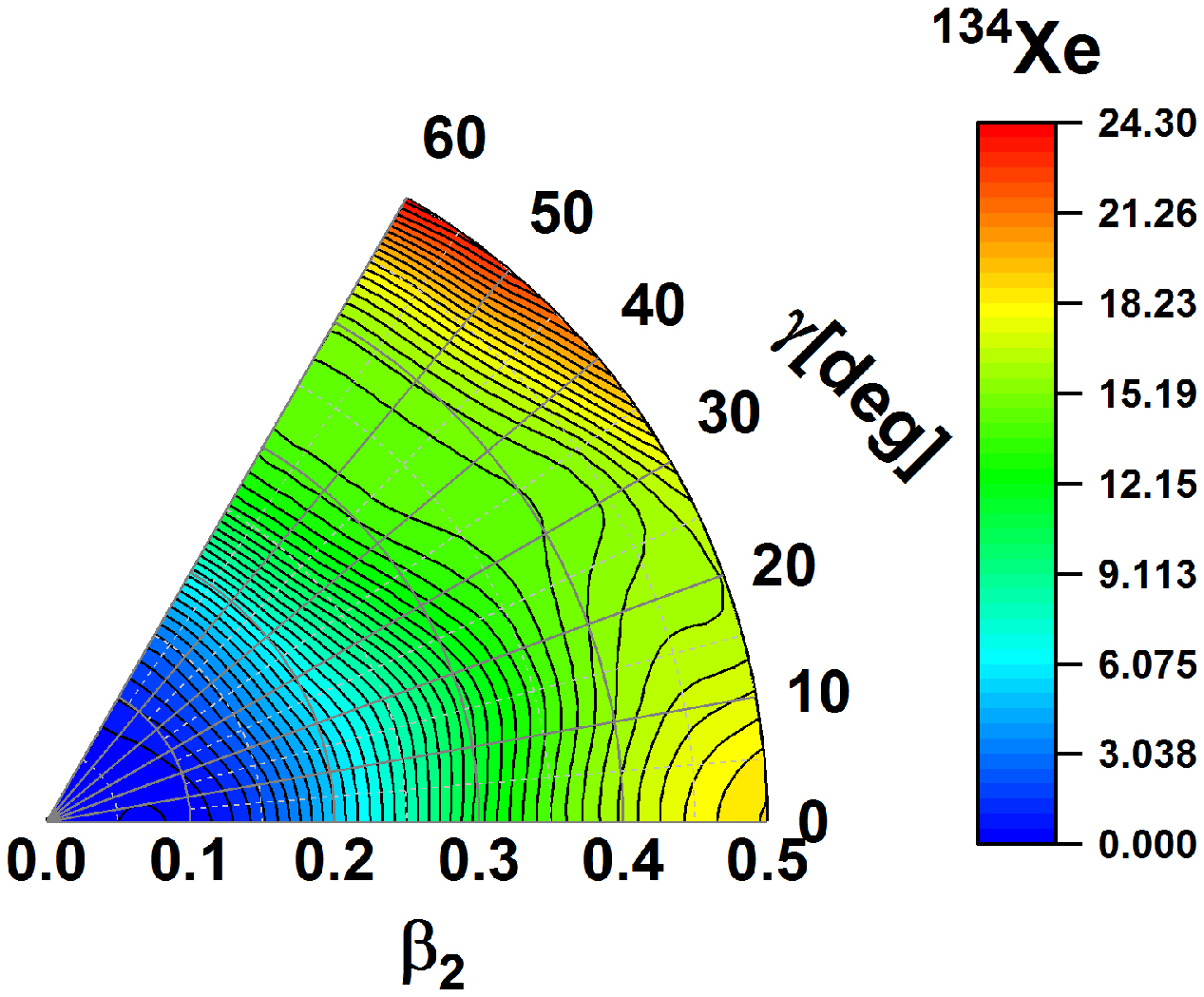}
\includegraphics[scale=0.28]{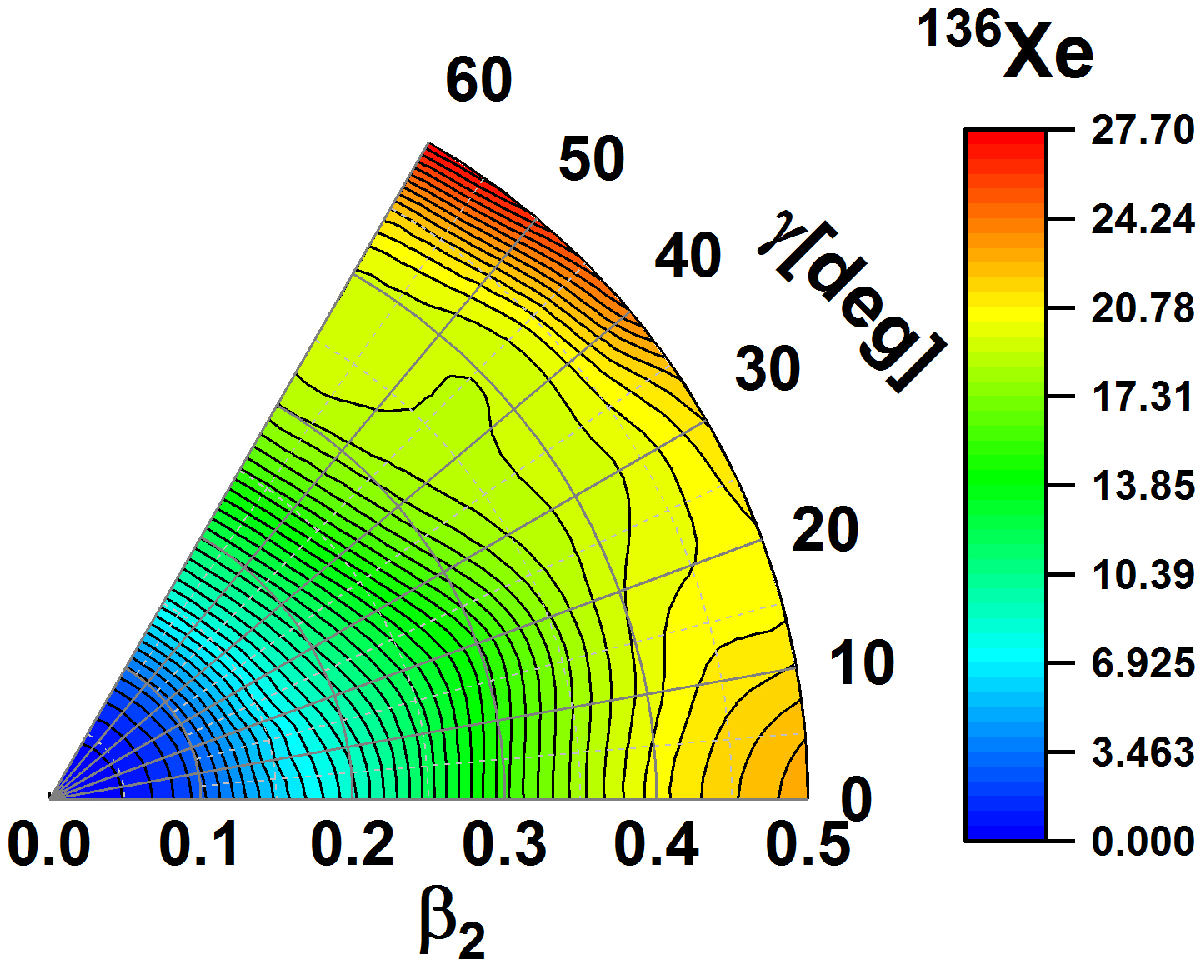}
\includegraphics[scale=0.28]{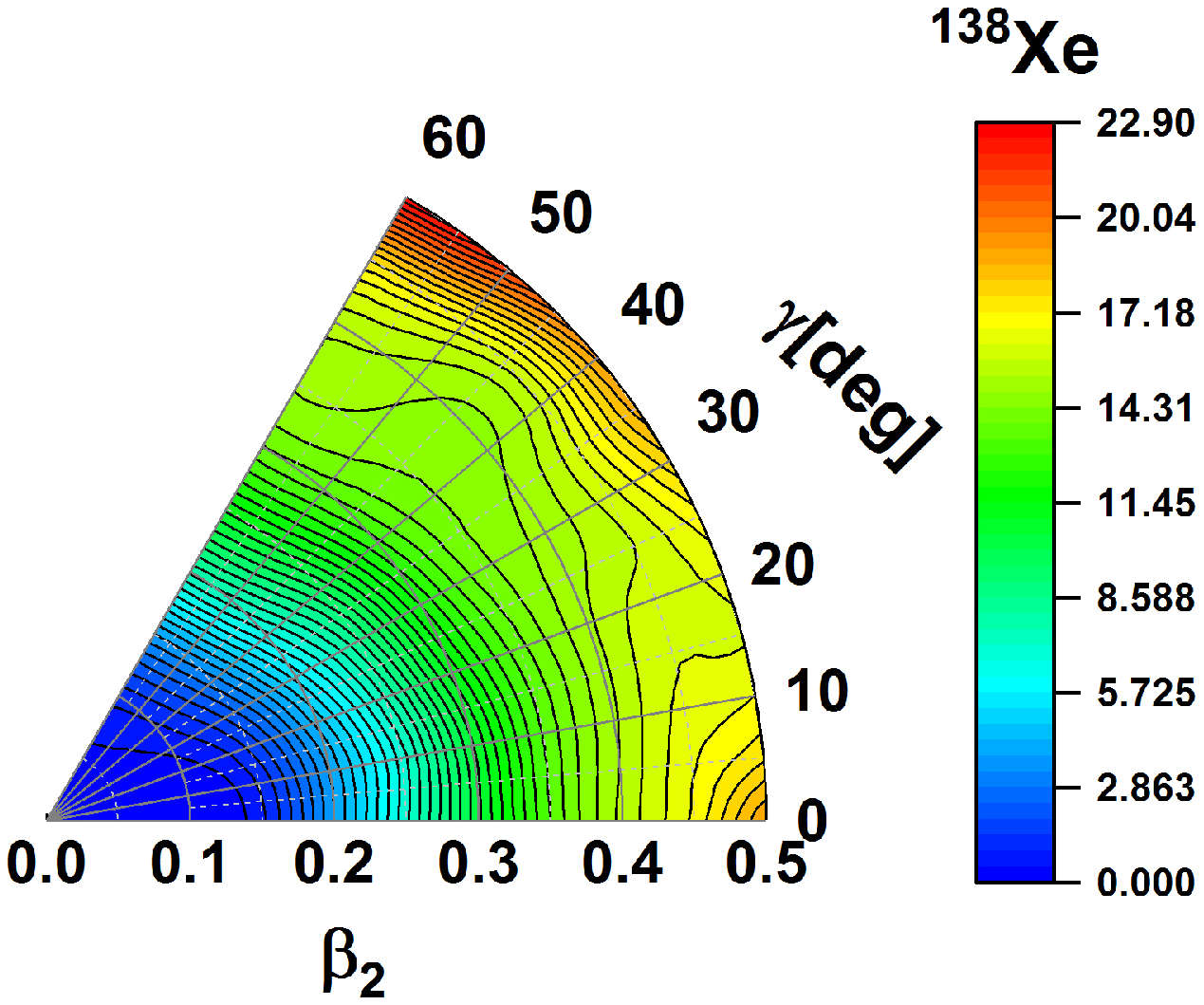}
\includegraphics[scale=0.28]{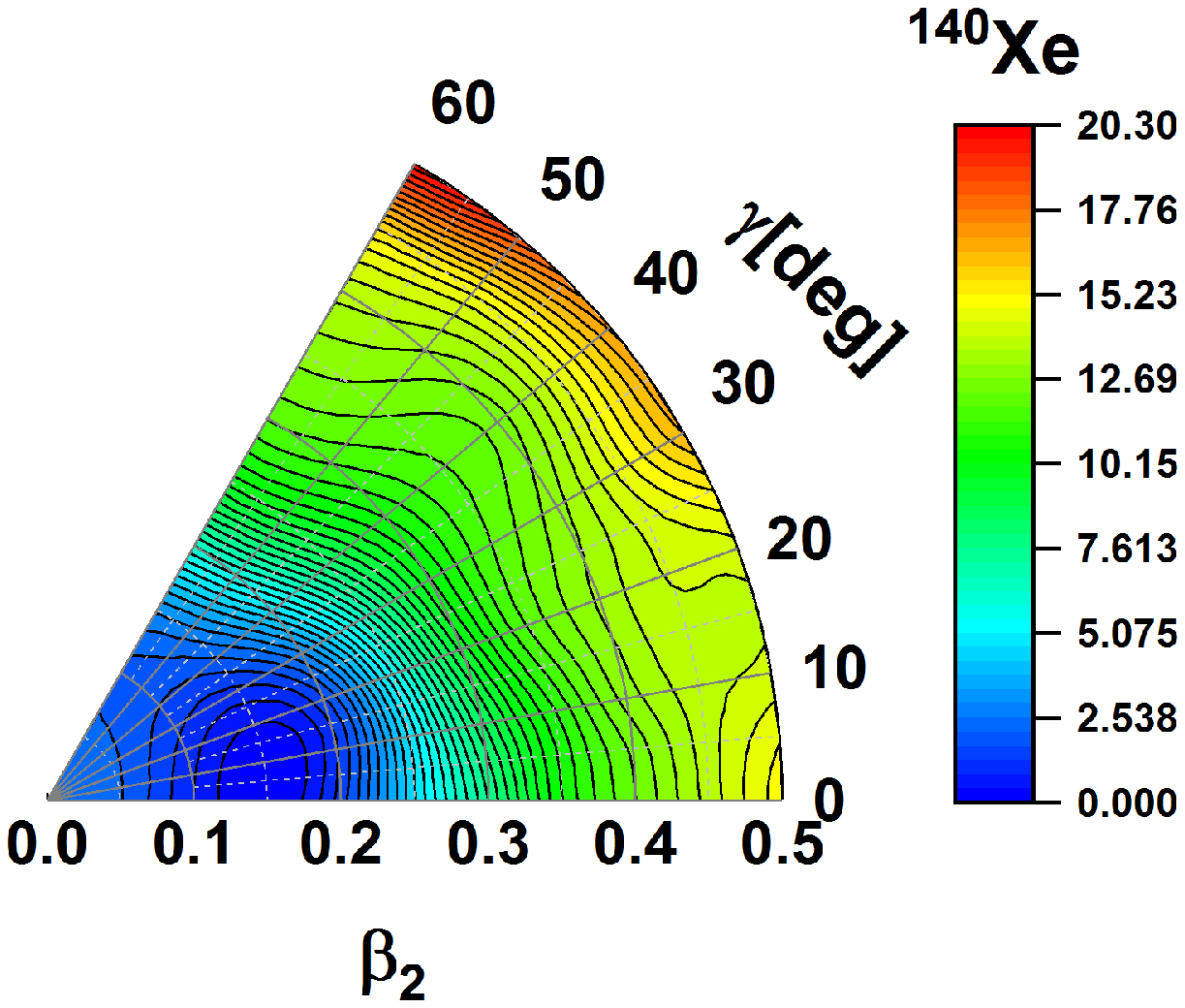}
\caption{\label{fig:figure22}(Color online) Same as Fig.\ref{fig:figure21} 
for the nuclei $^{128-140}$Xe.}
\end{figure}

\begin{figure}
\centering
\includegraphics[scale=0.28]{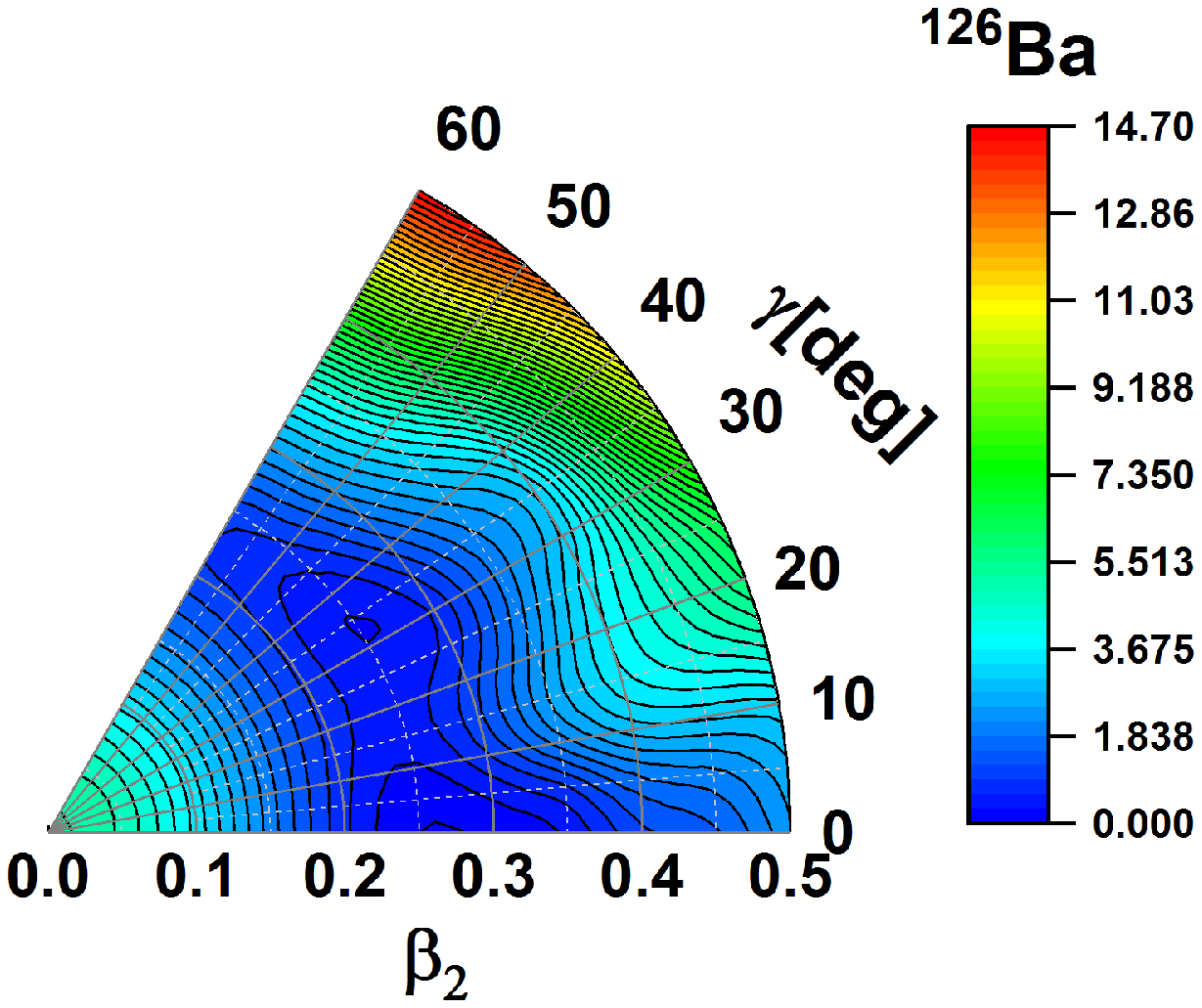}
\includegraphics[scale=0.28]{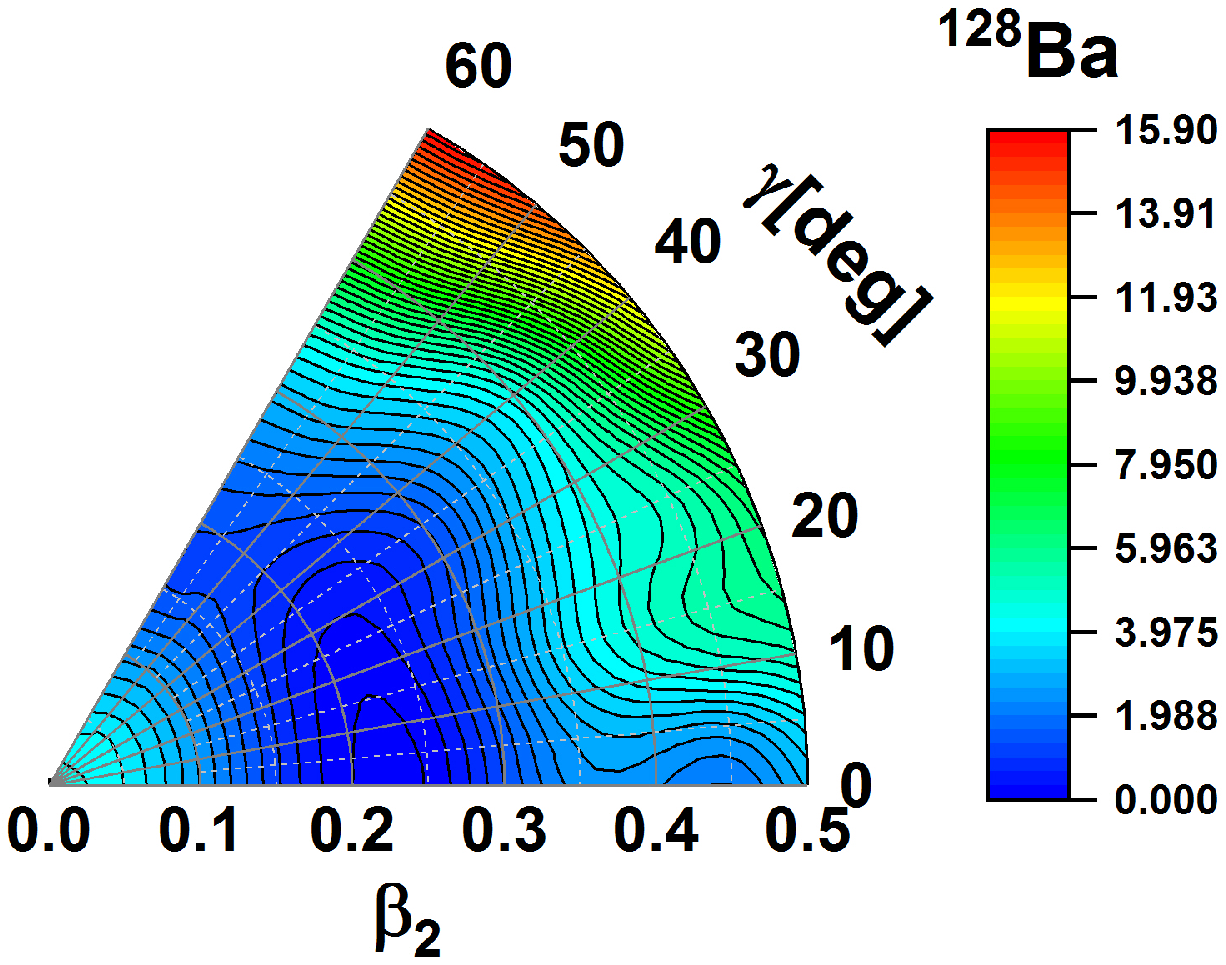}
\includegraphics[scale=0.28]{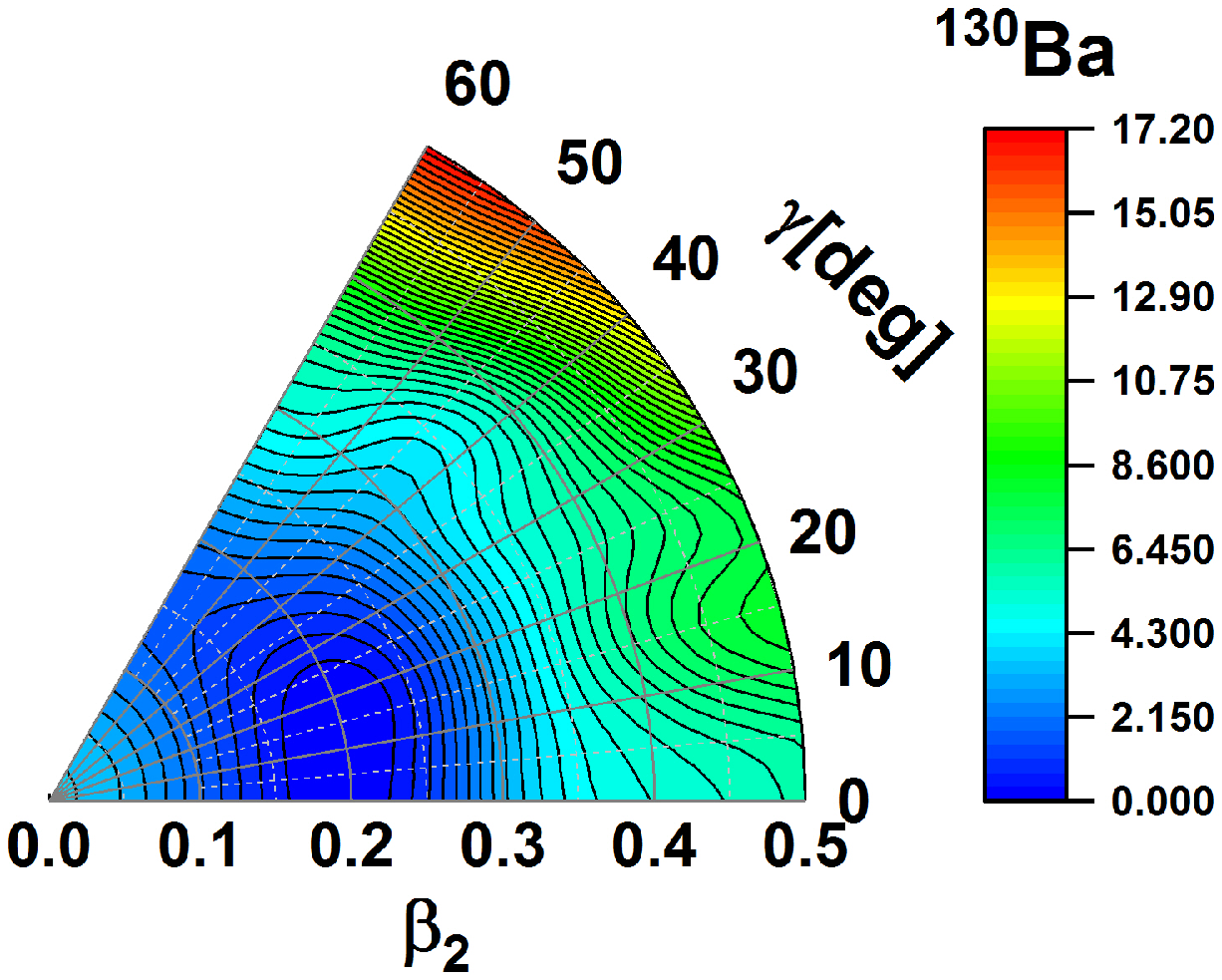}
\includegraphics[scale=0.28]{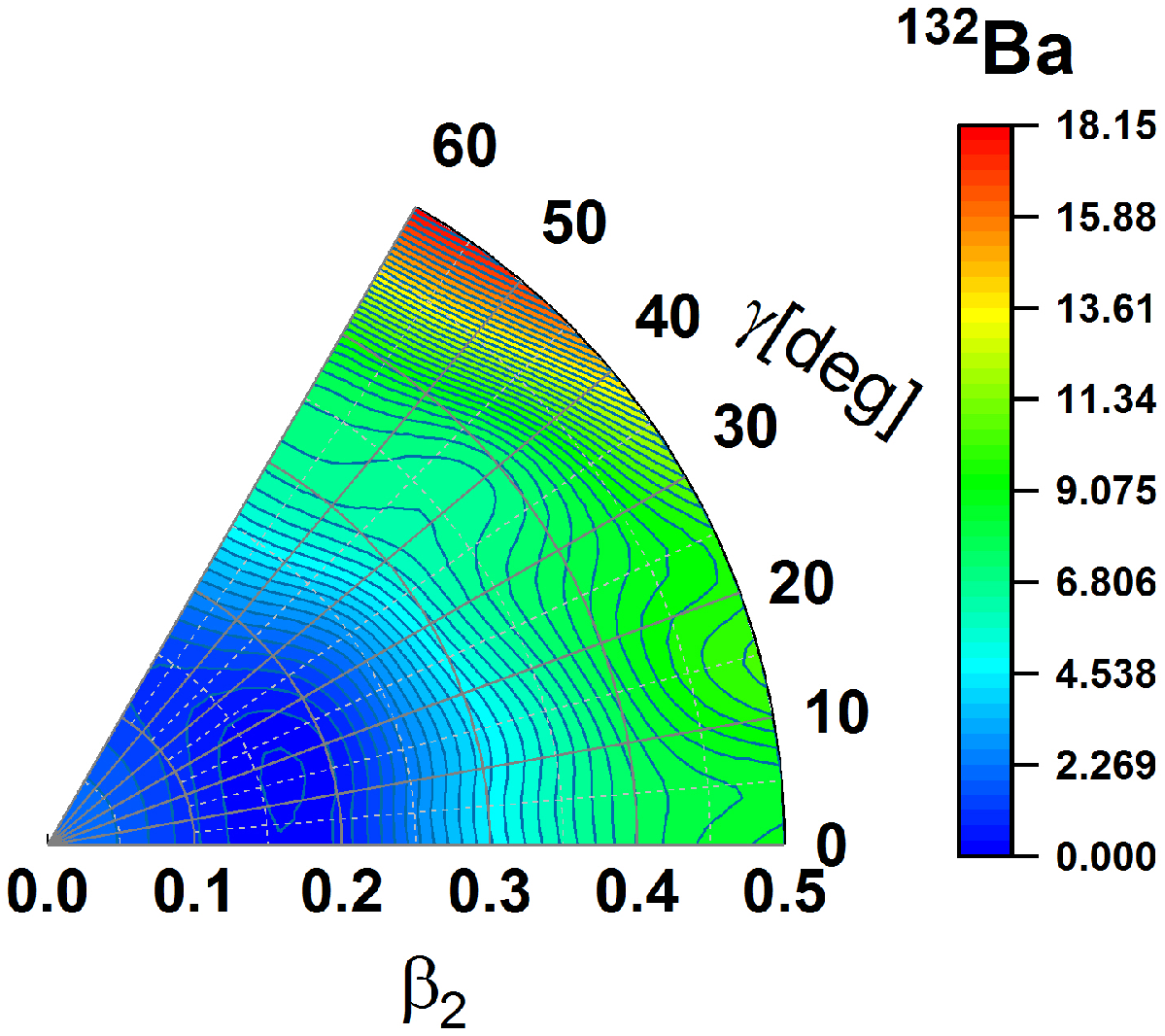}
\includegraphics[scale=0.26]{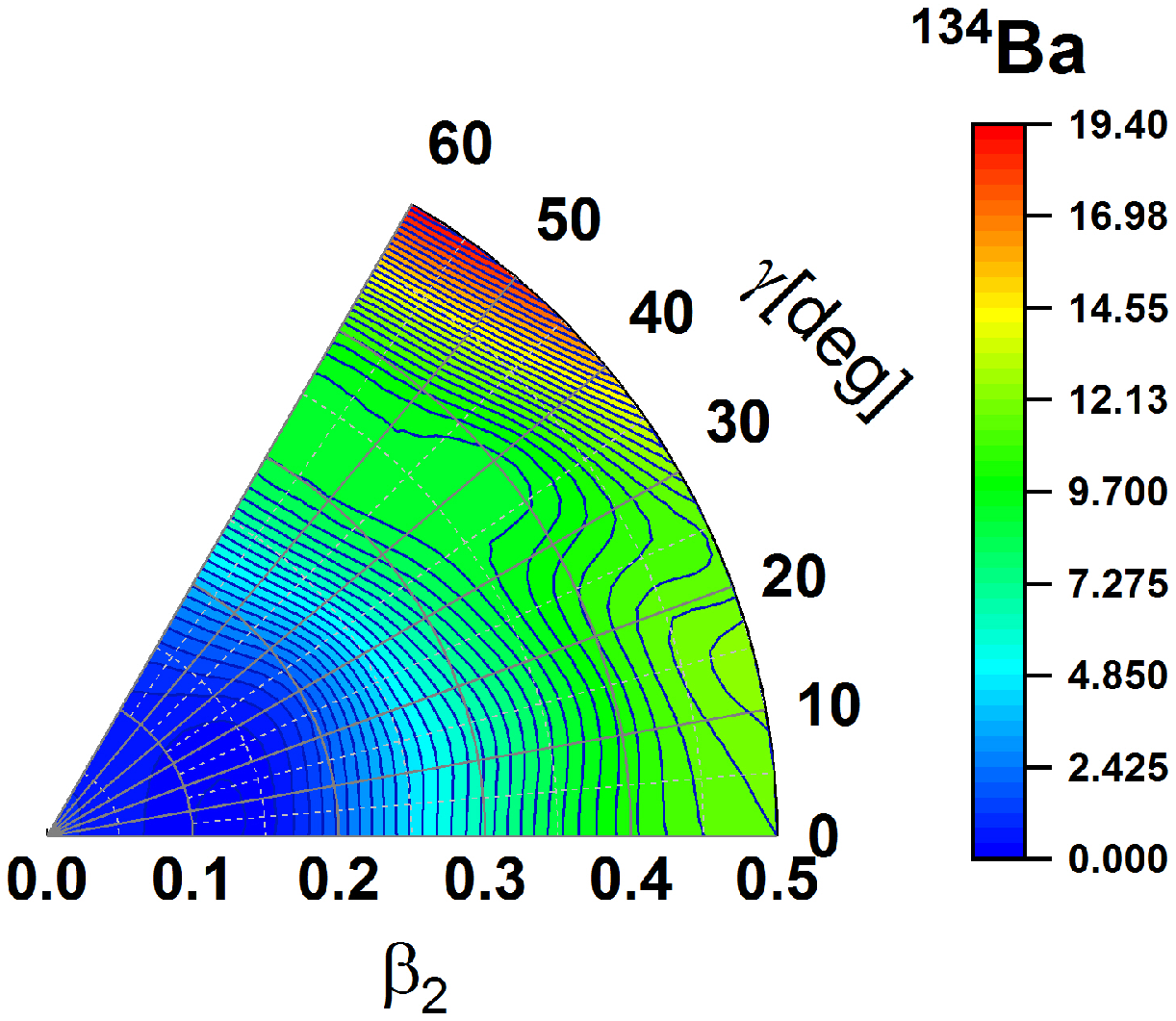}
\includegraphics[scale=0.26]{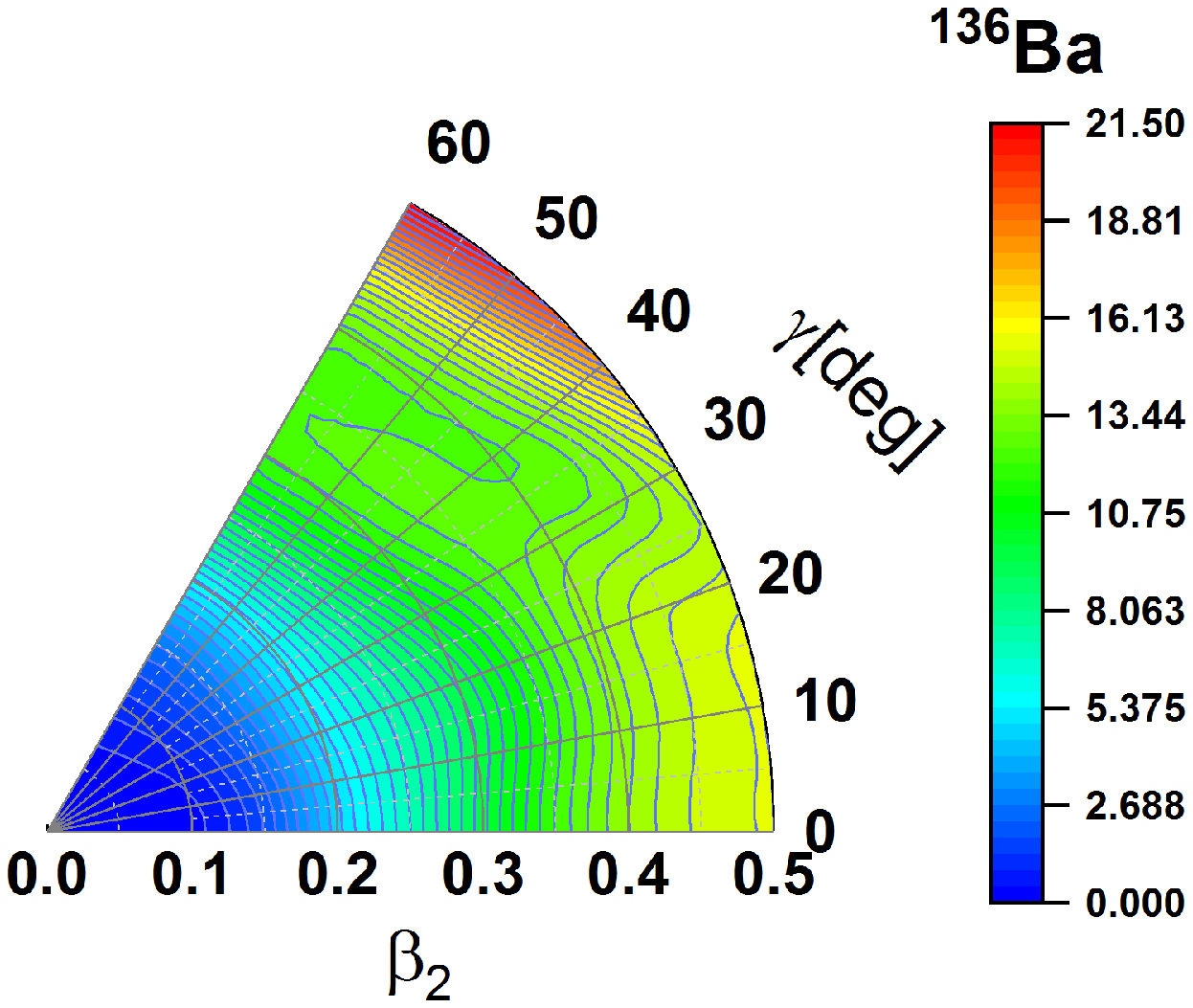}
\includegraphics[scale=0.26]{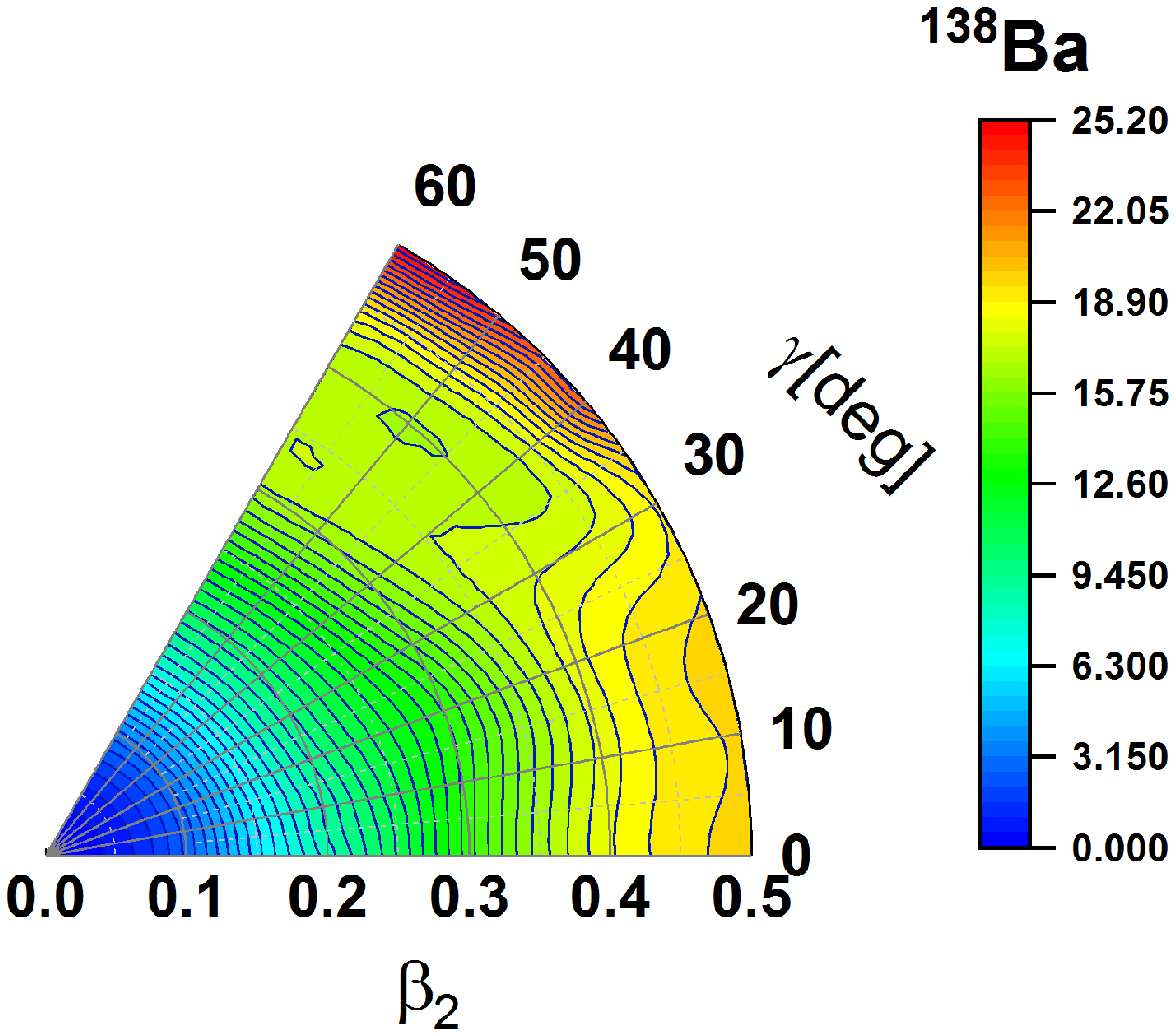}
\includegraphics[scale=0.26]{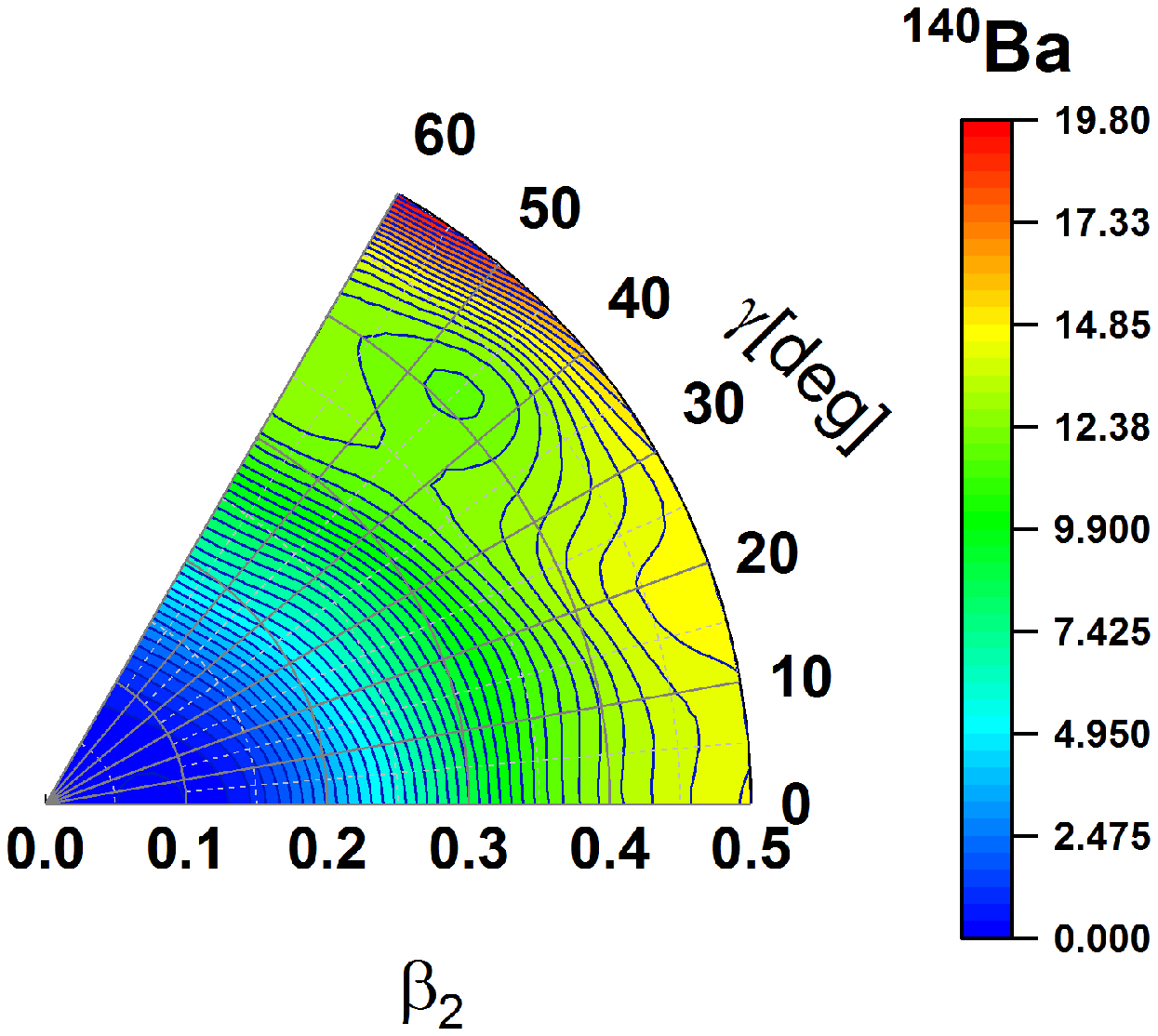}
\includegraphics[scale=0.26]{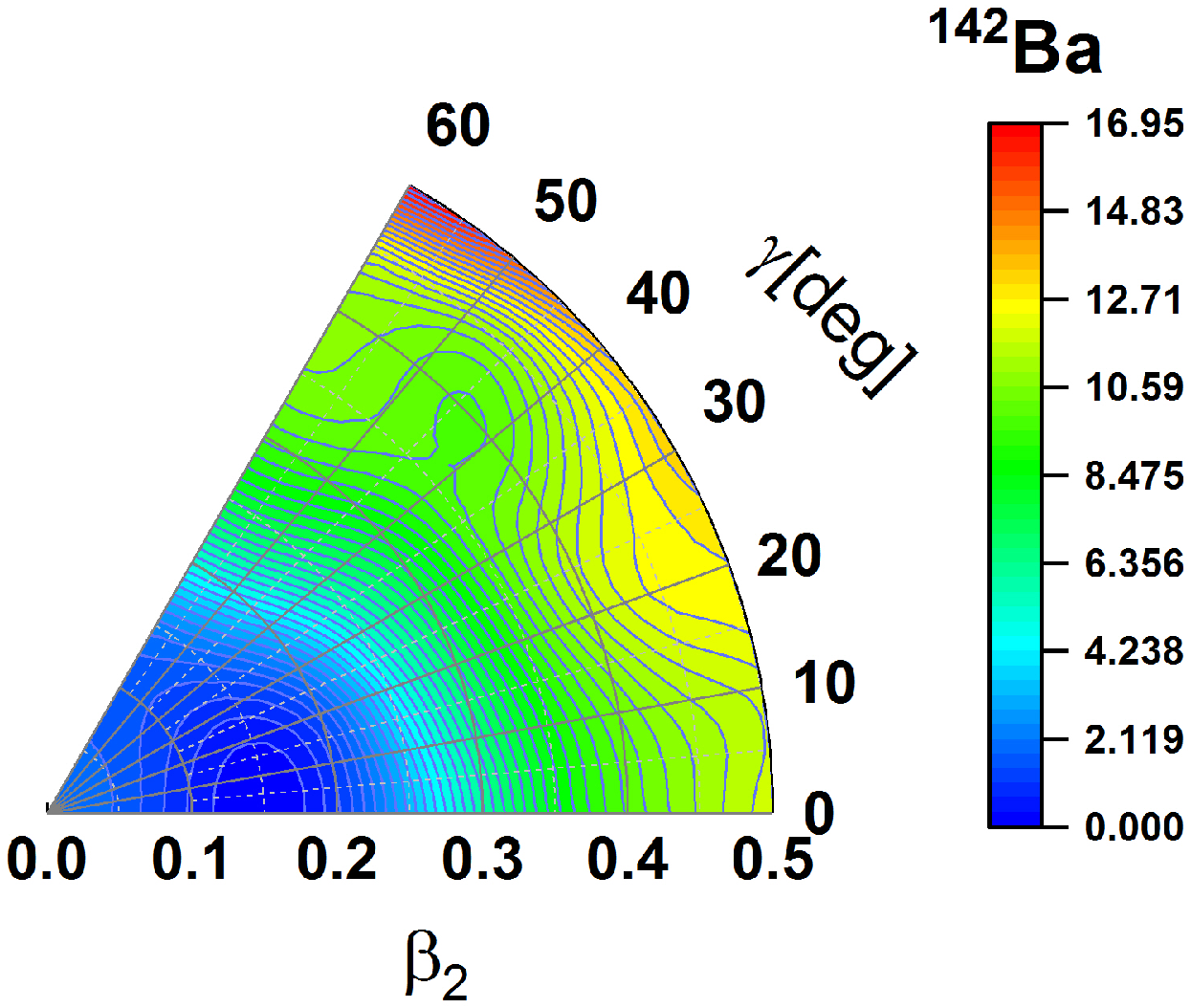}
\caption{\label{fig:figure23}(Color online) Same as Fig.\ref{fig:figure21} 
for the nuclei $^{126-142}$Ba.}
\end{figure}

The role of triaxiality becomes important due to the flatness in PESs leading to locate the exact global
minimum for such cases.
To study the dependency on $\gamma$, a systematic constrained triaxial calculation have
been done for mapping the quadrupole deformation space defined by
$\beta_2$ and $\gamma$ using DD-ME2  effective interaction. 
Energies are normalized with respect to the binding energy of the global minimum. 
In Fig.\ref{fig:figure21}, Fig.\ref{fig:figure22} and Fig.\ref{fig:figure23} 
we have displayed the contour plots for $^{96-114}$Pd, 
$^{128-140}$Xe, and $^{126-142}$Ba, respectively. The ground state quadrupole deformations in the 
$\beta_2$-$\gamma$ plane corresponding to the global minima of the triaxial PES 
are tabulated in Table~\ref{tab:table1}. Table~\ref{tab:table1}, also display the 
difference in the ground state energies ($\Delta E_{\mbox{tr}}$) as the triaxial 
deformation energy, with respect to the ground state energies ($E_{\mbox{ax}}$) 
corresponding to the axial symmetry.   
We can see the shape transition is very evident, starting from the spherical($^{96}$Pd)
to prolate($^{98-104}$Pd) deformation, and to triaxial($^{106-112}$Pd) deformation then,
shifted to oblate($^{114}$Pd) deformation.
The PESs appears to be quite $\gamma$-soft extended from prolate to oblate, and 
then starts to become slightly rigid in $\gamma$ direction towards oblate side. Further, the
softness in $\gamma$ direction is shifting to prolate side. 
The nucleus $^{106}$Pd has two minima at $(0.2, 15^o)$ and at $(0.4, 5^o)$ with energy
difference of 1.21MeV, and the deepest one is $(0.2, 15^o)$. Similarly, $^{108}$Pd has
two minima at $(0.25, 25^o)$ and at $(0.35, 15^o)$ with energy difference of 1.10MeV, and
$(0.25, 25^o)$ as the global minimum.
Finally, it becomes flat along
$\gamma$ direction towards prolate side at $^{106}$Pd.
The nucleus $^{108,110}$Pd besides triaxiality are showing the $\gamma$-soft nature too, 
favoring larger deformations $\beta$ on the prolate side ($\gamma=0^o$) and 
smaller on the oblate side ($\gamma=60^o$). In case of Xe(Z=54) isotopes, the deformation 
shifts from prolate to spherical at $^{136}$Xe, and then shifts to prolate deformation.
Here also the PESs appears to be soft in the $\gamma$-direction. The continuous $\gamma$-soft 
behaviour extended from prolate to oblate. These nuclei except $^{136}$Xe are not spherical 
but, rather, characterized by some degree of triaxiality.
For Ba(Z=56) isotopes, the nuclei $^{136}$Ba and $^{138}$Ba both are of spherical shape. 
But $^{138}$Ba show more spherical character because here the $\gamma$-softness is 
concentrated within smaller $\beta$-values. Either side of $^{138}$Ba, the deformation
is prolate, except at $^{132}$Ba, where, triaxiality appears at $(0.15, 15^o)$. 
The nucleus $^{126}$Ba being prolate at global minimum, also showing triaxial
character at $(0.25, 35^o)$ with energy difference of 0.287MeV. $^{128}$Ba nucleus is 
also showing two coexisting prolate minimum at $(0.2, 0^o)$ and $(0.45, 0^o)$ with
energy difference of 1.89MeV. 
\begin{table*}[htb]
 \caption{\label{tab:table1}
The quadrupole deformation ($\beta_2$, $\gamma$) of the global minima in Pd, Xe and Ba isotopes,
calculated within the RHB formalism with DD-ME2. $E_{ax}$ and $E_{tr}$ are the total energies for the
global minima under axial and triaxial symmetry, respectively.}
\centering
\begin{tabular}{rrrrrrrrrrrr}
\hline
\hline
\textrm{Nuclei}&
\textrm{$\beta_2$}&
\textrm{$\gamma$}&
\textrm{$E_{tr}$}&
\textrm{$E_{ax}$}&
\textrm{$\Delta E_{tr}$}&
\textrm{Nuclei}&
\textrm{$\beta_2$}&
\textrm{$\gamma$}&
\textrm{$E_{tr}$}&
\textrm{$E_{ax}$}&
\textrm{$\Delta E_{tr}$}\\
\hline
$^{96}$Pd&	0&$0^{\circ}$&815.274		&814.972&0.302	&$^{134}$Xe&	0.05&$0^{\circ}$&1127.397&1127.242&0.155\\
$^{98}$Pd&	0.15&$0^{\circ}$&834.281	&834.093&0.187&$^{136}$Xe&	0.0&$0^{\circ}$&1143.737&1143.495&0.242\\
$^{100}$Pd&	0.15&$0^{\circ}$&853.584	&853.418&0.165	&$^{138}$Xe&	0.1&$0^{\circ}$&1150.590&1150.329&0.261\\
$^{102}$Pd&	0.2&$0^{\circ}$&872.483		&872.518&-0.035	&$^{140}$Xe&	0.15&$0^{\circ}$&1158.992&1158.743&0.249\\
$^{104}$Pd&	0.2&$0^{\circ}$&889.450		&889.400&	0.050	&$^{126}$Ba&	0.25&$0^{\circ}$&1052.447&1052.811&-0.364\\
$^{106}$Pd&	0.2&$15^{\circ}$&905.683	&905.520&0.163	&$^{128}$Ba&	0.2&$0^{\circ}$&1071.578&1071.920&-0.342\\
$^{108}$Pd&	0.25&$25^{\circ}$&921.398	&920.673&0.725&$^{130}$Ba&	0.2&$0^{\circ}$&1090.329&1090.503&-0.174\\
$^{110}$Pd&	0.25&$35^{\circ}$&936.362	&935.772&0.589	&$^{132}$Ba&	0.15&$15^{\circ}$&1110.230&1108.204&2.026\\
$^{112}$Pd&	0.25&$40^{\circ}$&950.708	&950.433&0.274	&$^{134}$Ba&	0.16&$0^{\circ}$&1127.783&1125.765&2.018\\
$^{114}$Pd&	0.25&$60^{\circ}$&963.820	&964.191&-0.370	&$^{136}$Ba&	0.0&$0^{\circ}$&1143.900&1142.992&0.908\\
$^{128}$Xe&	0.2&$15^{\circ}$&1078.713&1078.666&0.047	&$^{138}$Ba&	0.0&$0^{\circ}$&1161.120&1161.884&-0.764\\
$^{130}$Xe&	0.15&$10^{\circ}$&1095.489&1095.467&0.022&$^{140}$Ba&	0.1&$0^{\circ}$&1170.622&1168.762&1.860\\	
$^{132}$Xe&	0.1&$0^{\circ}$&1111.620&1111.449&0.170&$^{142}$Ba&	0.15&$0^{\circ}$&1181.460&1178.140&3.320\\
\hline
\hline
\end{tabular}
\end{table*}

\begin{figure}
\centering
\includegraphics[scale=0.28]{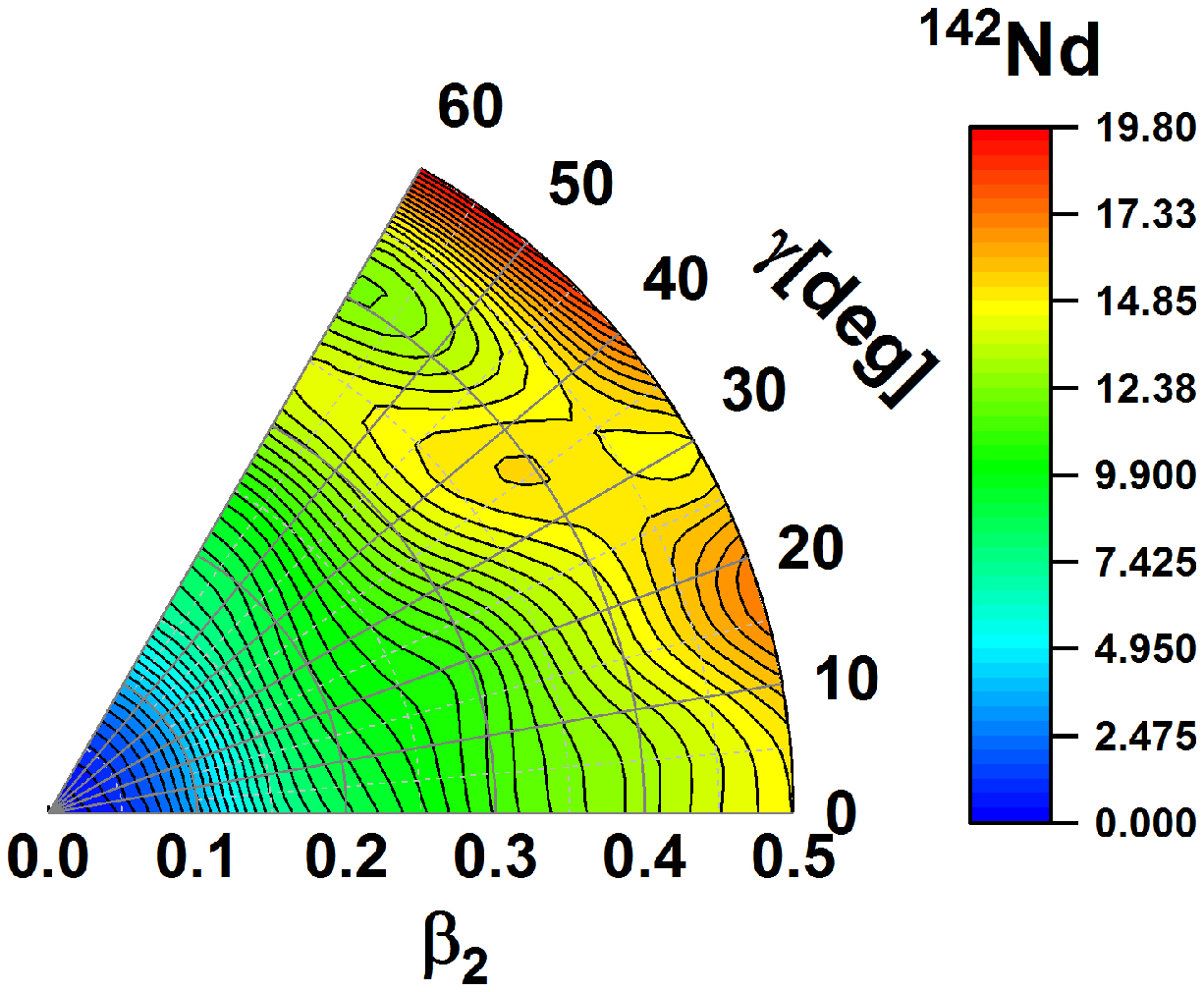}
\includegraphics[scale=0.28]{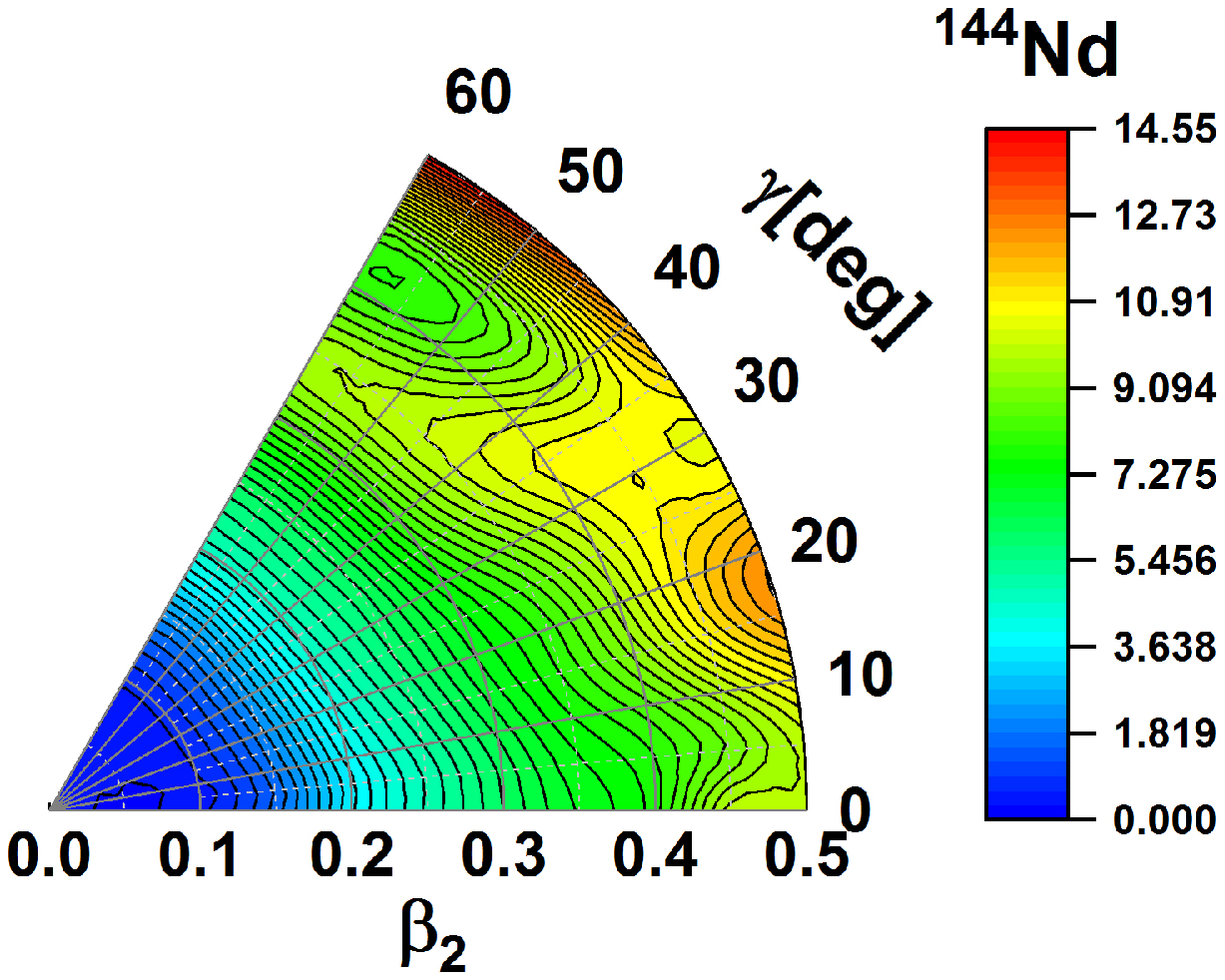}
\includegraphics[scale=0.28]{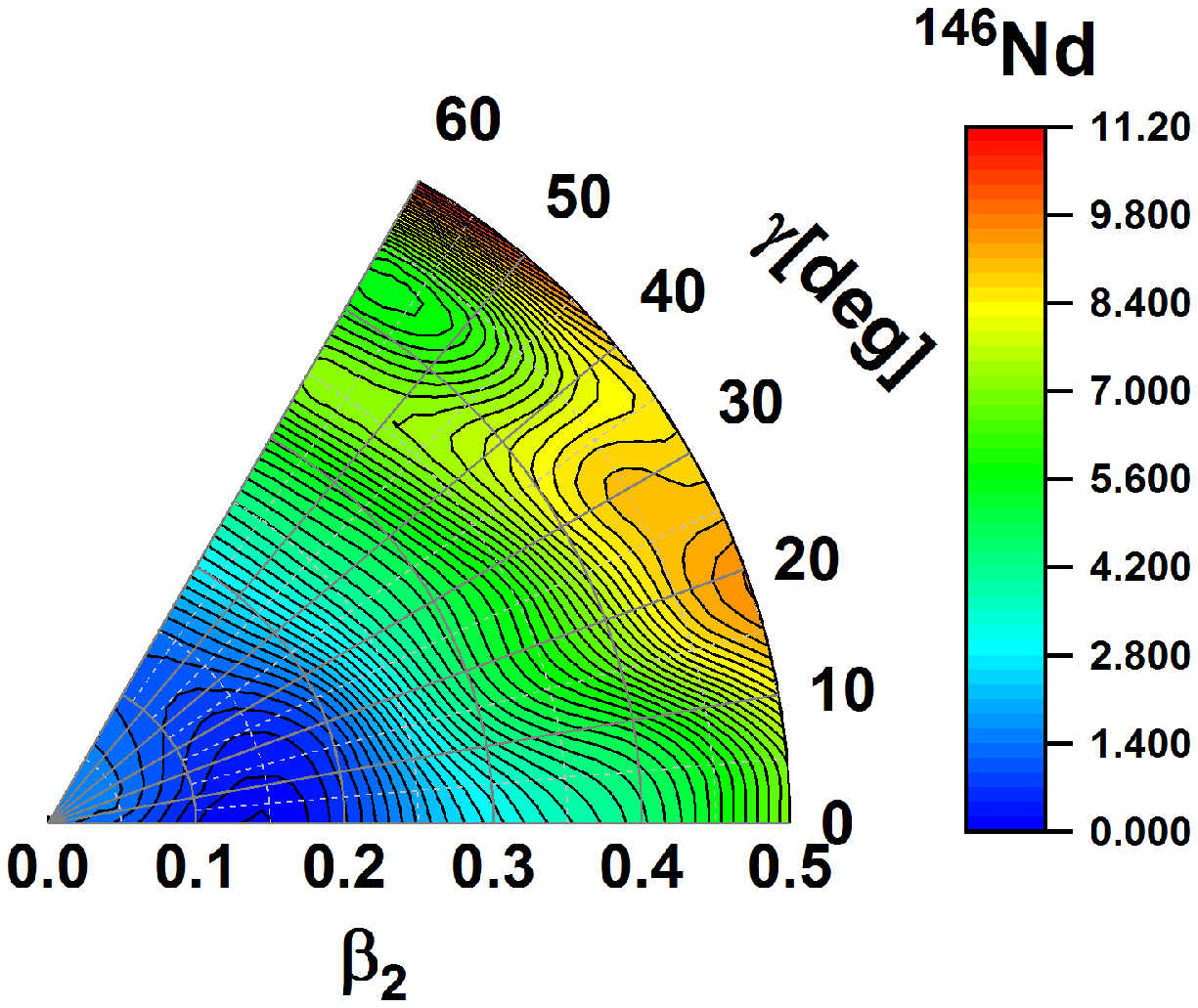}
\includegraphics[scale=0.28]{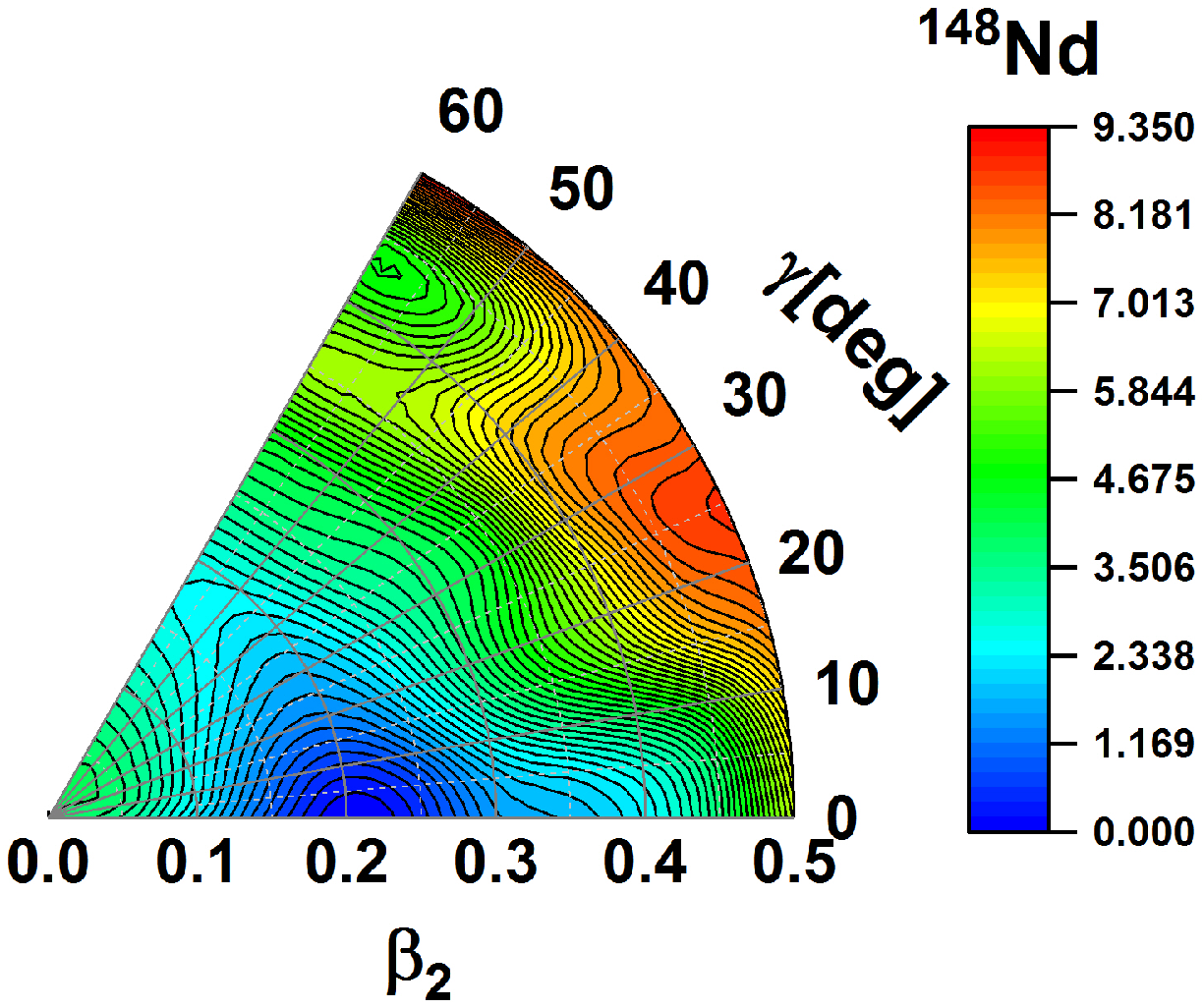}
\includegraphics[scale=0.28]{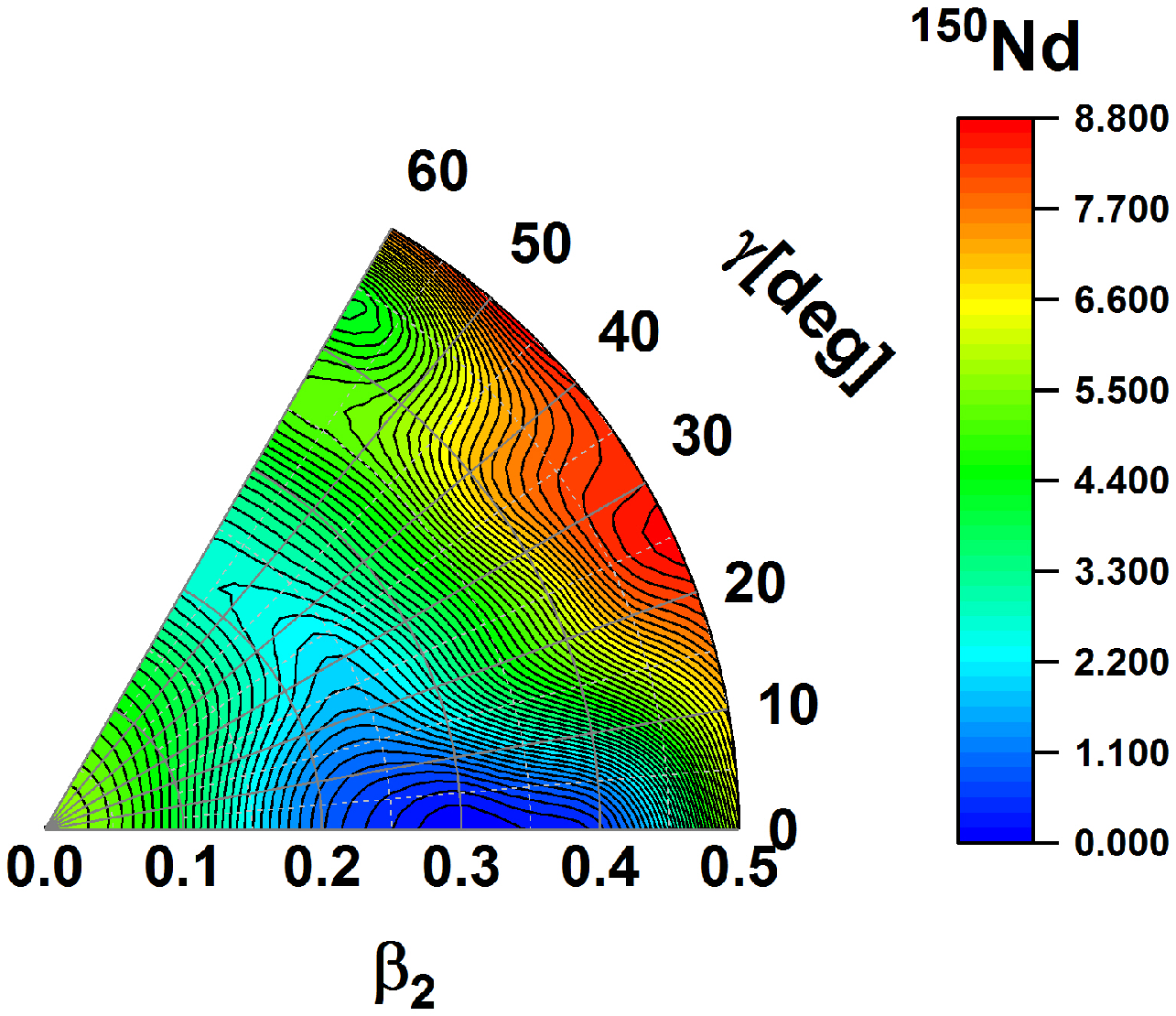}
\includegraphics[scale=0.28]{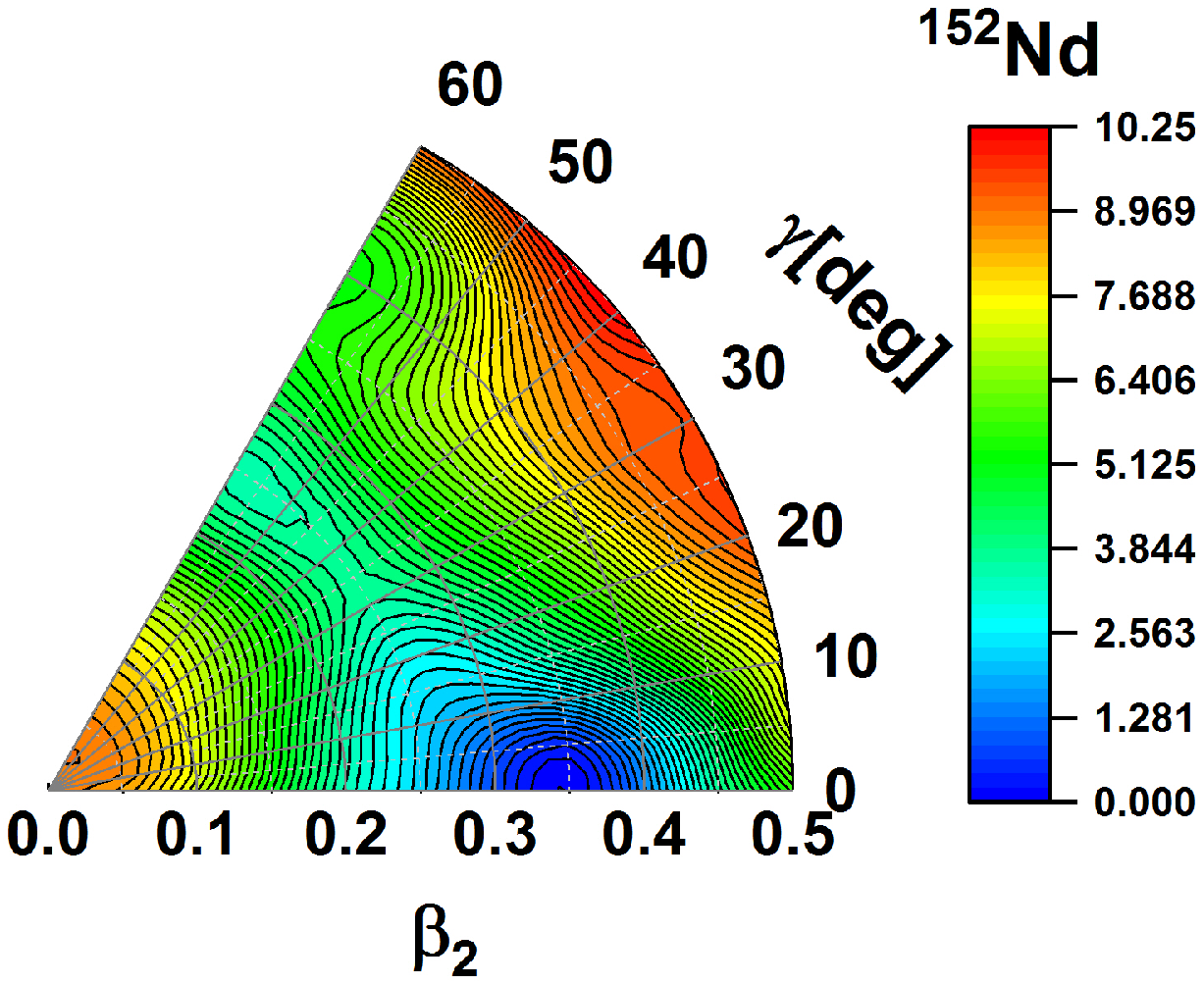}
\includegraphics[scale=0.28]{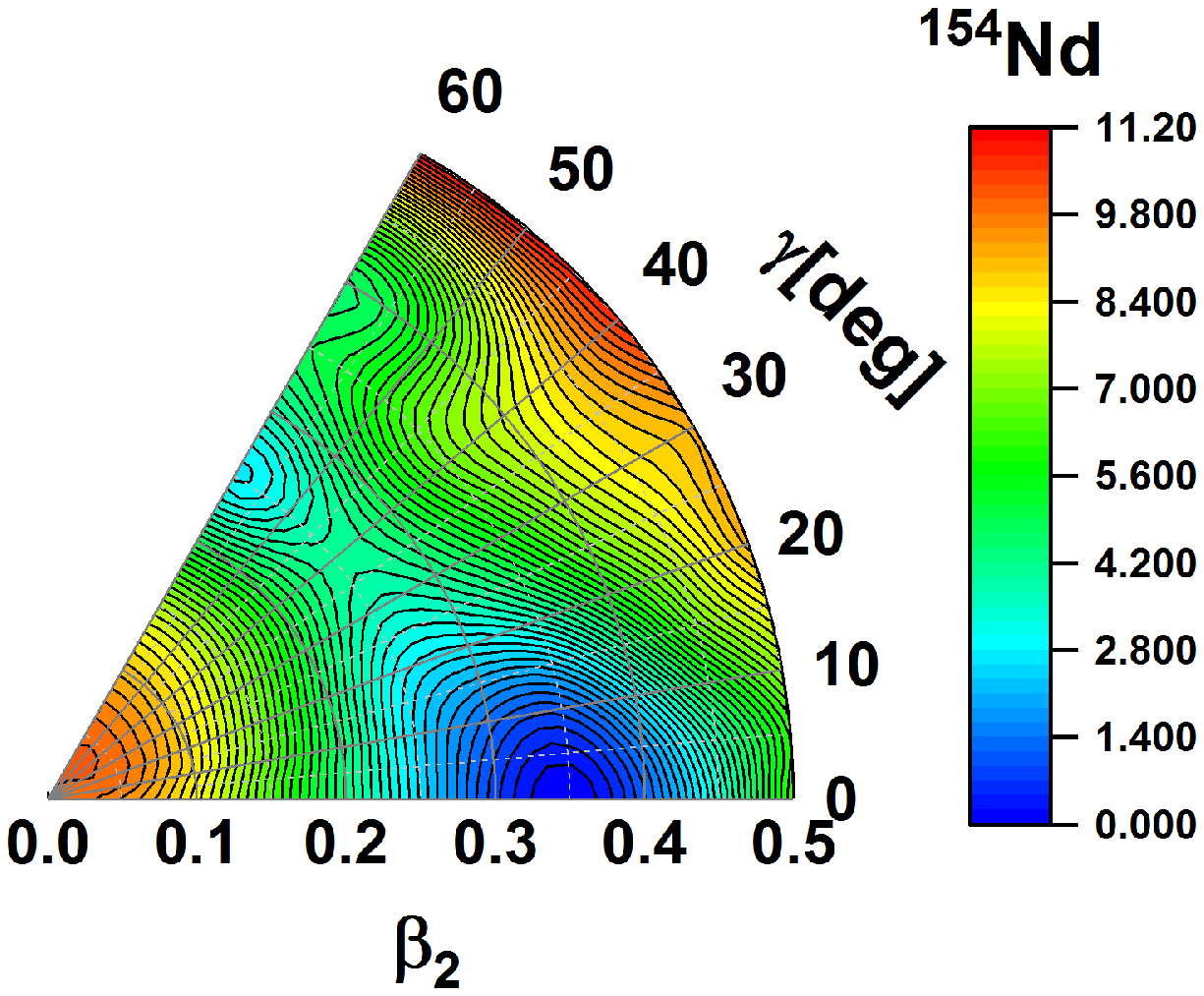}
\includegraphics[scale=0.28]{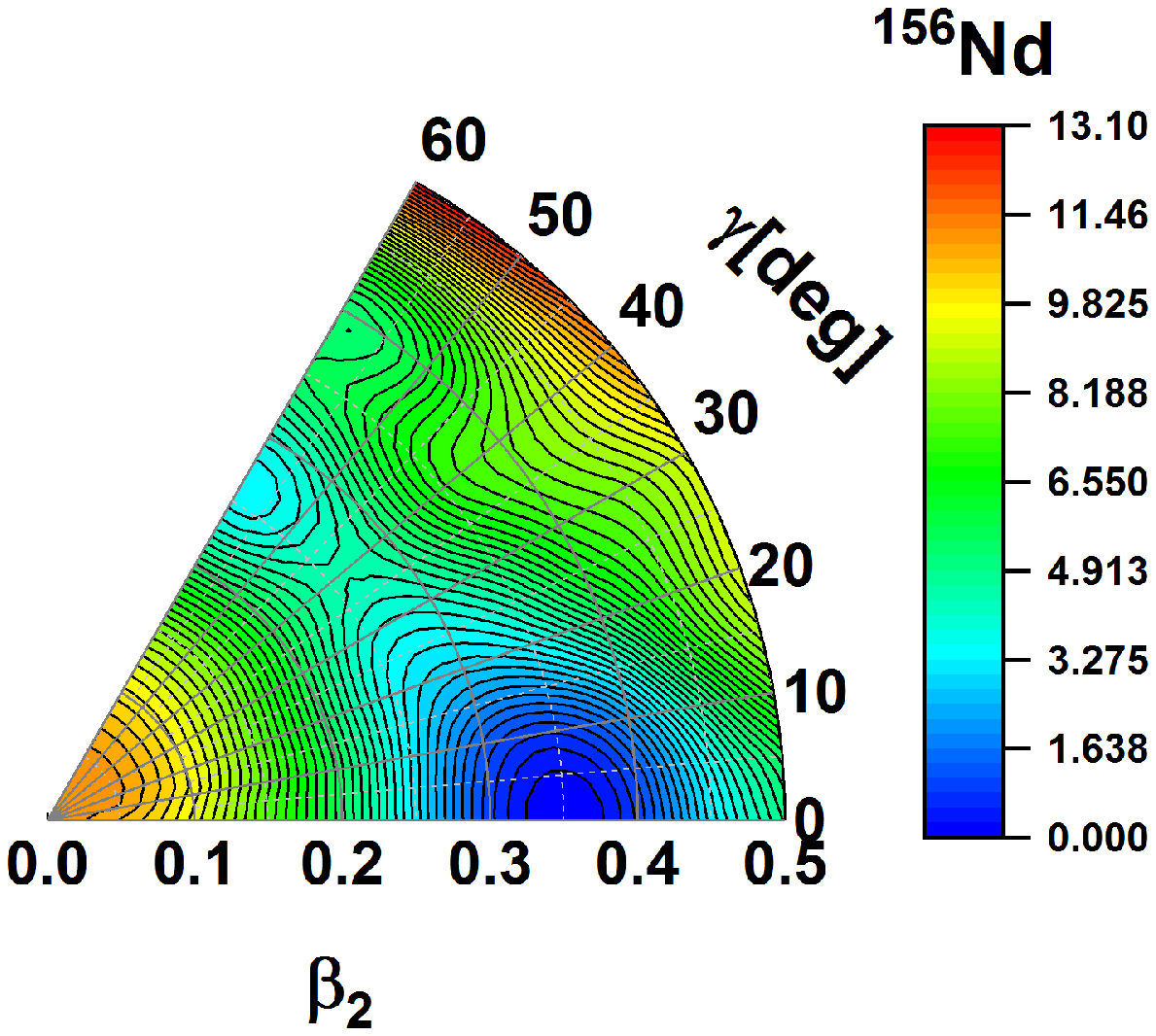}
\caption{\label{fig:figure51}(Color online) Same as Fig.\ref{fig:figure21} 
for the nuclei $^{142-156}$Nd.}
\end{figure}

\begin{figure}
\centering
\includegraphics[scale=0.28]{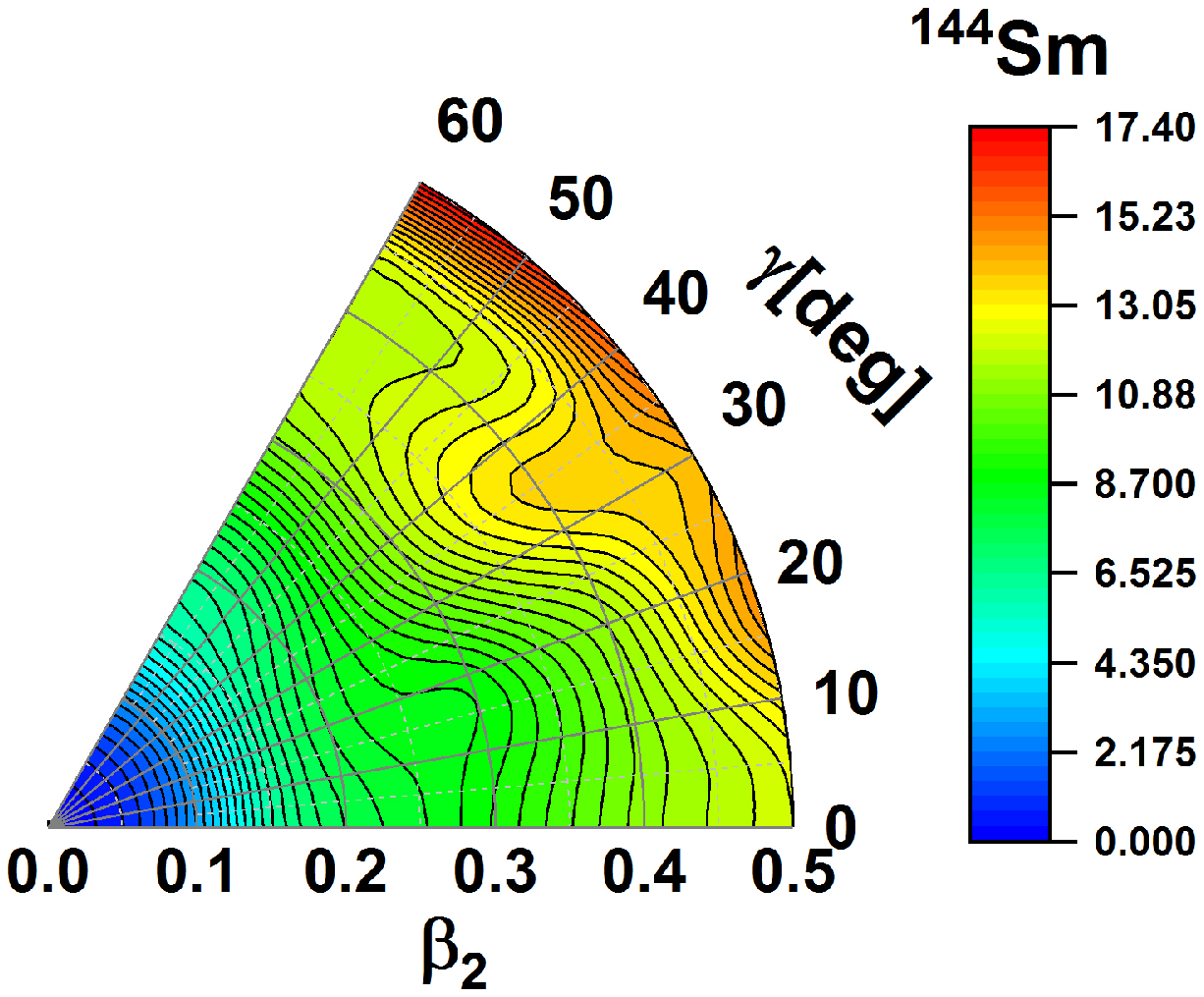}
\includegraphics[scale=0.28]{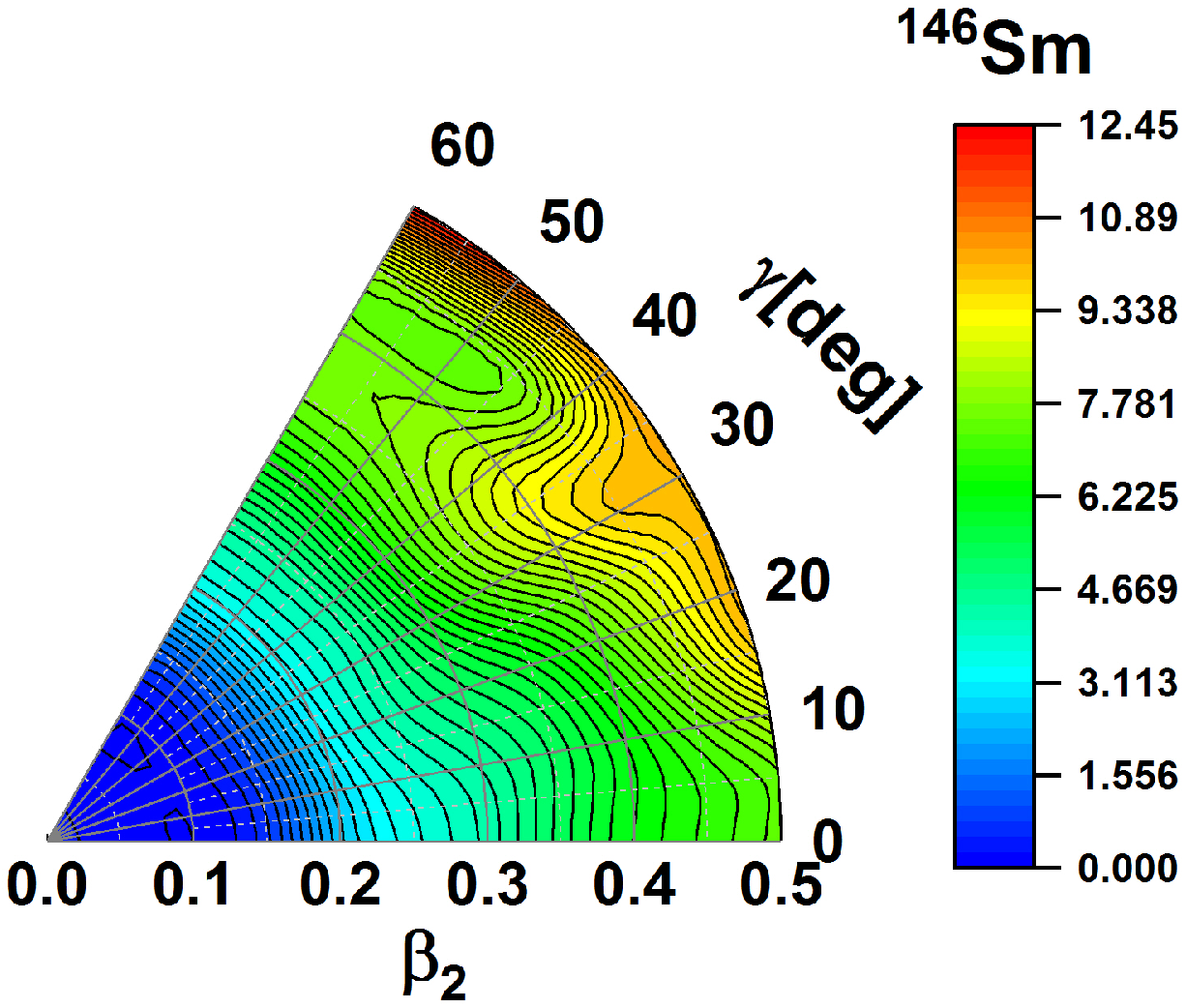}
\includegraphics[scale=0.28]{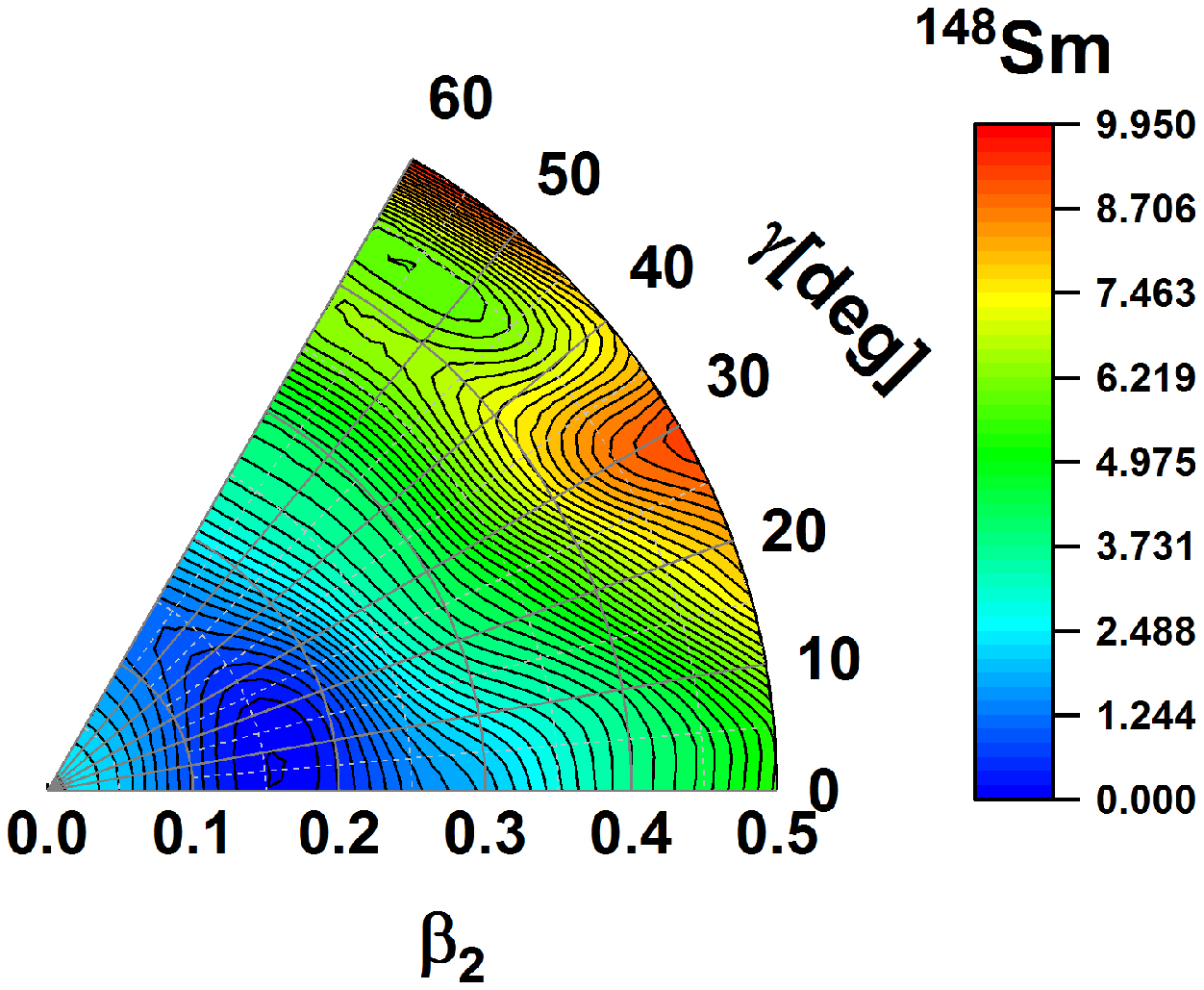}
\includegraphics[scale=0.28]{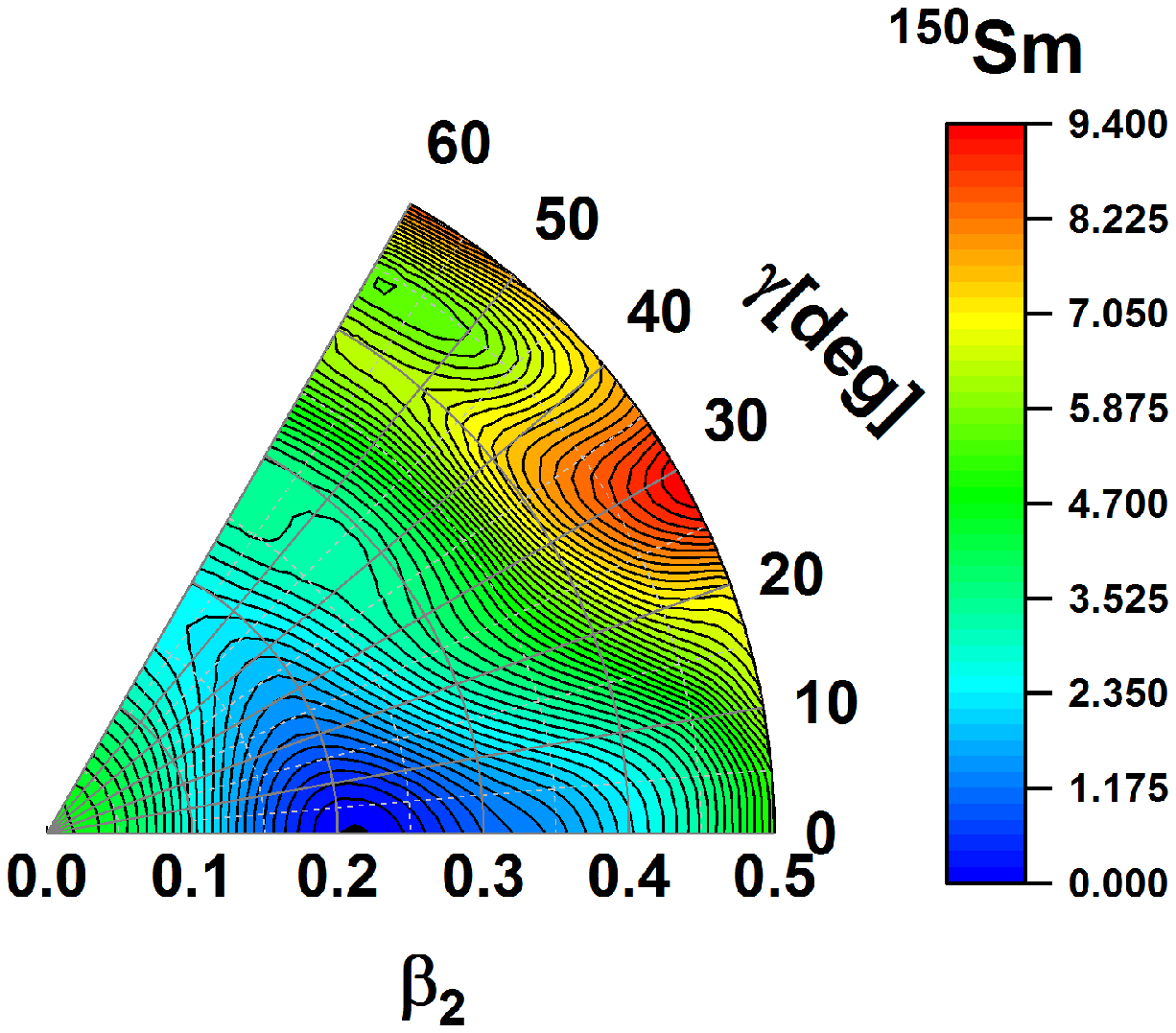}
\includegraphics[scale=0.28]{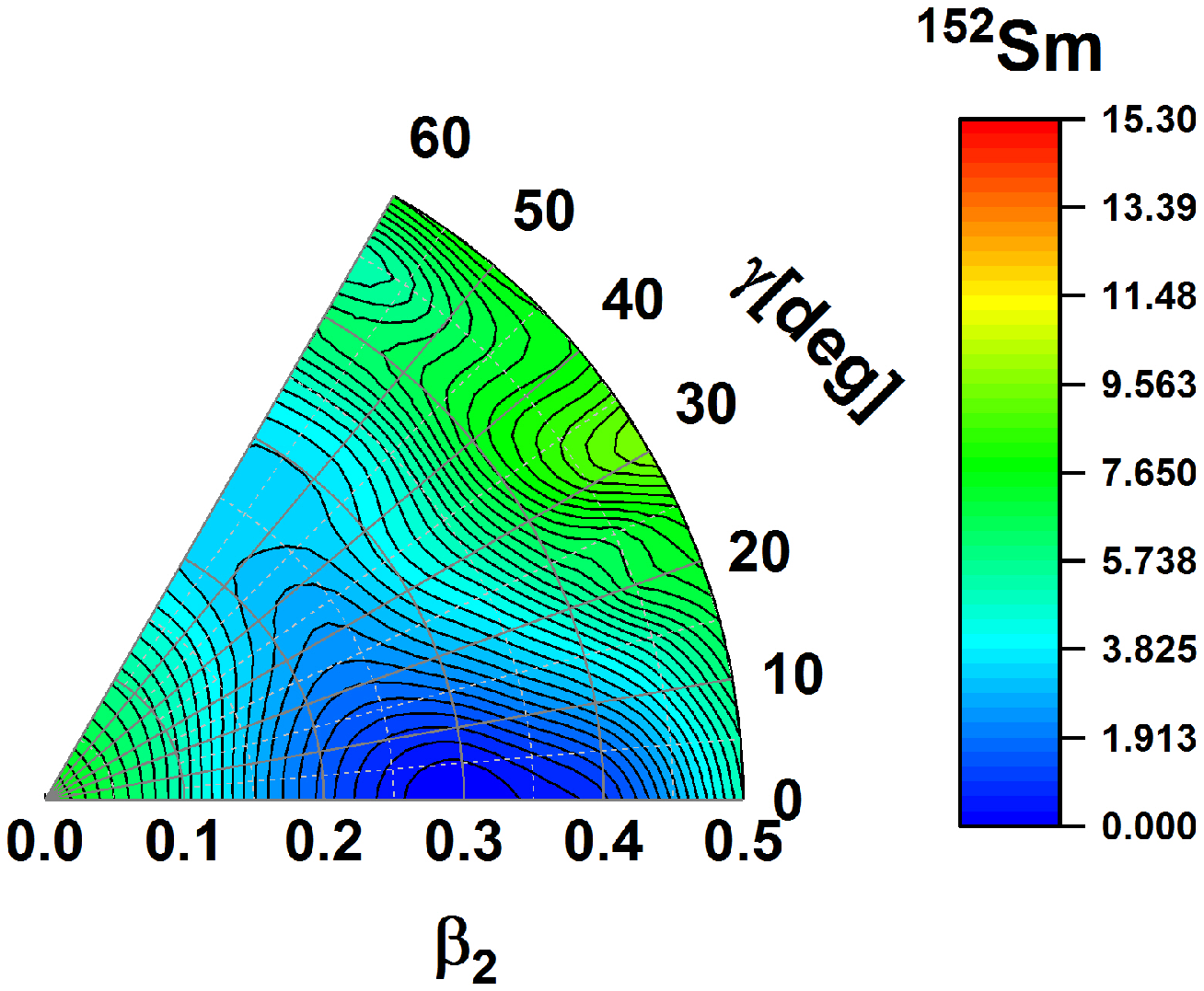}
\includegraphics[scale=0.28]{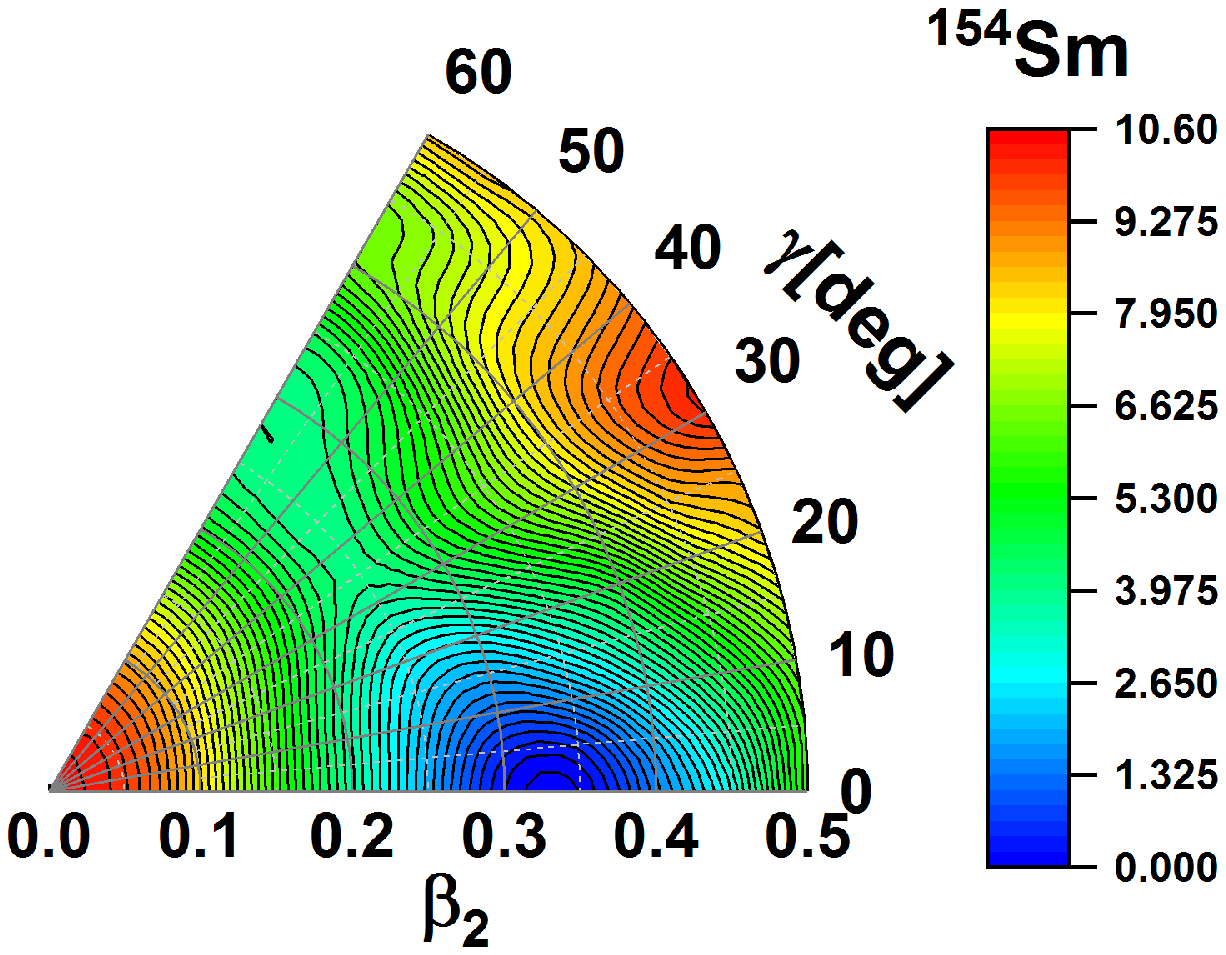}
\includegraphics[scale=0.28]{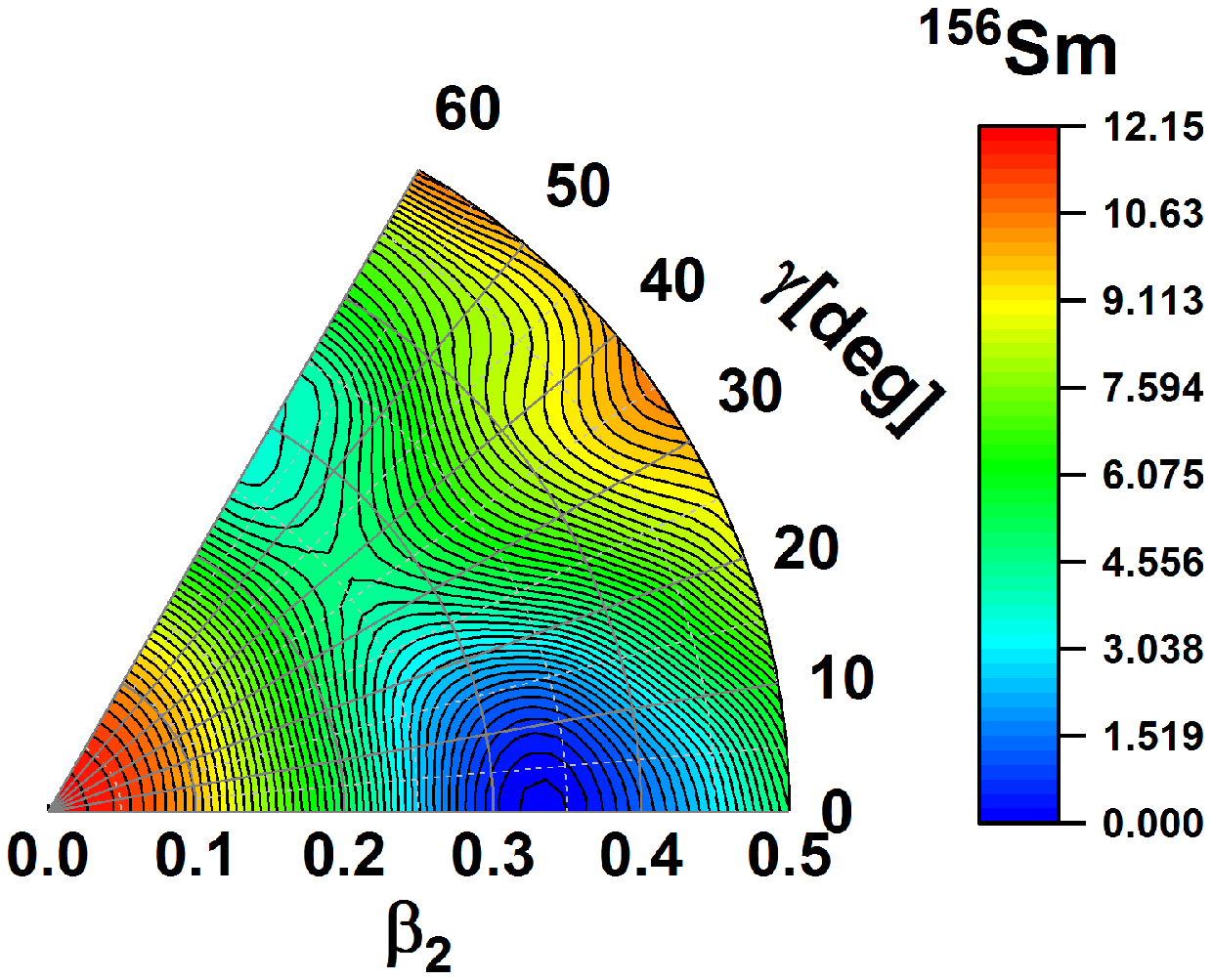}
\includegraphics[scale=0.28]{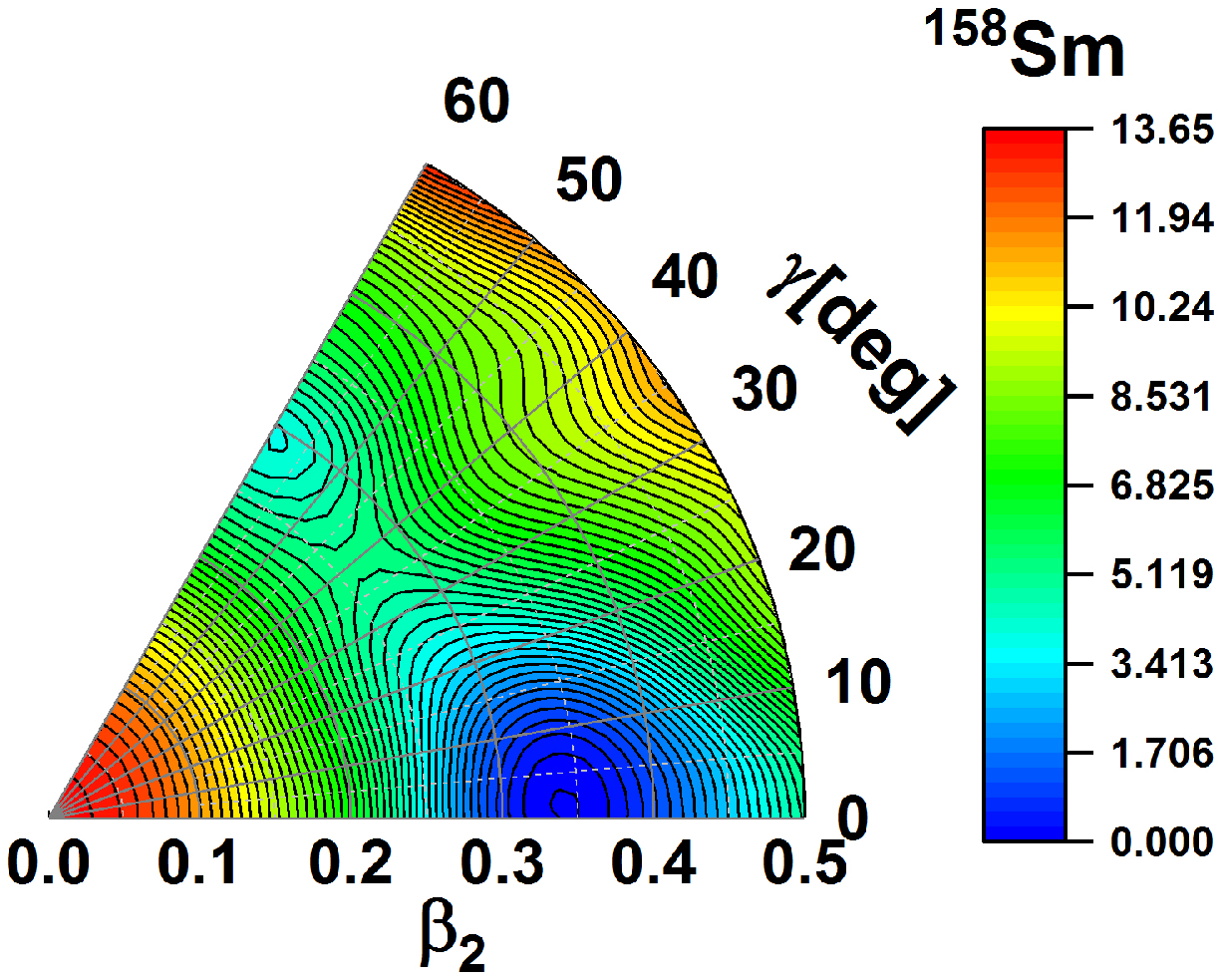}
\caption{\label{fig:figure52}(Color online) Same as Fig.\ref{fig:figure21} 
for the nuclei $^{144-158}$Sm.}
\end{figure}

\begin{figure}
\centering
\includegraphics[scale=0.28]{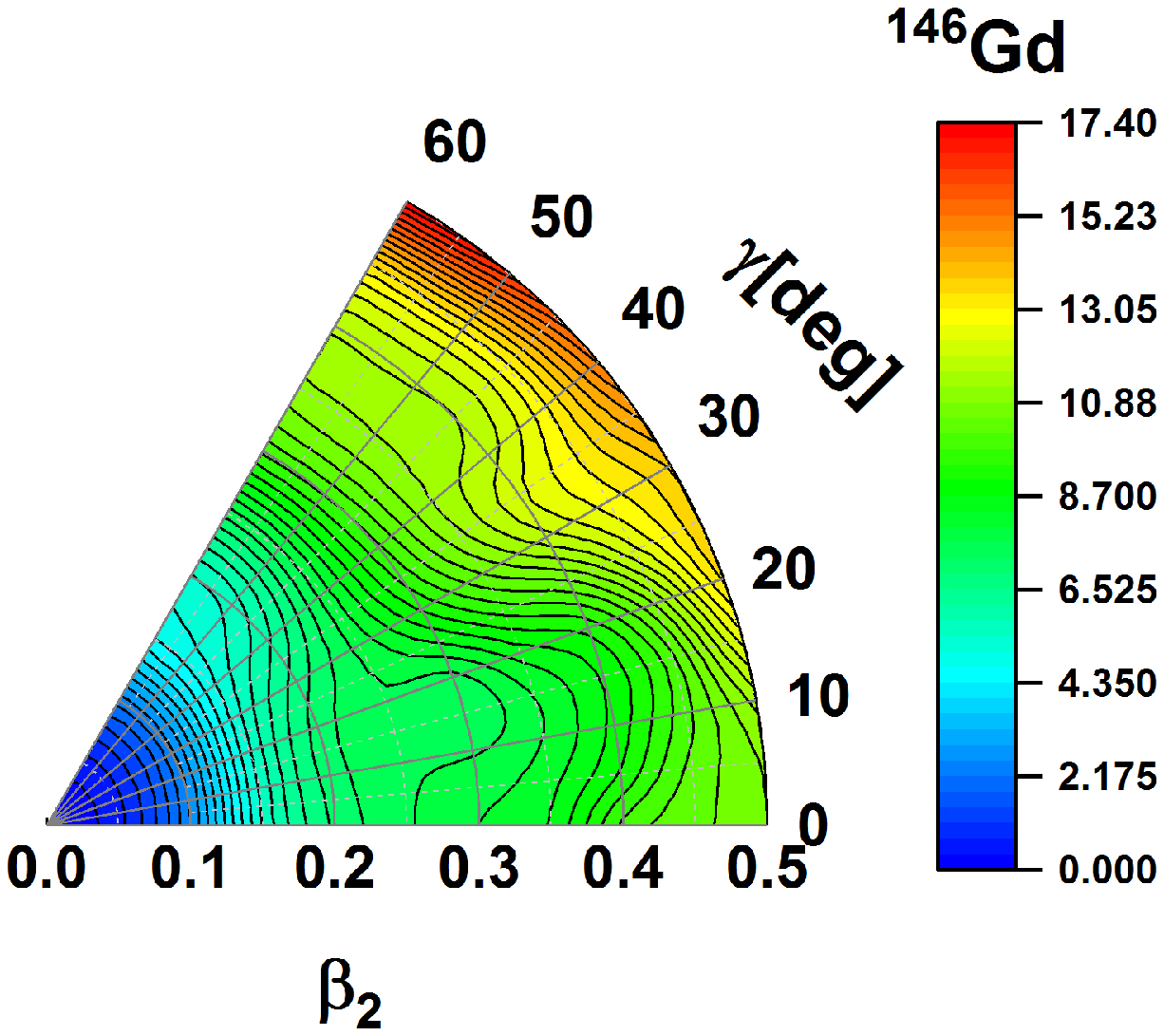}
\includegraphics[scale=0.28]{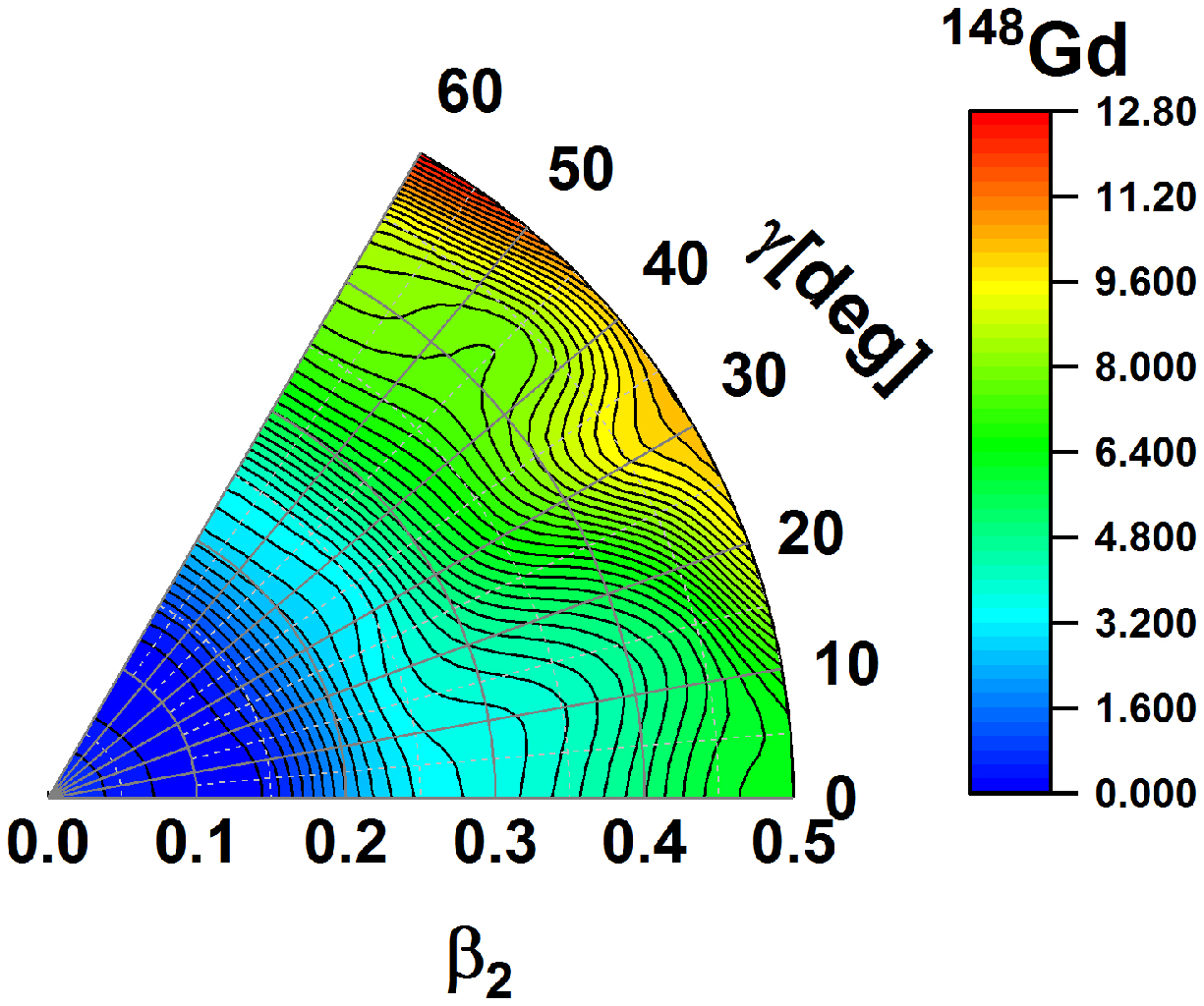}
\includegraphics[scale=0.28]{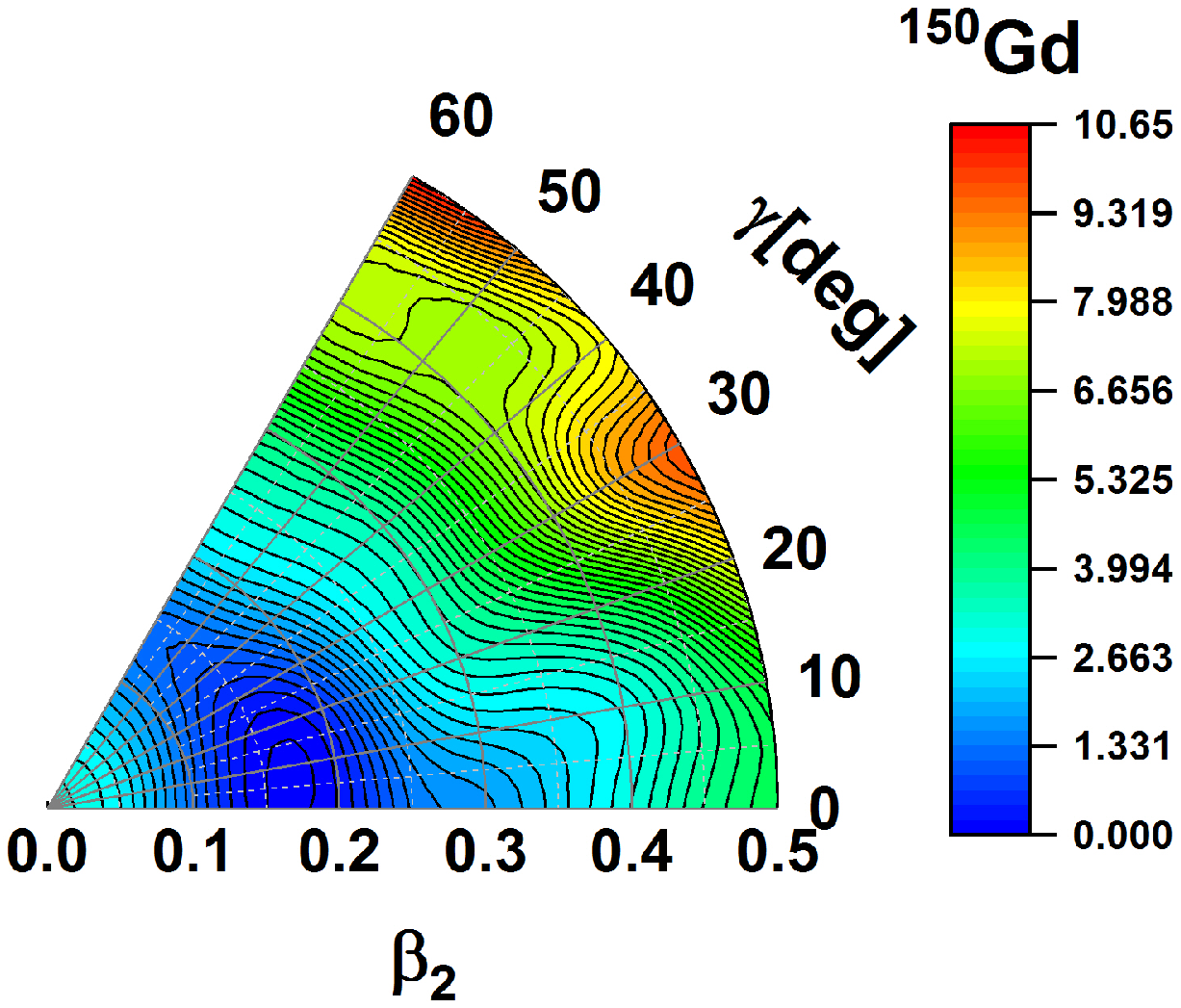}
\includegraphics[scale=0.28]{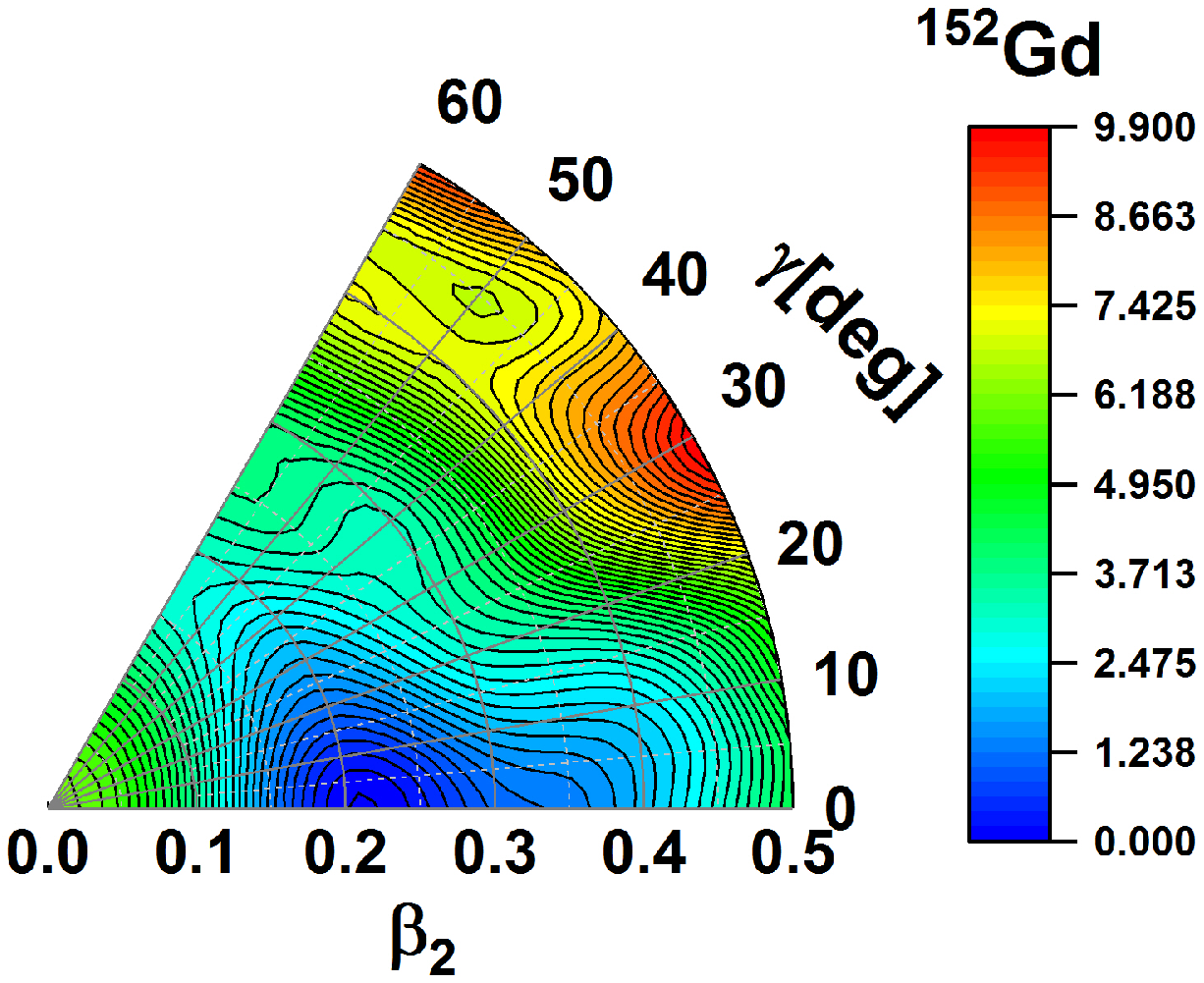}
\includegraphics[scale=0.28]{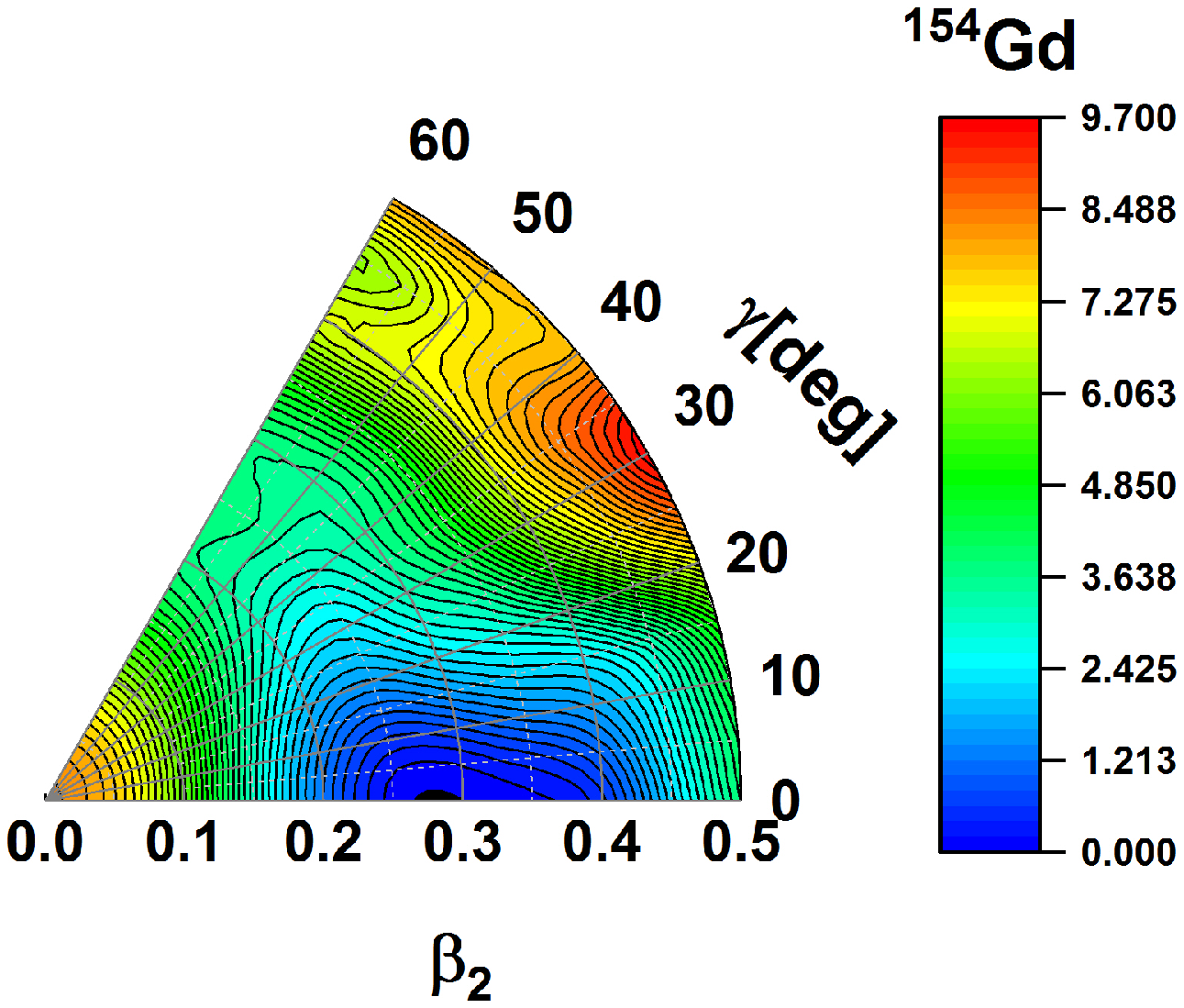}
\includegraphics[scale=0.28]{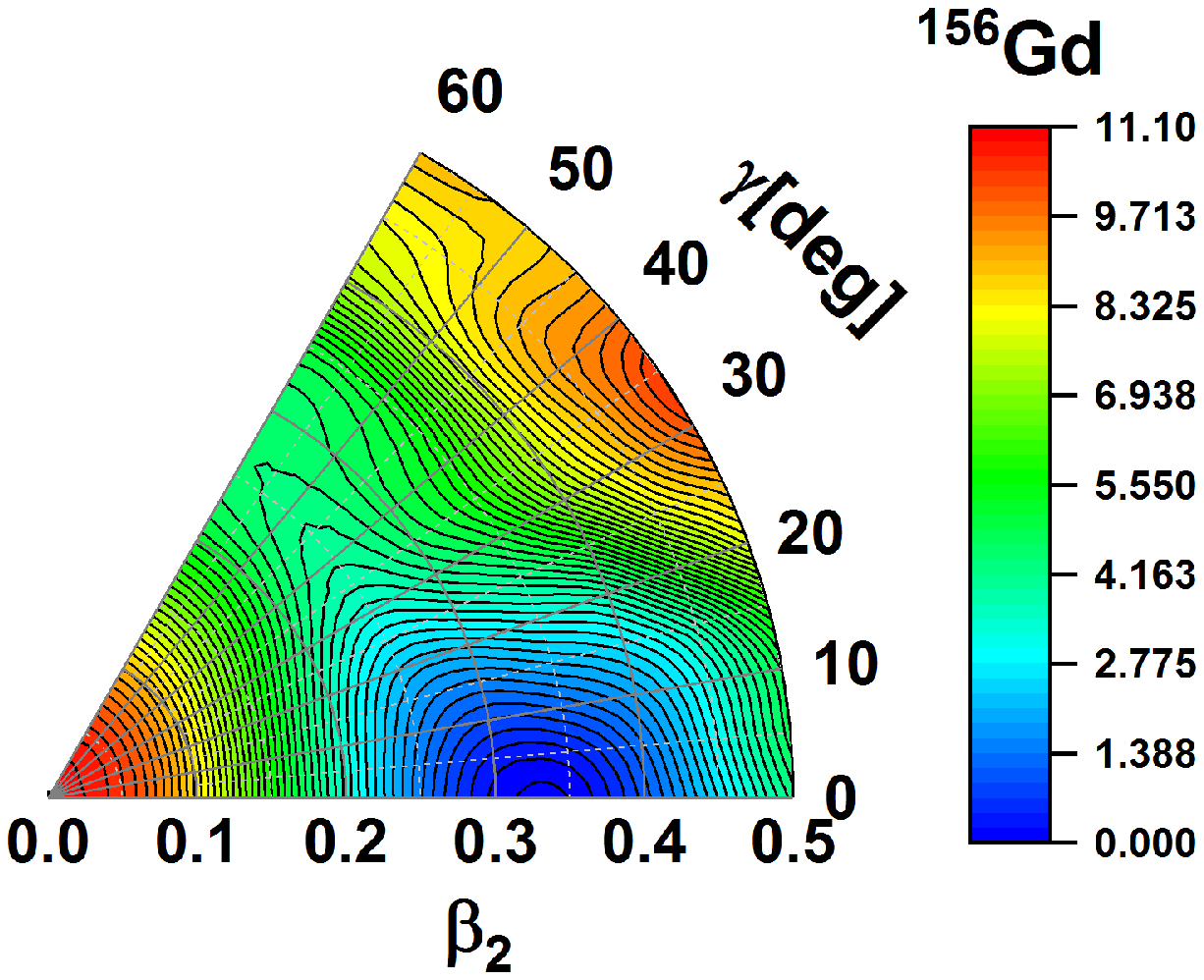}
\includegraphics[scale=0.28]{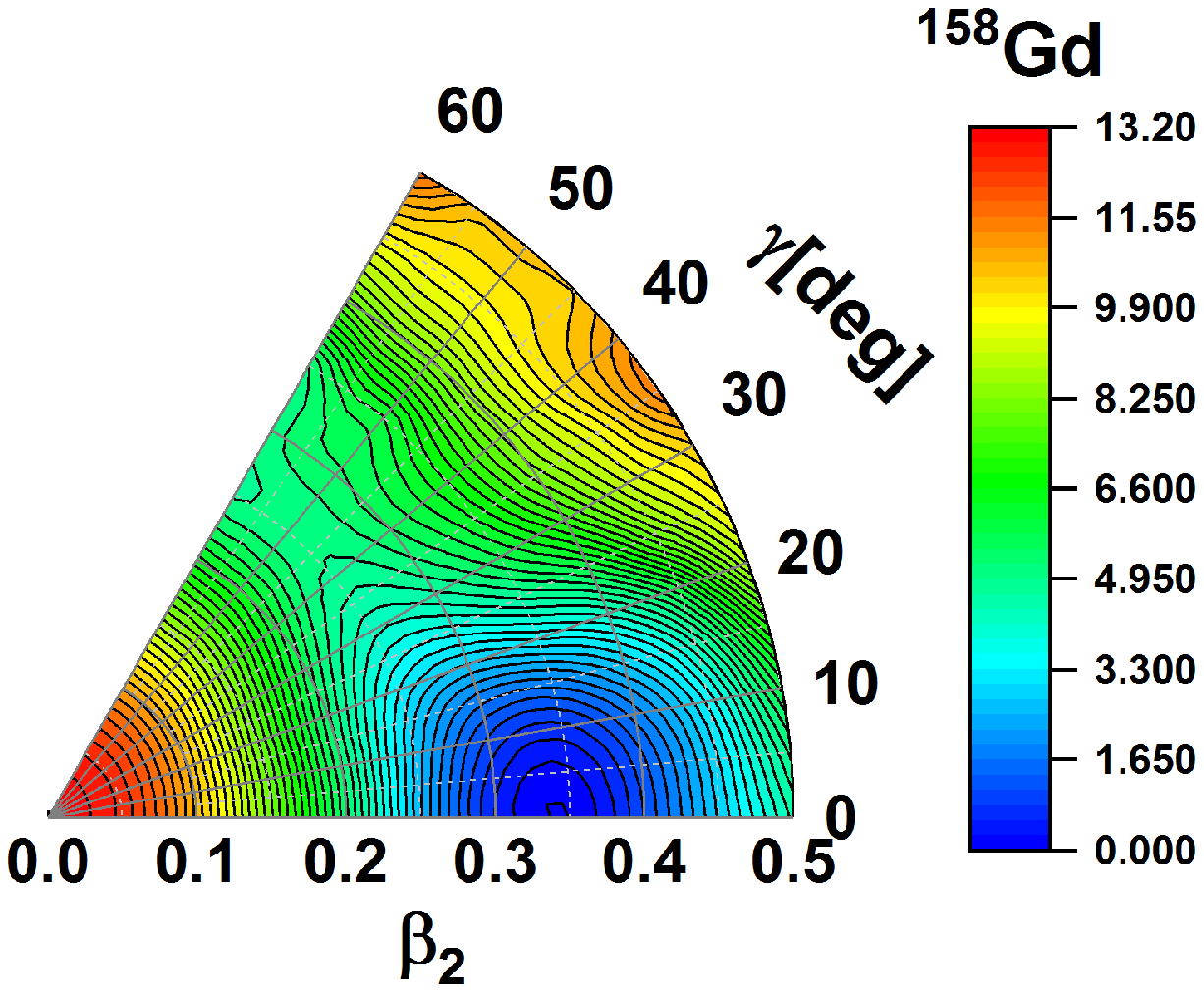}
\caption{\label{fig:figure53}(Color online) Same as Fig.\ref{fig:figure21} 
for the nuclei $^{146-158}$Gd.}
\end{figure}

\begin{figure}
\centering
\includegraphics[scale=0.28]{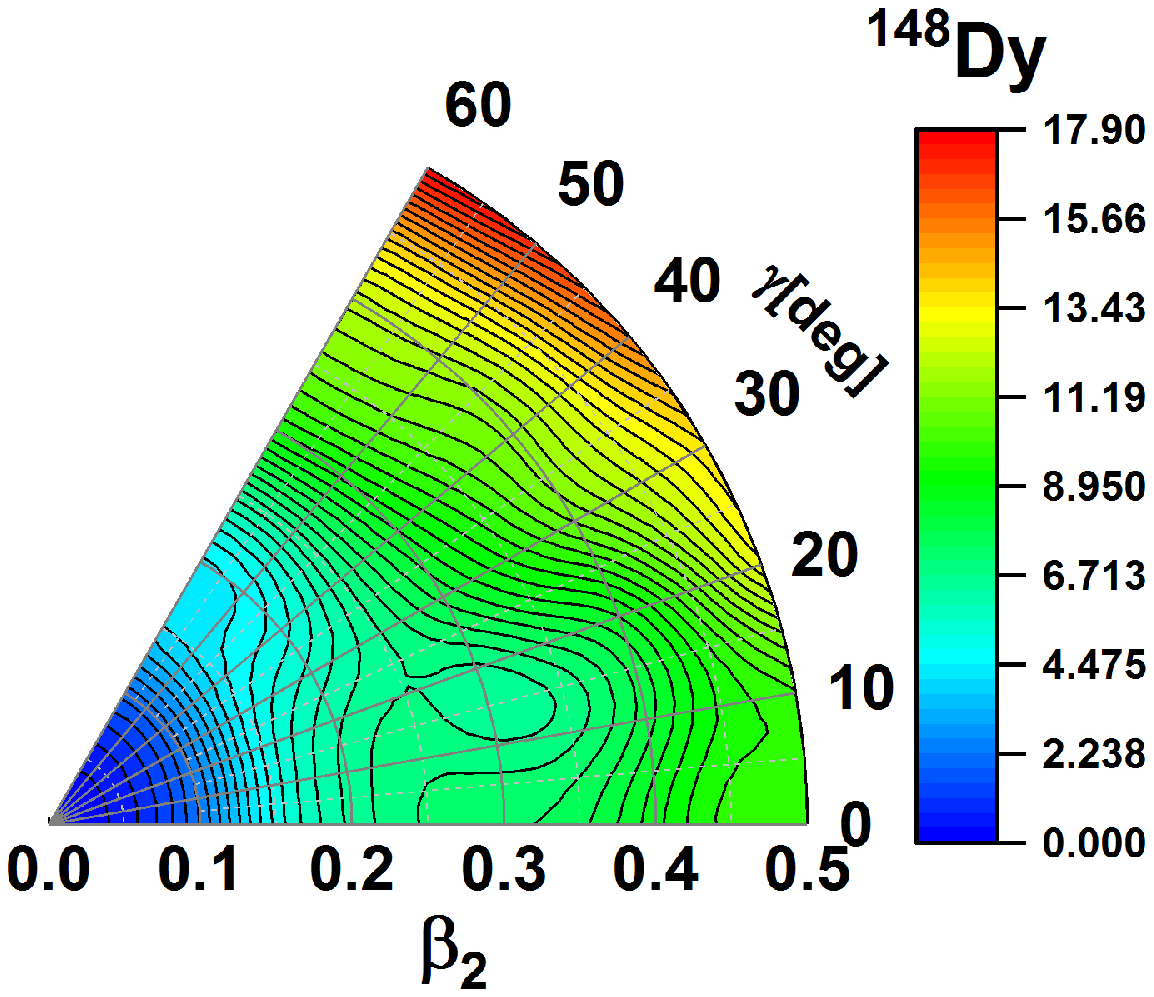}
\includegraphics[scale=0.28]{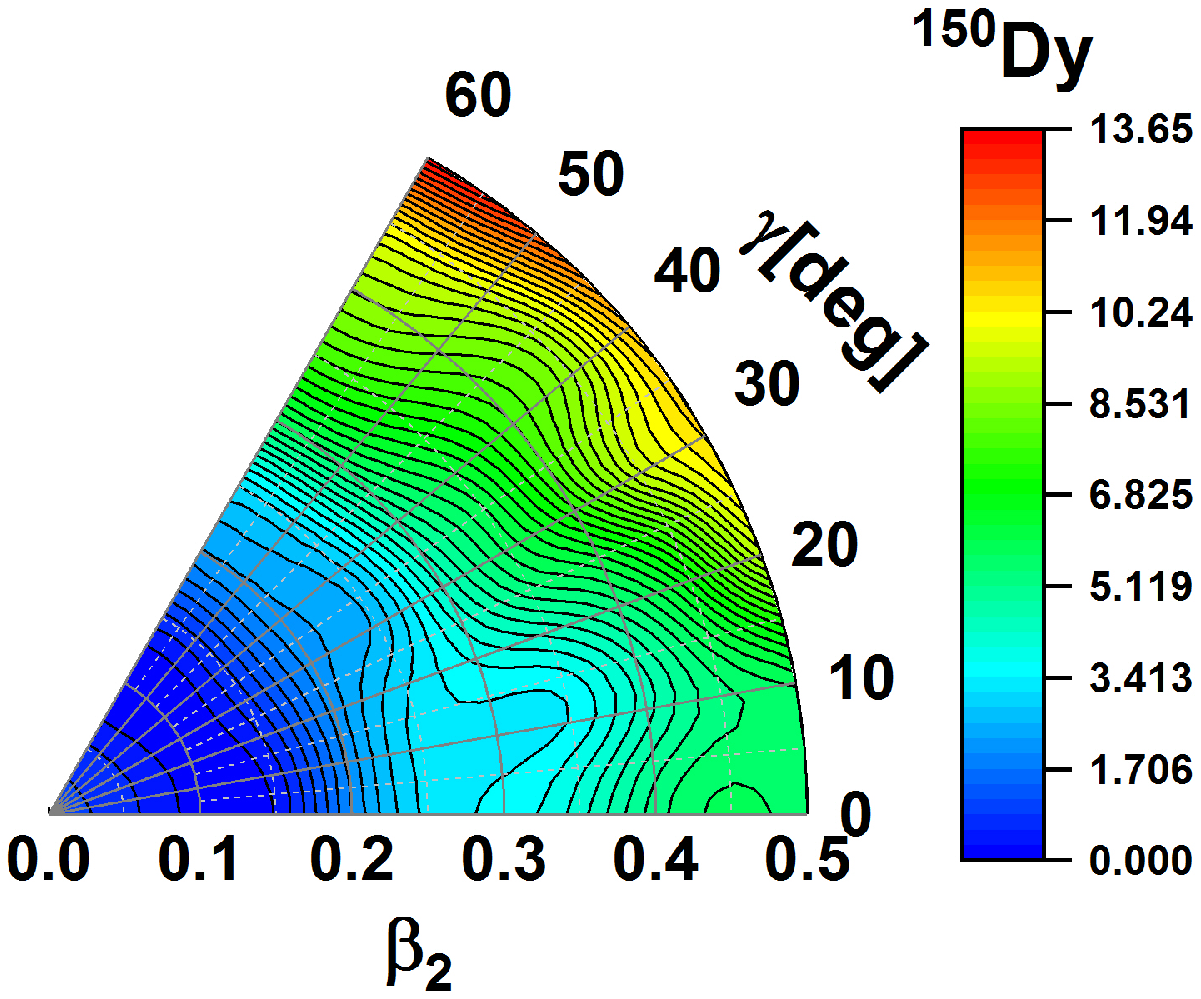}
\includegraphics[scale=0.28]{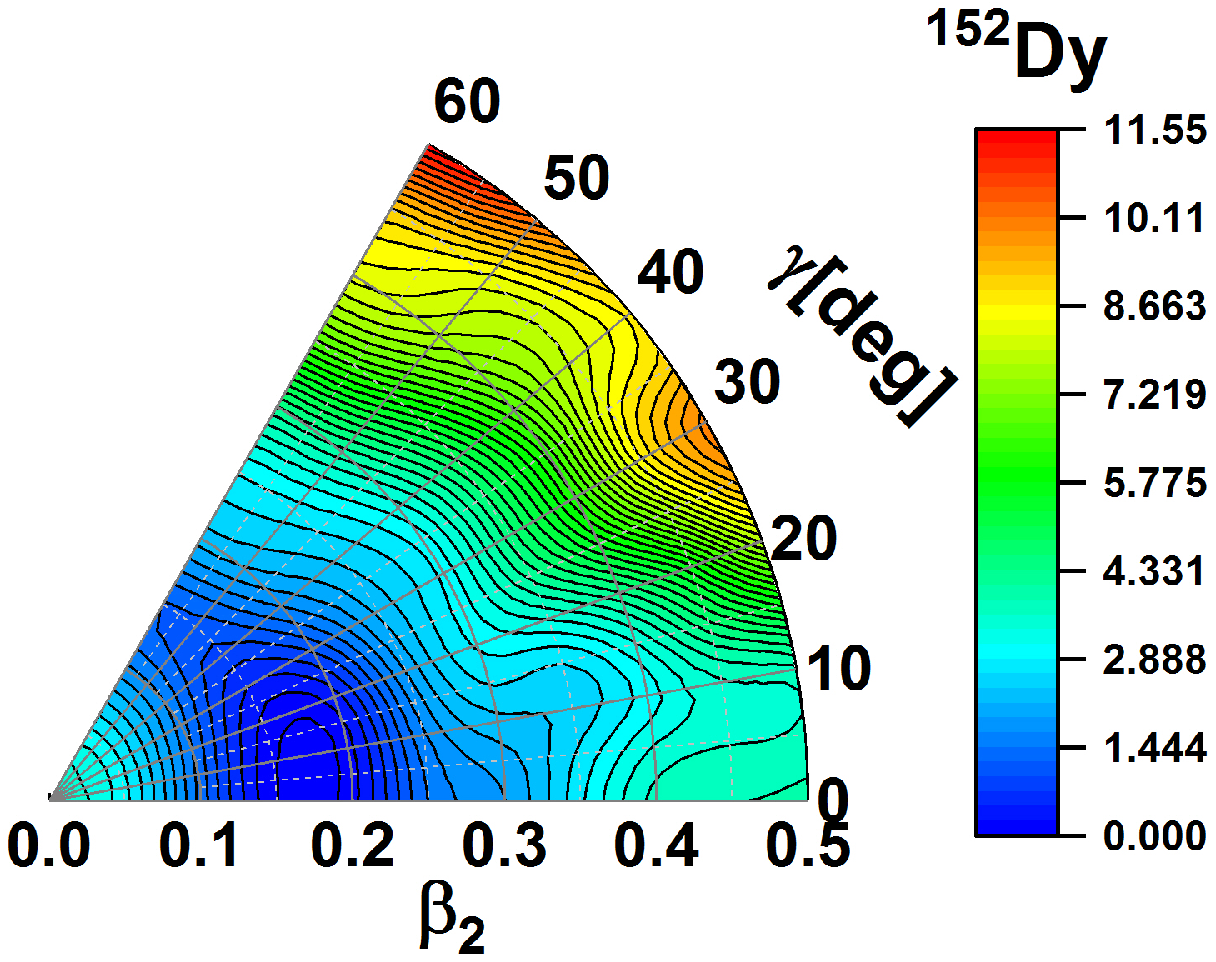}
\includegraphics[scale=0.28]{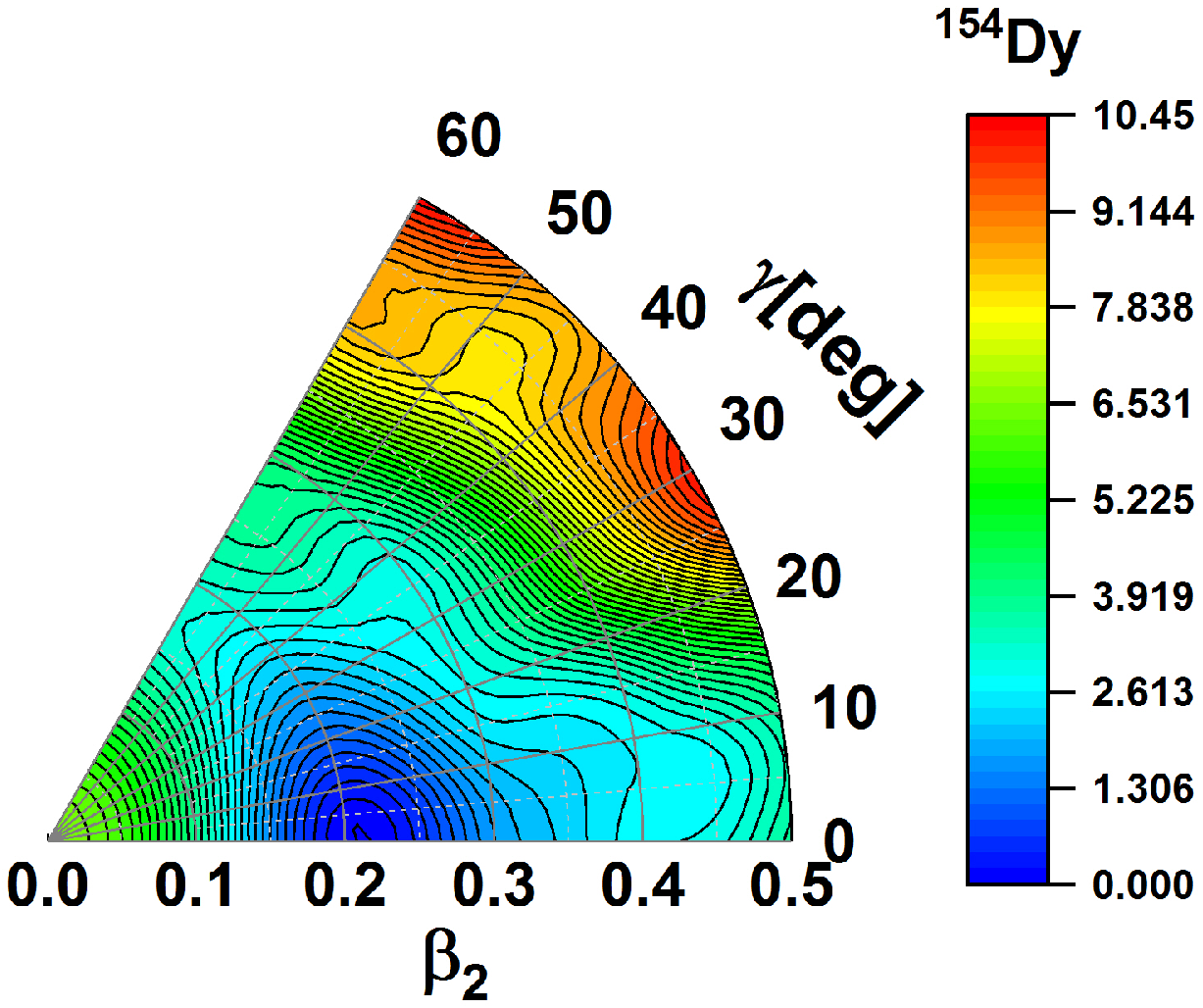}
\includegraphics[scale=0.28]{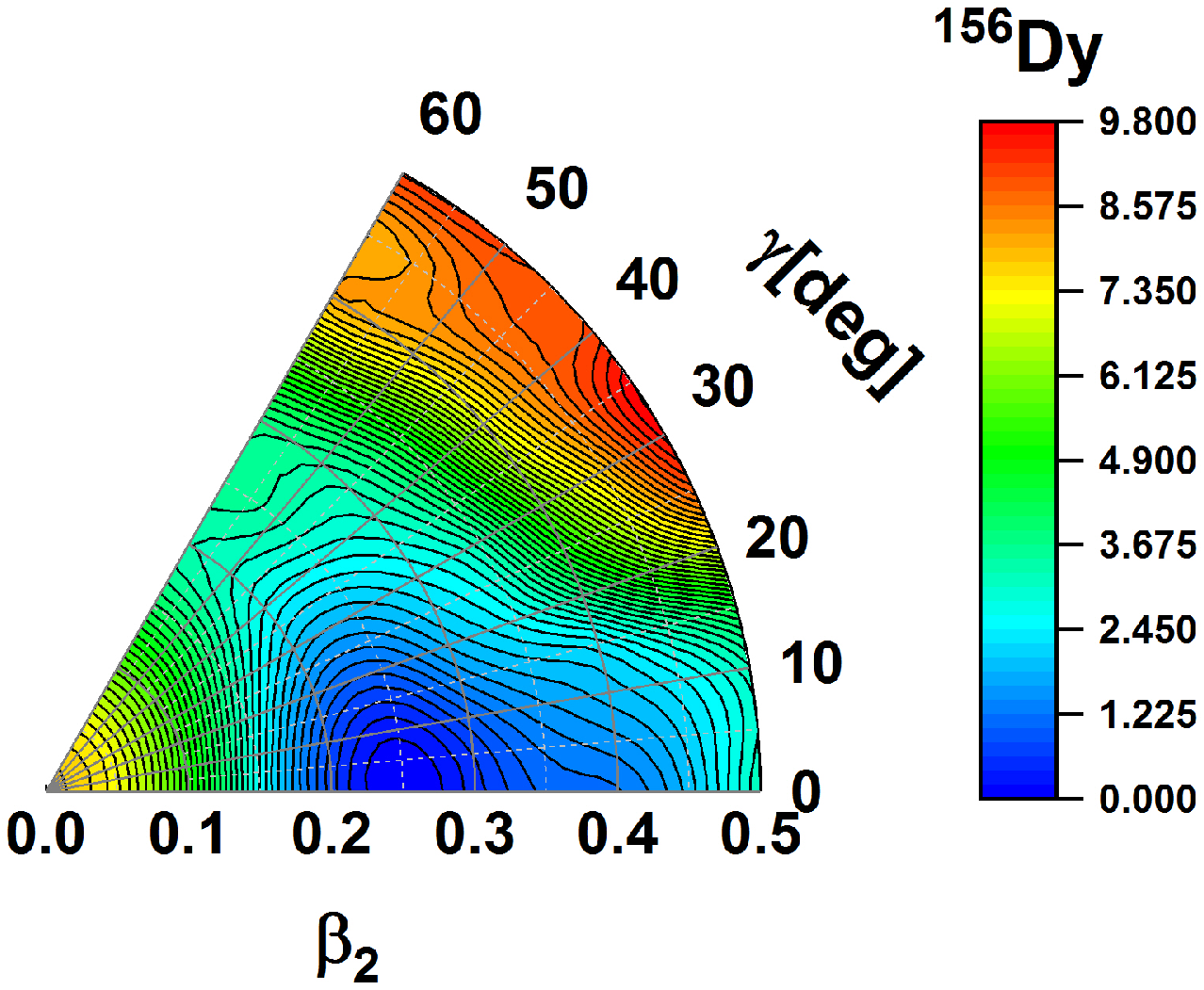}
\includegraphics[scale=0.28]{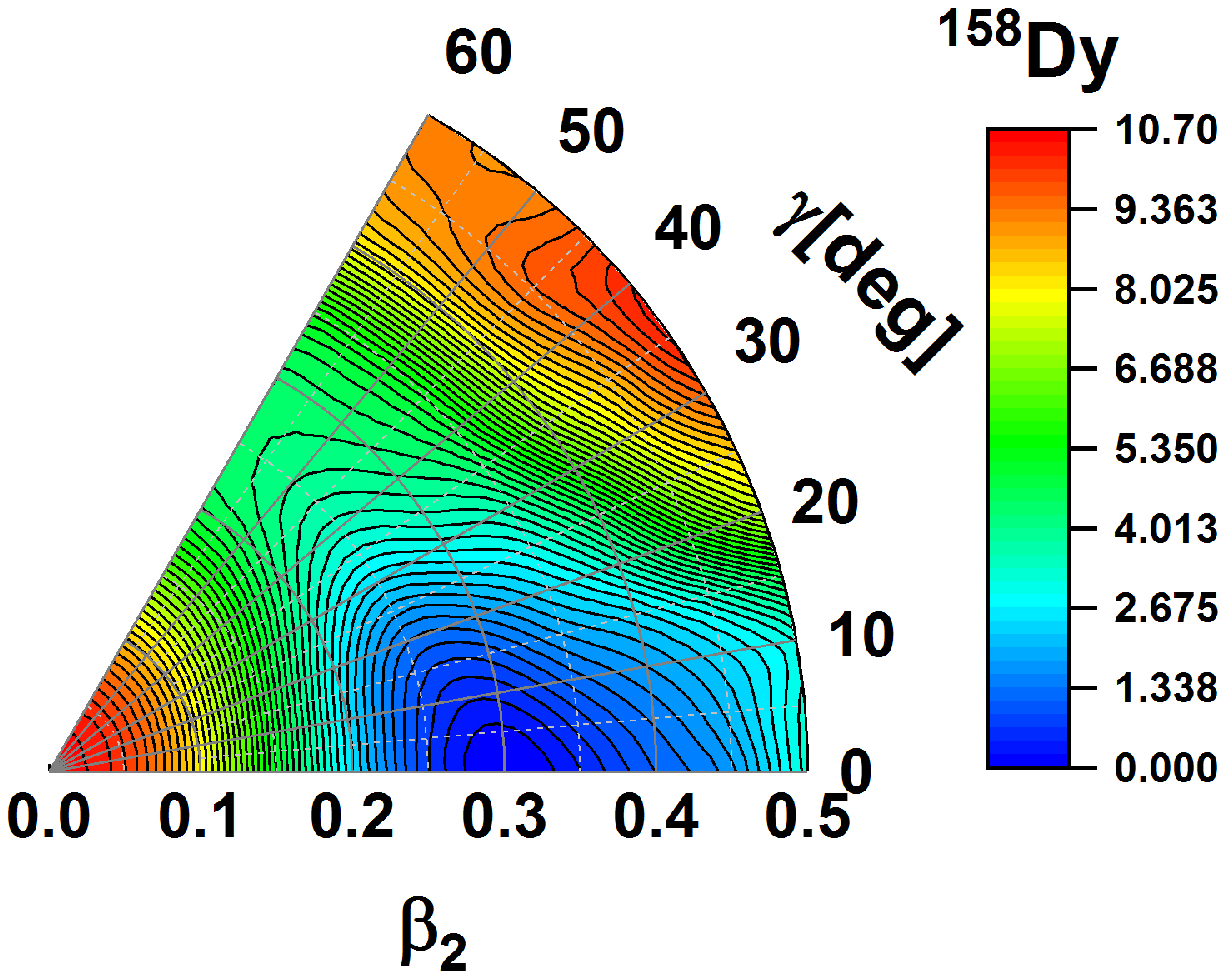}
\includegraphics[scale=0.28]{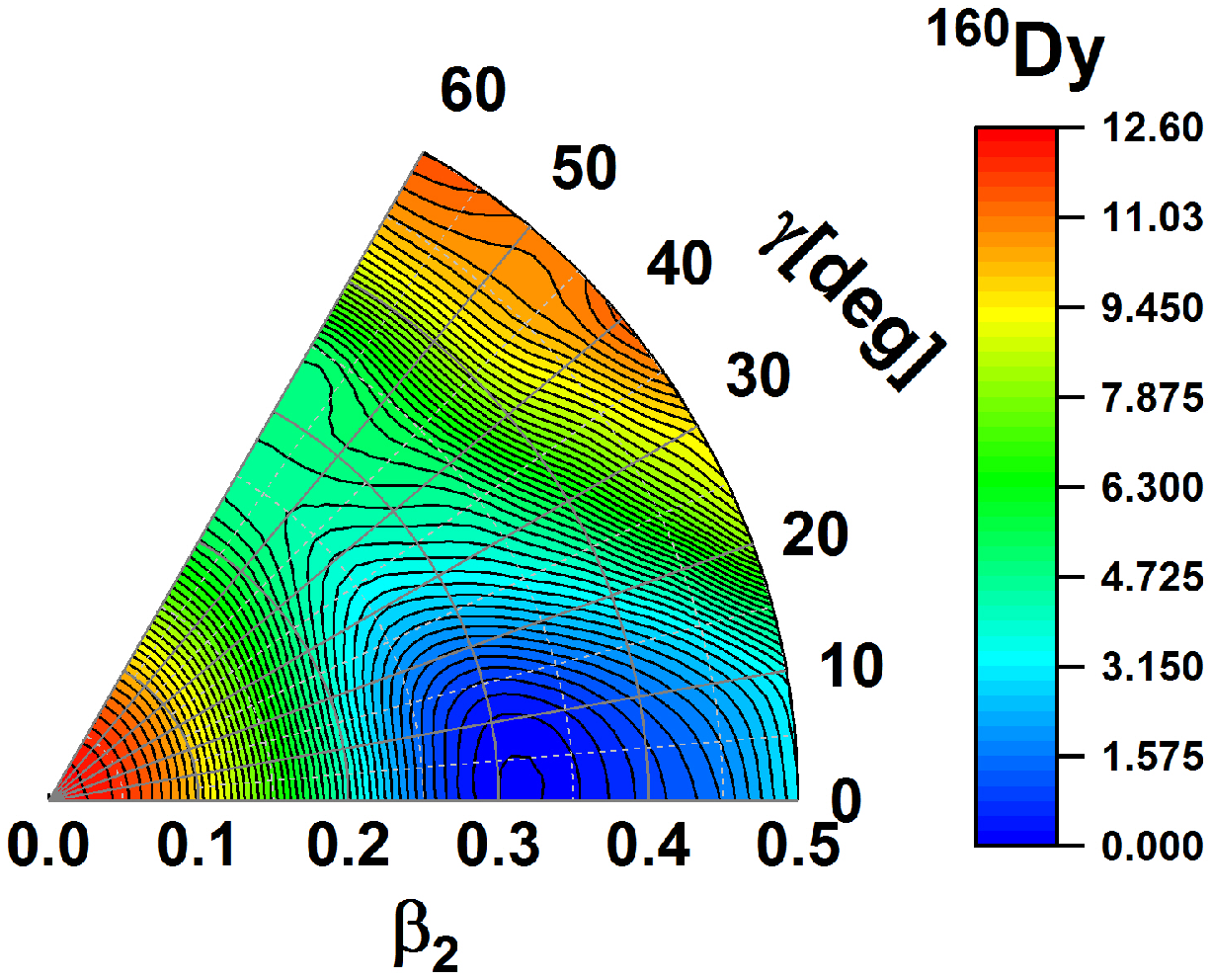}
\caption{\label{fig:figure54}(Color online) Same as Fig.\ref{fig:figure21} 
for the nuclei $^{148-160}$Dy.}
\end{figure}

In Fig.\ref{fig:figure51}, Fig.\ref{fig:figure52}, Fig.\ref{fig:figure53}, and Fig.\ref{fig:figure54}
we display the calculated PESs under triaxial symmetry for 
Nd(Z=60), Sm(Z=62), Gd(Z=64), and Dy(Z=66) isotopes, respectively.
We discuss the shape transition and the properties corresponding to the global minima 
for $^{142-156}$Nd, $^{144-158}$Sm, $^{146-158}$Gd, and $^{148-160}$Dy isotopes.
The ground state quadrupole deformations in the 
$\beta_2$-$\gamma$ plane corresponding to the global minima of the triaxial PESs 
are tabulated in Table~\ref{tab:table2}. 
The nucleus $^{142}$Nd is spherical, but $^{144}$Nd is slightly triaxial with its 
global minimum at $(0.05, 5^o)$. Further, it is shifting towards higher and higher
 prolate deformation. For $^{144}$Nd and $^{146}$Nd, the $\gamma$-softness is there 
covering prolate to oblate region. But as we move further the $\gamma$-softness shifted 
towards the prolate region and becomes rigid towards the oblate region. There is a flat 
character in $\gamma$-softness towards prolate region in $^{150}$Nd, after then it 
concentrates. The ground state of $^{150}$Nd nucleus is axially prolate$(0.3, 0^o)$. 
The convergence of circles around the ground state covers $\beta_2$ from 0.15 to 0.45, 
and the softness in $\gamma$ is just $10^o$ up to 1.2MeV of energy.
Similar behaviour can be seen in case Sm(Z=62) isotopes. $^{144}$Sm is spherical in shape, 
and as we move further the shape is shifting towards larger prolate deformation. 
$\gamma$-softness is there in $^{146}$Sm and $^{148}$Sm, but it shifted towards more 
prolate region and becomes rigid towards the oblate region. 
The nuclei, $^{150}$Sm$(0.2, 0^o)$,  $^{152}$Sm$(0.3, 0^o)$, 
have their ground state as axially prolate. The convergence of the circles around
their global minimum in general, covers the $\beta_2$ space from 0.15 to 0.45. 
But, the $\gamma$-softness is different for them. It is $35^o$ up to 1.86MeV of 
energy for $^{150}$Sm, $15^o$ up to 1.9MeV for $^{152}$Sm.
In case of Gd(Z=64) and Dy(Z=66) isotopes, we can see exactly the same behaviour in 
the shape transition as in case of Nd(Z=60) and Sm(Z=62).  The only difference is that 
after the spherical global minima in case of $^{146}$Gd and $^{148}$Dy, the nuclei 
$^{148}$Gd and $^{150}$Dy is having the oblate global minimum. The nuclei 
$^{154}$Gd$(0.3, 0^o)$, and $^{156}$Dy$(0.25, 0^o)$ have their ground state as 
axially prolate, and circles around their global minimum covers the $\beta_2$ 
space from 0.15 to 0.45. The $\gamma$-softness is
$15^o$ up to 1.55MeV for $^{154}$Gd, and $30^o$ up to 1.9MeV for $^{156}$Dy.
Here, the nuclei $^{150}$Sm and $^{156}$Dy show more $\gamma$-soft nature than others. 
It is in agreement with the experiment reporting more $\gamma$-soft behaviour of 
$^{156}$Dy amongst the N=90 isotones~\cite{Moller:2006yx}. 
In general all these nuclei show flatness around their global minimum 
within the energy range 0 - 2.5MeV approximately in the 
axially prolate ($0.15\leq\beta_2\leq 0.45$) regime effectively, 
thus reflecting the axial PES behaviour.   
\begin{table*}[htb]
 \caption{\label{tab:table2}
The quadrupole deformation ($\beta_2$, $\gamma$) of the global minima in Nd, Sm, Gd and Dy isotopes,
calculated within the RHB formalism with DD-ME2. $E_{ax}$ and $E_{tr}$ are the total energies for the
global minima under axial and triaxial symmetry, respectively.}
\centering
\begin{tabular}{rrrrrrrrrrrr}
\hline
\hline
\textrm{Nuclei}&
\textrm{$\beta_2$}&
\textrm{$\gamma$}&
\textrm{$E_{tr}$}&
\textrm{$E_{ax}$}&
\textrm{$\Delta E_{tr}$}&
\textrm{Nuclei}&
\textrm{$\beta_2$}&
\textrm{$\gamma$}&
\textrm{$E_{tr}$}&
\textrm{$E_{ax}$}&
\textrm{$\Delta E_{tr}$}\\
\hline
$^{142}$Nd&	0&$0^{\circ}$&1187.941&1187.647&0.294&$^{144}$Sm&	0&$0^{\circ}$&1197.203&1197.306&-0.103\\
$^{144}$Nd&	0.05&$5^{\circ}$&1198.508&1198.220&0.288	&$^{146}$Sm&	0.1&$0^{\circ}$&1209.314&1209.354&-0.039\\
$^{146}$Nd&	0.15&$0^{\circ}$&1210.364&1210.119&0.245&$^{148}$Sm&	0.15&$0^{\circ}$&1223.253&1223.204&0.049\\
$^{148}$Nd&	0.20&$0^{\circ}$&1223.080&1222.847&0.232&$^{150}$Sm&	0.2&$0^{\circ}$&1237.640&1237.350&0.290\\
$^{150}$Nd&	0.3&	$0^{\circ}$&1235.211&1235.002&0.200&$^{152}$Sm&	0.3&$0^{\circ}$&1251.460&1251.207&0.253\\
$^{152}$Nd&	0.35&$0^{\circ}$&1248.070&1247.891&0.178	&$^{154}$Sm&	0.35&$0^{\circ}$&1265.250&1264.900&	0.350\\
$^{154}$Nd&	0.35&$0^{\circ}$&1258.760&1258.550&0.210	&$^{156}$Sm&	0.35&$0^{\circ}$&1277.534&1277.171&	0.363\\
$^{156}$Nd&	0.35&$0^{\circ}$&1269.134&1268.802&0.332&$^{158}$Sm&	0.35&$0^{\circ}$&1289.398&1289.000&	0.397\\
$^{146}$Gd&	0&$0^{\circ}$&1204.987&1205.163&-0.175&$^{148}$Dy&	0&$0^{\circ}$&1211.099&1211.300&-0.200\\
$^{148}$Gd&	0.1&$60^{\circ}$&1218.817&1218.812&0.005&$^{150}$Dy&	0.1&$60^{\circ}$&1226.543&1266.496&0.046\\
$^{150}$Gd&	0.15&$0^{\circ}$&1234.344&1234.311&0.033&$^{152}$Dy&	0.15&$0^{\circ}$&1243.477&1243.483&-0.006\\
$^{152}$Gd&	0.2&$0^{\circ}$	&1250.100&1249.914&0.185&$^{154}$Dy&	0.2&$0^{\circ}$&1260.620&1260.453&0.167\\
$^{154}$Gd&	0.3&$0^{\circ}$&1264.934&1264.611&0.322	&$^{156}$Dy&	0.25&$0^{\circ}$&1276.499&1276.164&0.335\\
$^{156}$Gd&	0.35&$0^{\circ}$&1279.948&1279.533&0.415&$^{158}$Dy&	0.3&$0^{\circ}$&1292.187&1291.851&0.336\\
$^{158}$Gd&	0.35&$0^{\circ}$&1293.963&1293.516&0.446&$^{160}$Dy&	0.3&$0^{\circ}$&1307.106&1306.771&0.335\\
\hline\hline
\end{tabular}
\end{table*}

\subsubsection{Physical properties of global minima}
\begin{figure}
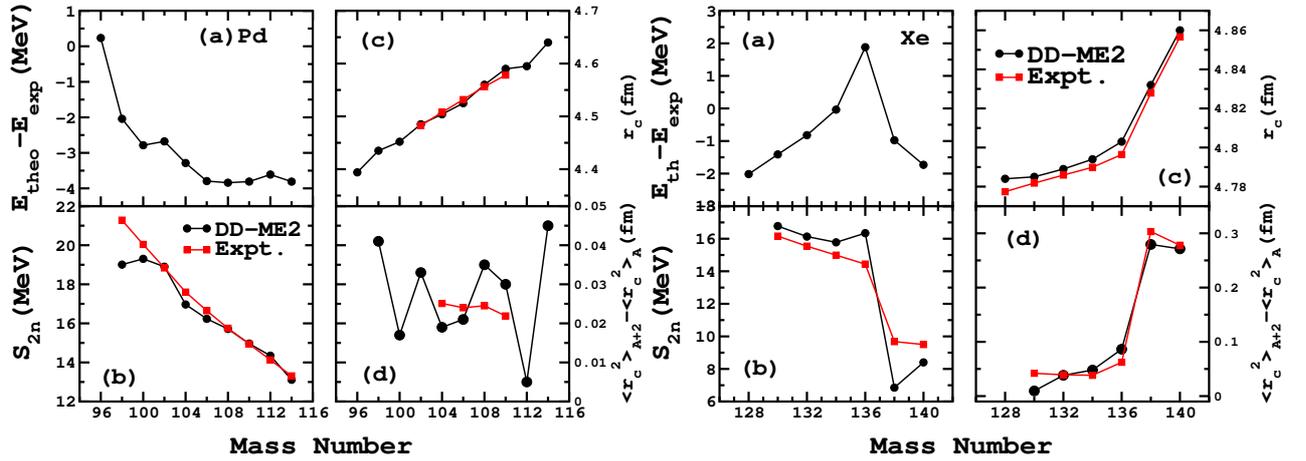

\includegraphics[scale=0.3]{pd.eps}
\includegraphics[scale=0.3]{xe.eps}
\caption{\label{fig:pd}(Color online) For Pd(Z=46) and Xe(Z=54) isotopes (a) Binding energy deviation from experimental data
(b) Two-neutron separation energies $S_{2n}$ (c)  root-mean square charge radii $r_c$,
and (d) isotopic shifts of the ground state charge radii. Comparison is made with the
available experimental data.}
\end{figure}

\begin{figure}
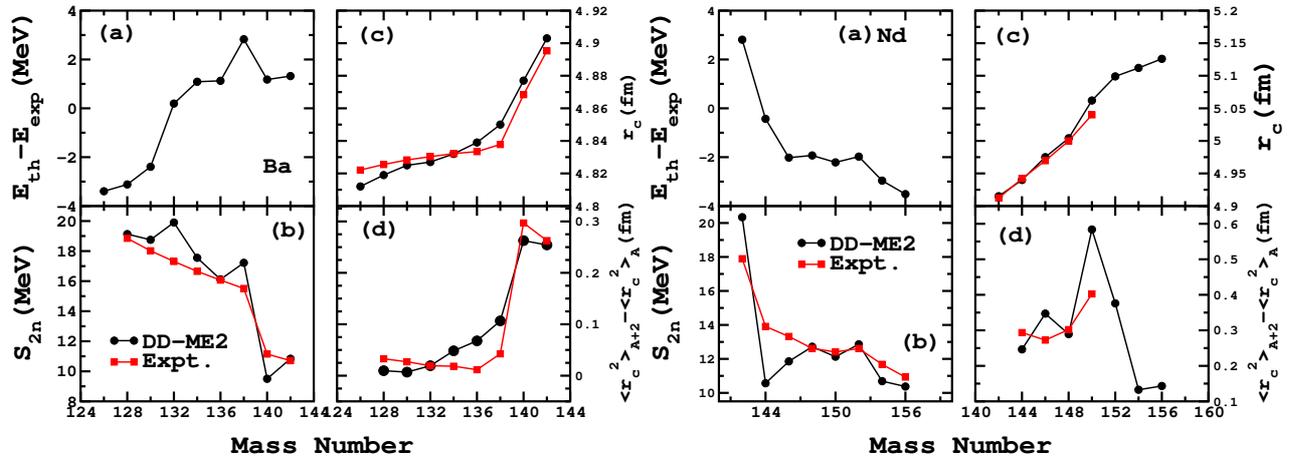

\includegraphics[scale=0.3]{ba.eps}
\includegraphics[scale=0.3]{nd.eps}
\caption{\label{fig:Ba}(Color online) Same as Fig.\ref{fig:pd} but for Ba(Z=56) and Nd(Z=60) isotopes.}
\end{figure}

\begin{figure}
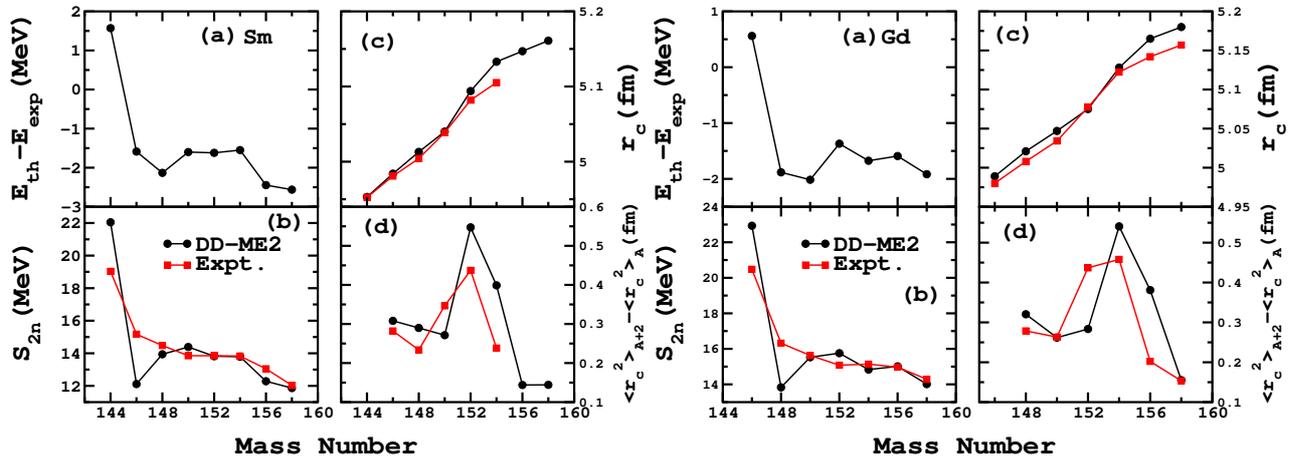

\includegraphics[scale=0.3]{sm.eps}
\includegraphics[scale=0.3]{gd.eps}
\caption{\label{fig:sm}(Color online) Same as Fig.\ref{fig:pd} but for Sm(Z=62) and Gd(Z=64) isotopes.}
\end{figure}

\begin{figure}
\centering
\includegraphics[scale=0.3]{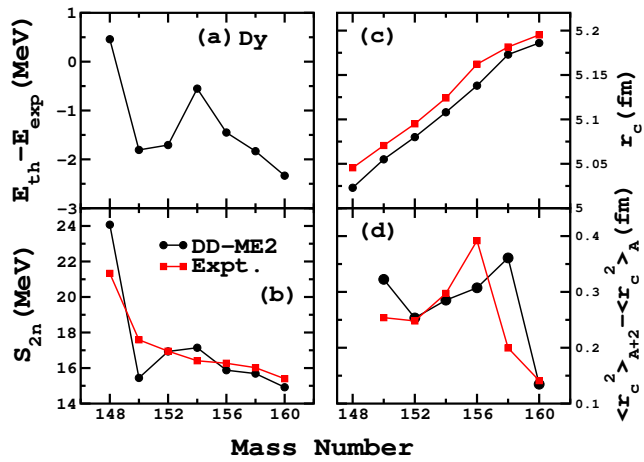}
\caption{\label{fig:dy}(Color online) Same as Fig.\ref{fig:pd} but for Dy(Z=66) isotopes.}
\end{figure}

In Figures \ref{fig:pd}, \ref{fig:Ba}, \ref{fig:sm},  
and \ref{fig:dy}, we present the physical properties such as binding energies, two neutron 
separation energies, charge radii and isotopic shifts of the ground state charge radii. 
These observables can be measured experimentally, and we have compared these observables 
with experimental values.  The binding energies have been presented in Table:\ref{tab:table1} 
and Table:\ref{tab:table2} calculated using axial and triaxial symmetry. In these figures, 
physical properties presented are by the triaxial calculations. From these figures, we can see 
the deviation of the calculated triaxial binding energies with respect to the experimental 
values are around ~0.5MeV for most of the cases to a maximum of around 3.5MeV. In case of Xe, 
ba, Nd, Sm, Gd and Dy isotopes, the sharp kink at N=82 can be seen in two neutron separation energy, 
as N=82 being neutron closed shell number. It is also observed in charge radii as well as in 
the isotopic shifts  of the ground state charge radii. In general, we can say that the 
calculated numerical values as well their behaviour is overall in good agreement with available 
experimental data. 
\subsubsection{E(5) and X(5) Critical Point Symmetry}
The possible candidates of the E(5) critical-point symmetry based on the calculations
assuming axial symmetry can be explored by analyzing the flat regions in the axial PESs.
In our axial calculations, the relatively flat axial PESs exhibited by $^{108,110}$Pd, $^{132,134,138}$Xe, and $^{134}$Ba
as shown in Fig.~\ref{fig:figure1}, it is evident that,
these are transitional nuclei, and can be described assuming infinite square well potential in $\beta$-variable.
These can be the possible candidates for the E(5) critical-point symmetry.
Our findings of $^{108,110}$Pd, $^{132,134,138}$Xe, and $^{134}$Ba nuclei to be possible
E(5) candidates are in agreement with many theoretical and experimental 
studies~\cite{Casten:2000zz,Zhang:2002zu,Fossion:2006xg,RodriguezGuzman:2007tq,Li:2010qu}.
However, in the present calculation, $^{102}$Pd is not showing a flat PES. 
So, it cannot be considered as a possible candidate for E(5).  
This is not in agreement with the earlier predictions~\cite{Zamfir:2002dk,Clark:2004xb,Fossion:2006xg}.
But, our result is in agreement with the very recent experiment based on lifetime measurements
of yrast and non-yrast states of $^{102}$Pd through a Recoil Distance 
Doppler Shift(RDDS)~\cite{Konstantinopoulos:2016his}. 
We also know that, from N=90 nuclei, $^{150}$Nd, $^{152}$Sm, $^{154}$Gd and $^{156}$Dy have been
predicted and identified to exhibit the best possible candidates for X(5) 
critical-point symmetry~\cite{ca3,mac,za,cl,rb,dlz0,r.kr,Gupta:2017fyl,dt,dewald,Moller:2006yx}. 
The X(5) symmetry corresponds to first order phase transition
from a spherical shape to a well-deformed prolate ($\gamma$-unstable) shape.
The present observations from the axial PESs for $^{142-156}$Nd, $^{144-158}$Sm, $^{146-158}$Gd, 
and $^{148-160}$Dy isotopes shown in Fig.~\ref{fig:figure3}, we can notice the possible candidates 
for X(5) critical-point symmetry. These are $^{150}$Nd, $^{150,152}$Sm, $^{154}$Gd and $^{156}$Dy. 
The present results agree qualitatively with earlier theoretical~\cite{Fossion:2006xg,RodriguezGuzman:2007tq,jm,jyz}, 
and experimental studies~\cite{r.kr,Gupta:2017fyl,dewald,Moller:2006yx}.
 
\begin{figure}[t]
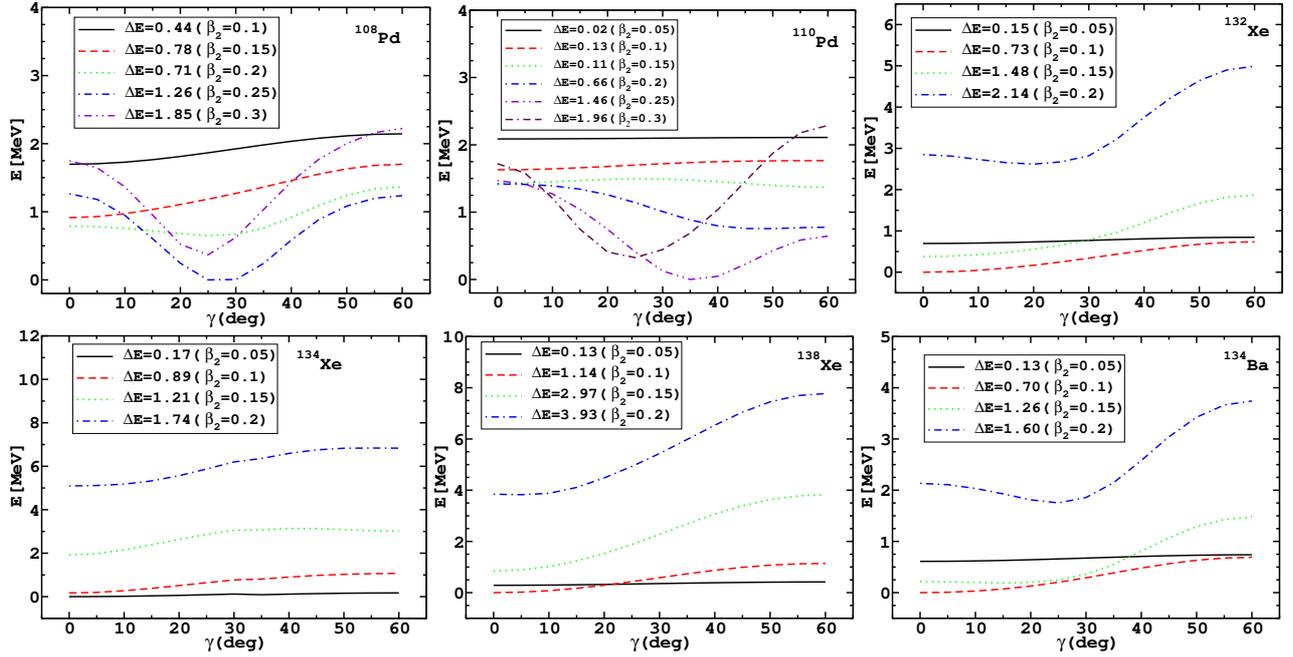

\includegraphics[scale=0.2]{betagama_108pd.eps}  
\includegraphics[scale=0.2]{betagama_110pd.eps} 
\includegraphics[scale=0.2]{betagama_132xe.eps} 
\includegraphics[scale=0.2]{xe134.eps} 
\includegraphics[scale=0.2]{betagama_138xe.eps}
\includegraphics[scale=0.2]{134ba.eps}
\caption{\label{fig:figure6}(Color online) Binding energy curves of the nuclei $^{108,110}$Pd, $^{132,134,138}$Xe, and $^{134}$Ba
as functions of the deformation parameter $\gamma$, for fixed values of axial deformation($\beta_2$)
with the DD-ME2 parameter set.}
\end{figure}

The dependency on the triaxial parameter($\gamma$) is an important investigation for such cases.
In order to check the properties of $\gamma$-dependence for E(5) symmetry in these nuclei, we have plotted the
energy curves as a function of $\gamma$-variable for fixed values of $\beta_2$. The results are shown in Fig.\ref{fig:figure6}.
Here $\Delta$E is the energy difference between the minimum and maximum energy from $\gamma$ = $0^\circ$ to $\gamma$ = $60^\circ$ for fixed $\beta_2$.  
Out of the possible E(5) candidates of Xe-isotopes, we can see that, $^{132,134}$Xe are showing a weak dependence
on $\gamma$ for 0.05$\leq|\beta_2|\leq$0.2 than $^{138}$Xe. In case of Pd-isotopes, both the $^{108,110}$Pd show weaker $\gamma$-dependence
for 0.05$\leq|\beta_2|\leq$0.3. Similarly, the negligible dependence on the $\gamma$-variable is found in $^{134}$Ba.
Therefore, we can say that, $^{108,110}$Pd, $^{132,134}$Xe, and $^{134}$Ba could be suitable candidates to look
for E(5) critical-point symmetry.
\begin{figure}
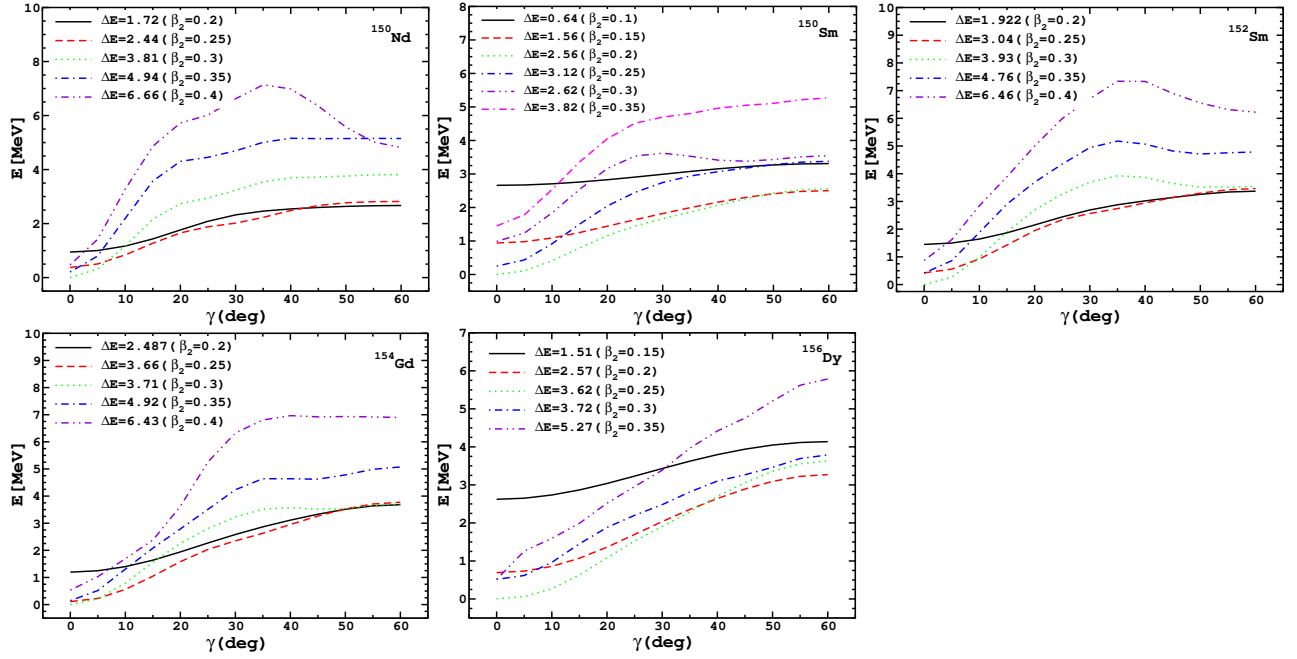

\includegraphics[scale=0.2]{betagama_nd.eps}  
\includegraphics[scale=0.2]{betagama_150sm.eps} 
\includegraphics[scale=0.2]{betagama_sm.eps}  
\includegraphics[scale=0.2]{betagama_gd.eps}  
\includegraphics[scale=0.2]{betagamma_dy.eps}  
\caption{\label{fig:figure66}(Color online) Binding energy curves of the nuclei $^{150}$Nd, $^{150,152}$Sm, $^{154}$Gd, and $^{156}$Dy
as functions of the deformation parameter $\gamma$, for fixed values of axial deformation($\beta_2$)
with the DD-ME2 parameter set.}
\end{figure}
Further, in case of X(5) symmetry, for a more clear picture of the $\gamma$ dependence, 
we have plotted the energy as a function of $\gamma$ for
the fixed values of $\beta_2$ in Fig.~\ref{fig:figure66}.  
A clear indication of strong $\gamma$-dependence can be seen from Fig.~\ref{fig:figure66}. 
However, there is an almost gradual increase in energy with the increase in $\gamma$ for 
$^{156}$Dy up to $\gamma$ = $55^\circ$ above which it remains constant.
From the above discussion, it is evident that in the present calculations, N=90 isotopes
do not show a flat PESs, and $^{150}$Nd, $^{150,152}$Sm, $^{154}$Gd, and $^{156}$Dy are
possible candidates for X(5) critical-point symmetry. Qualitatively, this is in agreement 
with earlier calculations~\cite{Fossion:2006xg,RodriguezGuzman:2007tq,Li:2010qu,jm,zq}.
\subsection{TPSM Results}
\label{TPSM-res}
In order to probe the high-spin properties of Dy, Gd, Sm, Nd, Pd, Xe and Ba isotopes,
TPSM calculations have been performed with the basis deformation
values chosen close to those obtained in the previous section. The
variation in the deformations is justified as the two models employ very different
Hamiltonian and the configuration spaces.  What was noted that the
axial deformations obtained in the previous section, don't require
any major modifications, however, the non-axial deformation in the TPSM analysis
needed readjustments to reproduce the $\gamma$ bandhead energies.  
The deformation values used in the TPSM study are provided in Table \ref{tab:tpsmtab}. 

TPSM study of the spectroscopic properties of atomic nuclei proceeds
in several stages. In the first stage, the deformations listed
in Table \ref{tab:tpsmtab} are used to solve the
three-dimensional Nilsson potential. The wave functions of this
potential form the intrinsic basis functions in the TPSM approach.
We would like to mention that, in principle, the basis functions with 
arbitrary deformation values can be used.
However, optimum deformation values, which are close to the
expected deformation of the system, are employed so that a small window
of the basis states are chosen for the diagonalization of the shell
model Hamiltonian. In the present work, an energy window containing about
40 basis states has been selected for all the nuclei studied.
\begin{table}[htb]
\centering
\caption{\label{tab:tpsmtab} Axial and triaxial quadrupole deformation parameters
$\beta$ and $\gamma$ employed in the TPSM calculation
for $^{156}$Dy,$^{154}$Gd, $^{150,152}$Sm $^{150,156}$Nd $^{102,108}$Pd, $^{132,134}$Xe and $^{134}$Ba isotopes.}
\resizebox{0.75\textwidth}{!}{
\begin{tabular}{cccccccccccc}
\hline\hline
  \textrm{} & 
\textrm{$^{156}$Dy} &
\textrm{$^{154}$Gd} & 
\textrm{$^{150}$Sm}&
\textrm{$^{152}$Sm}&
\textrm{$^{150}$Nd} & 
\textrm{$^{154}$Nd}&
\textrm{$^{102}$Pd}&
\textrm{$^{108}$Pd}& 
\textrm{$^{132}$Xe}&
\textrm{$^{134}$Xe}&
\textrm{$^{134}$Ba}\\
\hline
 $\beta$ & 0.29 &  0.3 &    0.20    &  0.30  & 0.30    & 0.30     & 0.20  &0.25  & 0.10 &0.10  &0.10  \\
 $\gamma$&20     & 19   &    22       &  19      & 20       & 19       & 20 &22  &20  & 20 &20\\
\hline\hline
\end{tabular}}
\end{table}

\begin{figure}
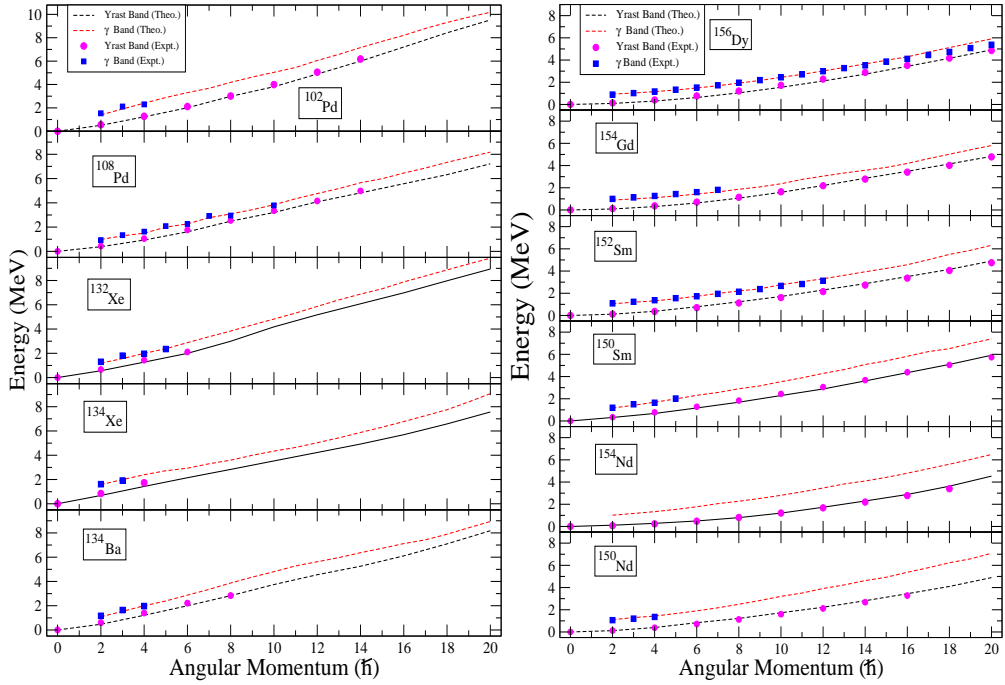

\centering
\includegraphics[width=6.5cm,height=9.cm]{tabasum_tpsm222_kashmiruni.eps} 
\includegraphics[width=6.5cm,height=9.cm]{tabasum_tpsm_kashmiruni.eps}  
\caption{\label{fig:figure7}(Color online) Comparison of experimental and the calculated band
energies for Pd, Xe, Ba, Dy, Gd, Sm, and Nd-isotopes. (Data  taken from Refs.~~\cite{102}-\cite{156}.)}
\end{figure}

In the second stage, the deformed basis states are projected onto good
angular momentum states using the explicit three-dimensional angular momentum
projection technique. In the third and final stage, these projected
states are used to diagonalize the shell model given in section-\ref{TPSM}.
%
Calculated energies
for yrast and the $\gamma$ bands for $^{156}$Dy,$^{154}$Gd, $^{150,152}$Sm
$^{150,156}$Nd $^{102,108}$Pd, $^{132,134}$Xe and $^{134}$Ba  are compared with the available
experimental data in Fig.~\ref{fig:figure7}. It is quit evident from
the results that TPSM calculations reproduce the experimental energies
reasonably well. 

\begin{figure}
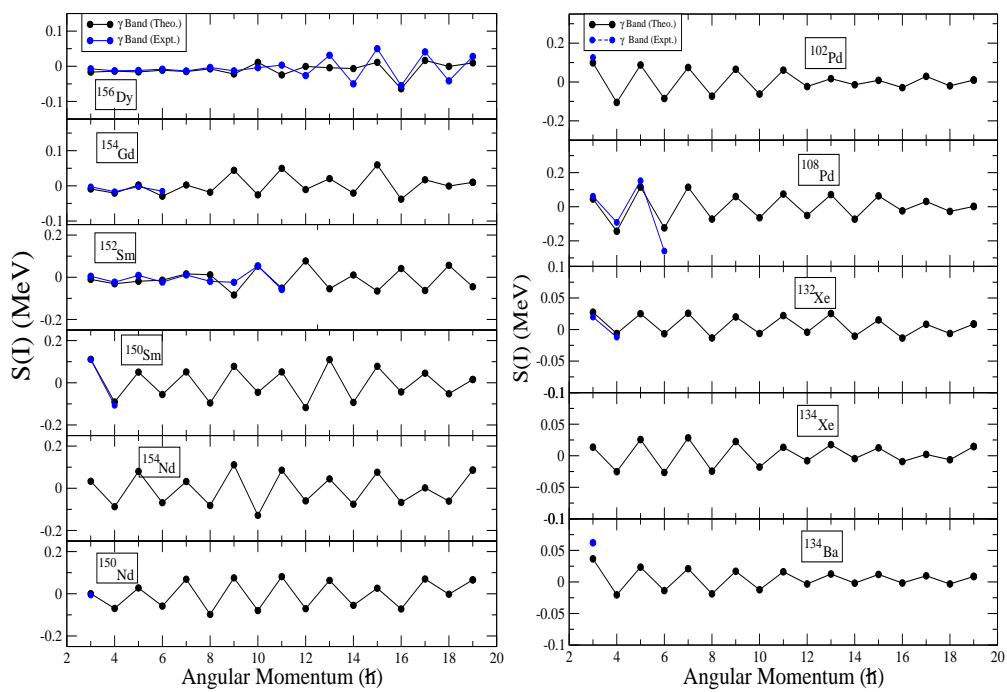

\centering
\includegraphics[width=6.5cm,height=9.cm]{stag_amu1.eps}  
\includegraphics[width=6.5cm,height=9.cm]{stag_amu2.eps} 
\caption{\label{fig:figure992}(Color online) Comparison of observed and TPSM
calculated staggering parameter~\ref{eq:staggering} for the $\gamma$-band
in $^{156}$Dy~\cite{156}, $^{154}$Gd~\cite{154}, $^{150,152}$Sm~\cite{150,152}, 
$^{150,154}$Nd~\cite{150,154}, $^{102,108}$Pd~\cite{102,108},
$^{132,134}$Xe~\cite{132,134}, and $^{134}$Ba~\cite{134}.}
\end{figure}

In order to explore whether nuclei, under study, are $\gamma$-rigid or soft,
odd-even staggering parameter defined as :
\begin{equation}\label{eq:staggering}
 S(I)= \frac{E(I)-\left(E(I-1)+E(I+1)\right)/2}{E(2^{+}_1)},
\end{equation}
is plotted in Fig.~\ref{fig:figure992} for the $\gamma$ bands. 
It is known that phenomenological Davydov-Filippov and Wilets-Jean potentials belonging to 
$\gamma$-rigid and $\gamma$-soft limits, respectively, give rise to similar excitation 
spectra for the ground-state band \cite{AS58,LW56}. Hence, it is not possible to separate the 
two limiting cases from the ground-state properties.  However, it has been demonstrated 
that energy staggering, $S(I)$, in the $\gamma$-band  may provide information on the nature
of the $\gamma$-motion. It is shown that \cite{nvz,eam} in case of $\gamma$-rigid, odd-spin 
values are favored as compared to the even-spin members and for $\gamma$-soft case it is opposite. 
It is evident from the two staggering figures that in all
the nuclei, except for $^{152}$Sm, even-spin states are lower than the
odd-spin states, implying that all these nuclei are $\gamma$-soft.
In the case of $^{152}$Sm, the staggering parameter is quite small 
for low-spin states and it is not possible to make any statement
regarding $\gamma$ nature of these states. However, for high-spin states beyond I=8,
odd-spin states are lower than even-spin states, indicating that this 
nucleus at high-spin has $\gamma$-rigid character.

\begin{figure}
\centering
\includegraphics[width=6.5cm,height=11.cm]{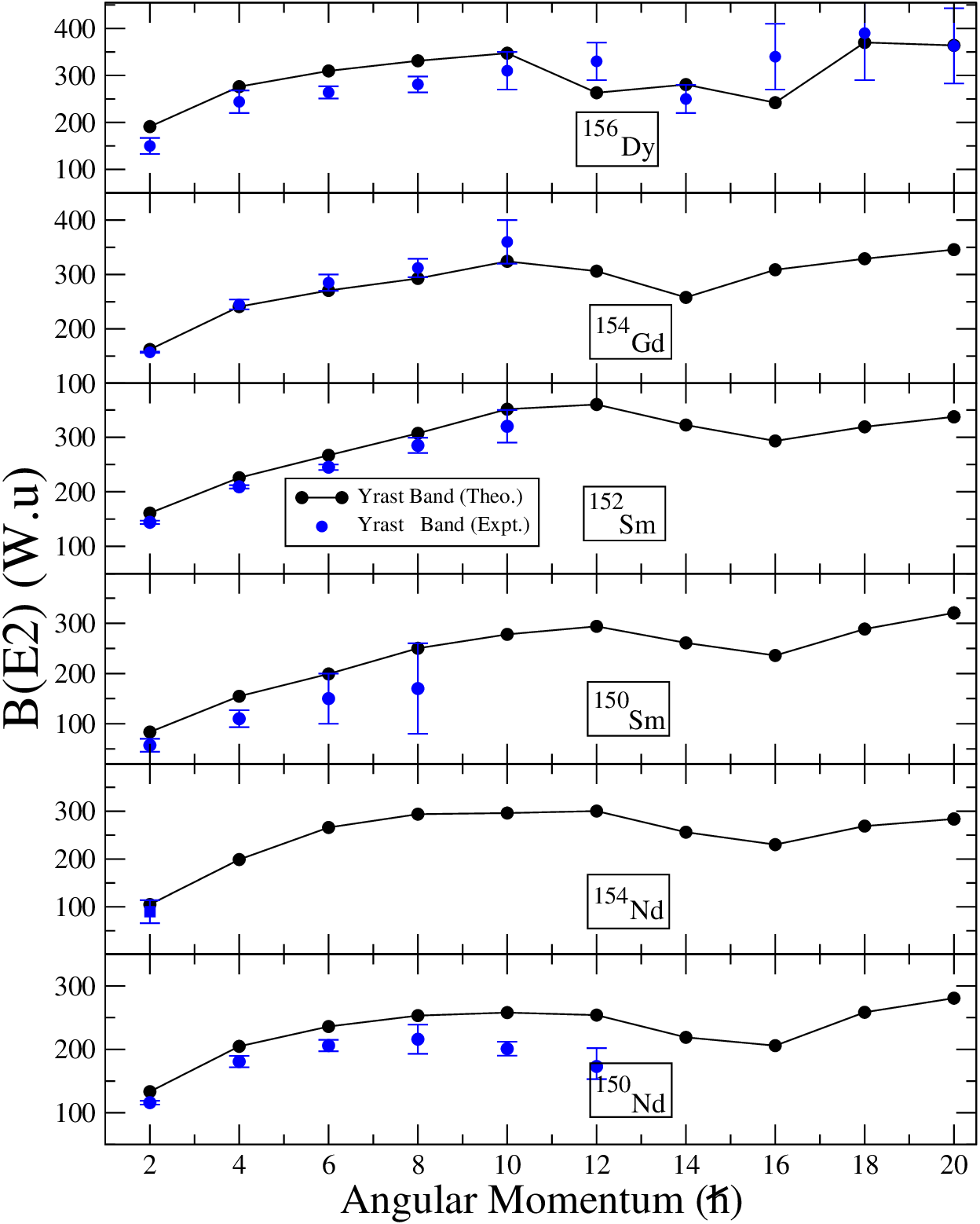}  
\includegraphics[width=6.5cm,height=11.cm]{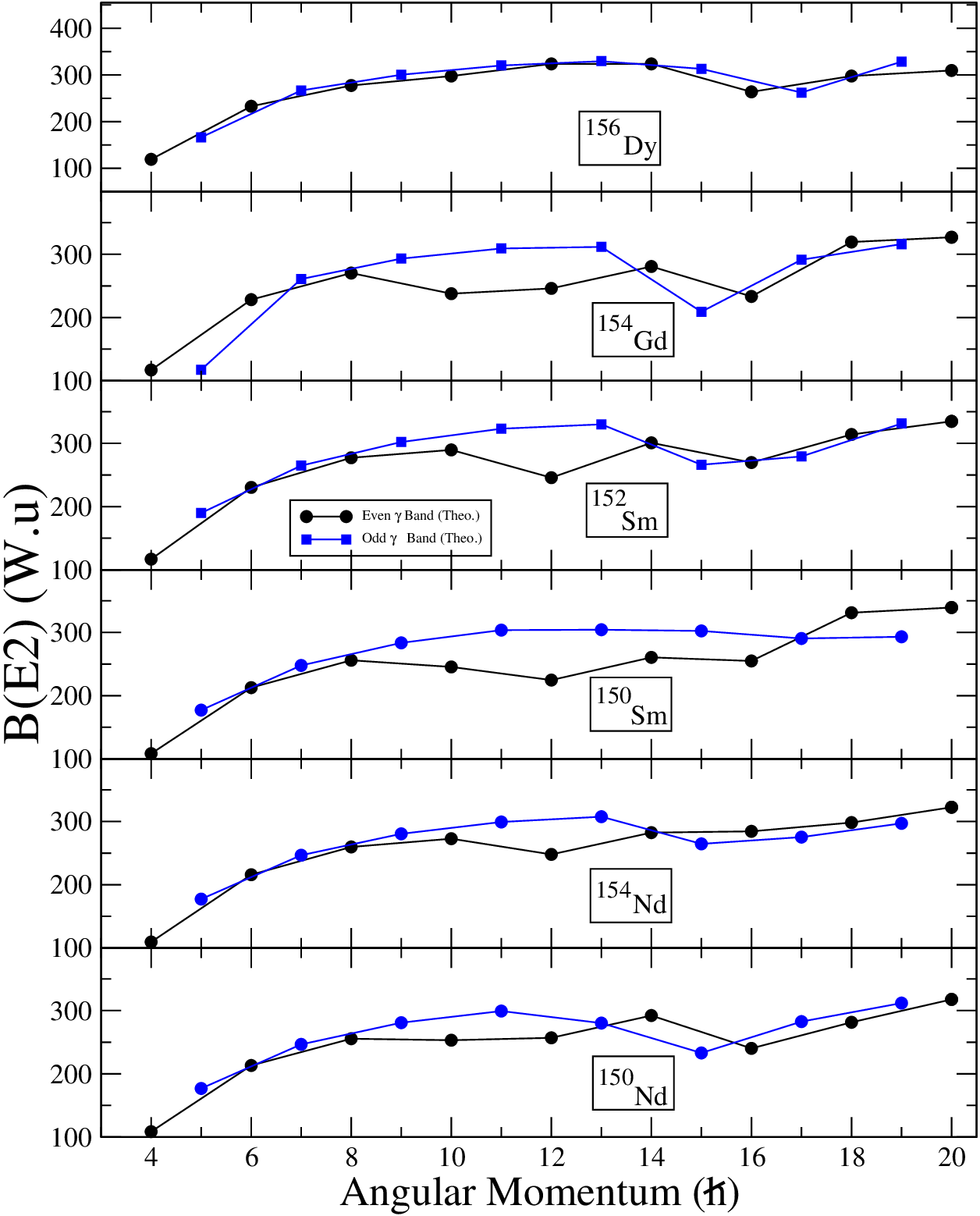}
\caption{\label{figure99}(Color online) Calculated B(E2) transition probabilities for
the ground state bands (left panel) and $\gamma$-bands (right panel) using TPSM
approach. The experimental values have been taken from Ref.~~\cite{102}-\cite{156}.}
\end{figure}

\begin{figure}
\centering
\includegraphics[width=6.5cm,height=9.cm]{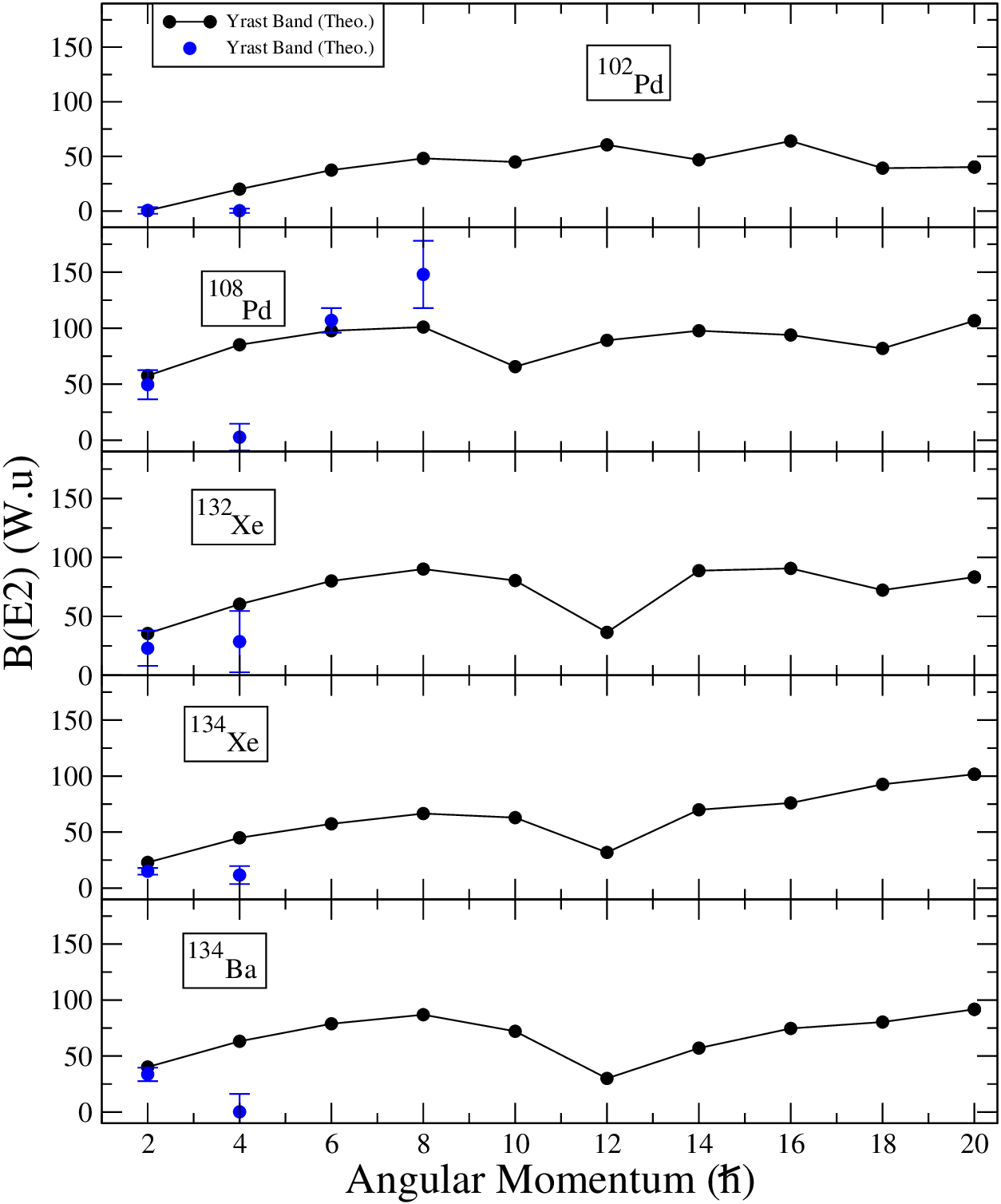}  
\includegraphics[width=6.5cm,height=9.cm]{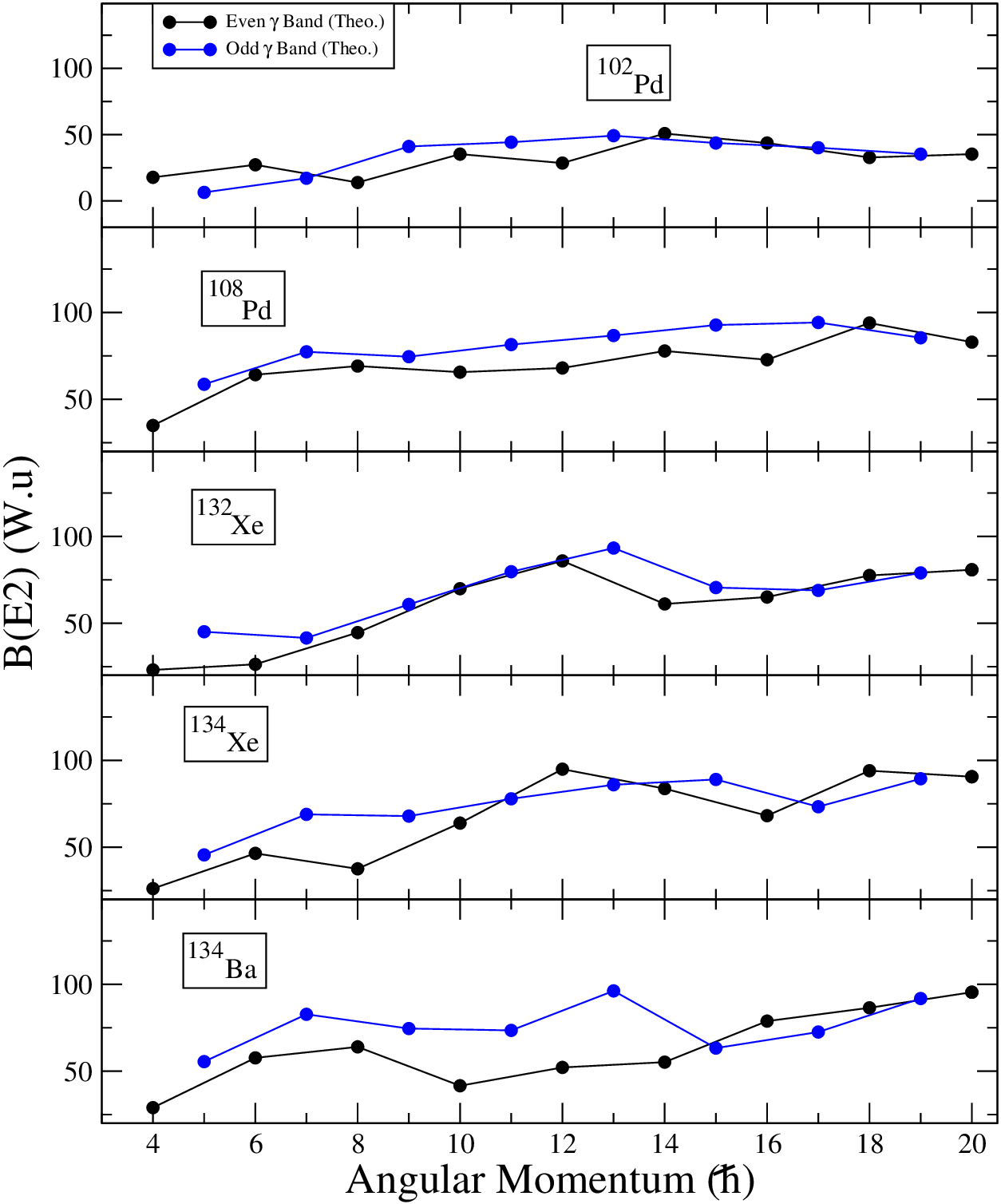} 
\caption{\label{figure991}(Color online) Calculated B(E2) transition probabilities for
the ground state bands (left panel) and $\gamma$-bands (right panel) using TPSM
approach. The experimental values have been taken from Refs.~\cite{102}-\cite{156}.}
\end{figure}

The electromagnetic transition probabilities  have also been studied in the present
work using the expressions published in the earlier work
\cite{ja}. The calculated BE2 transitions for the yrast and the $\gamma$ bands
are displayed in Figs.~\ref{figure99} and ~\ref{figure991}. In these figures, the available
measured have also been plotted for a comparison. In most of the nuclei, a drop in BE2 is
observed between I=12 -16, and is due to the rotation alignment of
two-neutrons in the $i_{13/2}$ orbital. BE2 transitions for the
$\gamma$-band also depict structural changes due to the crossing of
the $\gamma$-band based on aligned two-quasiparticle configuration
as discussed in our earlier work \cite{ds}. However, there is no
experimental data available to compare with our theoretical predictions.
\section{Conclusion}
\label{conclusion}
In the present calculations, we have investigated PESs for 
transitional nuclei $^{96-114}$Pd, $^{128-140}$Xe, and $^{126-138}$Ba, $^{142-156}$Nd, 
$^{144-158}$Sm, $^{146-158}$Gd and $^{148-160}$Dy. In this paper, we have searched
for the structural evolution with the increase of number of neutrons and protons within the isotopic/isotonic chains.
We have also searched for the possible candidates that exhibit E(5) and X(5) critical-point symmetry behaviour
at the critical point of the shape transition.
The self-consistent RHB formalism with DD-ME2 force has been used.
The analysis of the structural evolution is done on the basis of the evolution of the ground state 
shapes located within the axial potential energy surface calculations. It is further
boosted by the triaxial calculations.
The shape transition from spherical to deformed shape manifest themselfs in a very clear manner in almost
all the isotopic chains.Shape coexistence as well as the triaxial character is also observed.
The calculations of the properties corresponding to the triaxial global minimum 
reproduces the experimental data very well. Based on the behaviour of the axial  
potential energy surface and further supported by the $\gamma$-dependence with the 
triaxial calculations, we have found that, $^{108,110}$Pd, $^{132,134}$Xe and $^{134}$Ba 
are suitable candidates to look for E(5) critical-point symmetry which have been suggested by earlier studies as examples of E(5) symmetry.
We do not find $^{102}$Pd as a possible candidate for E(5),   
contradicting some earlier predictions~\cite{Zamfir:2002dk,Clark:2004xb,Fossion:2006xg},
but, in agreement with the very recent experiment for $^{102}$Pd through Recoil Distance Doppler Shift(RDDS)~\cite{Konstantinopoulos:2016his} measurement. 
The isotopes $^{150}$Nd, $^{152}$Sm and $^{154}$Gd are found to be good candidates while 
$^{150}$Sm and $^{156}$Dy are poor candidates of X(5) critical-point symmetry.  
Triaxial projected shell model has also been employed to study band structures
in this work. The TPSM calculation for yrast and
$\gamma$-band energy produces the experimental results. Through the analysis of the $\gamma$-band
and the staggering, it has been shown that all the nuclei, except
for $^{152}$Sm are $\gamma$-soft at high spin. 
\section{Acknowledgement}
S. Ahmad would like to thank the Inter-University Accelerator 
Centre (IUAC), New Delhi for providing the High Performance Computing 
Facility (HPC) at IUAC.

\end{document}